\documentclass{article}
\usepackage{amsmath}
\usepackage{amssymb}

\newtheorem{theorem}{Theorem}

\newtheorem{axiom}[theorem]{Axiom}

\newtheorem{claim}[theorem]{Claim}
\newtheorem{conclusion}[theorem]{Conclusion}

\newtheorem{definition}[theorem]{Definition}

\newtheorem{notation}[theorem]{Notation}

\newtheorem{remark}[theorem]{Remark}

\input{tcilatex}

\begin{document}

\bigskip

\bigskip

\bigskip

\bigskip

\bigskip

\subsection{\protect\bigskip\ \textbf{Bimetric Theory of
Gravitational-Inertial Field in Riemannian and in Finsler-Lagrange
Approximation.}}

\ \ \ \ \ \ \ \ \ \ \ \ \ \ \ 

\ \ \ \ \ \ \ \ \ \ \ \ 

\ \ \ \ \ \ \ \ \ \ \ \ \ \ \ \ \ \ \ 

\ \ \ \ \ \ \ \ \ \ \ \ \ \ \ \ \ J.Foukzon$^{1}$,S.A.Podosenov$^{2}$,
A.A.Potapov$^{3}$, E.Menkova$^{4}$.

\bigskip\ \ \ \ \ \ \ \ \ \ \ \ \ \ \ \ \ \ \ \ \ \ \ \ \ \ \ \ \ \ \ \ \ \
\ \ \ \ jaykovfoukzon@list.ru

$\bigskip $

$^{1}$Israel Institute of Technology.

$^{2}$All-Russian Scientific-Research Institute

\ \ for Optical and Physical Measurements.

$^{3}$IREE RAS.

$^{4}$All-Russian Scientific-Research Institute

\ \ for Optical and Physical Measurements.

\ \ \ \ \ \ \ \ \ \ \ \ \ \ \ \ \ \ \ \ \ \ \ \ \ \ \ \ \ \ \ \ \ \ \ \ \ \
\ \ \ \ \ \ \ \ \ \ \ \ \ \ \ \ \ \ \ \ \ \ \ \ \ \ \ \ \ \ \ \ \ \ \ \ \ \
\ \ \ \ \ \ \ \ \ \ \ \ \ \ \ \ \ \ \ \ \ \ \ \ \ \ \ \ \ \ \ \ \ \ \ \ \ \
\ \ \ \ \ \ \ \ \ \ \ \ \ \ \ \ \ \ \ \ \ \ \ \ \ \ \ 

\ \ \ \ \ \ \ \ \ \ \ \ \ \ \ \ \ \ \ \ \ \ \ \ \ \ \ \ \ \ \ \ \ \ \ \ \ \
\ \ \ \ \ \ \ \ \ \ \ \ \ 

\ \ \ \ \ \ \ \ \ \ \ \ \ \ \ \ \ \ \ \ \ \ \ \ \ \ \ \ \ \ \ \ \ \ \ \ \ \
\ \ \ \ \ \ \ \ \ \ \ \ \ \ \ \textbf{Abstract}

\ \ \ \ \ \ \ \ \ \ \ \ \ \ \ \ \ \ \ \ \ \ \ \ \ \ \ \ \ \ \ \ \ \ \ \ \ \
\ \ \ \ \ \ \ \ \ \ \ \ \ \ \ \ \ \ \ \ \ \ \ \ \ \ \ \ \ \ \ \ \ \ \ \ \ \
\ \ \ \ \ \ \ \ \ \ \ \ \ \ \ \ \ \ \ \ \ \ \ \ \ \ \ \ \ \ \ \ \ \ \ \ \ \
\ \ \ \ \ \ \ \ \ \ \ \ \ \ \ \ \ \ \ \ \ \ \ \ \ \ \ \ \ \ \ \ \ \ \ \ \ \
\ \ \ \ \ \ \ \ \ \ \ \ \ \ \ \ \ \ \ \ \ \ \ \ \ \ \ \ \ \ \ \ \ \ \ \ \ \
\ \ \ \ \ \ \ \ \ \ \ \ \ \ \ \ \ \ \ \ \ \ \ \ \ \ \ \ \ \ \ \ \ \ \ \ \ \
\ \ \ \ \ \ \ \ \ \ \ \ \ \ \ \ \ \ \ \ \ \ \ \ \ \ \ \ \ \ \ \ \ \ \ \ \ \
\ \ \ \ \ \ \ \ \ \ \ \ \ \ \ \ \ \ \ \ \ \ \ \ \ \ \ \ \ \ \ \ \ \ \ \ \ \
\ \ \ \ \ \ \ \ \ \ \ \ \ \ \ \ \ \ \ \ \ \ \ \ \ \ \ \ \ \ \ \ \ \ \ \ \ \
\ \ \ \ \ \ \ \ \ \ \ \ \ \ \ \ \ \ \ \ \ \ \ \ \ \ \ \ \ \ \ \ \ \ \ \ \ \
\ \ \ \ \ \ \ \ \ \ \ \ \ \ \ \ \ \ \ \ \ \ \ \ \ \ \ \ \ \ \ \ \ \ \ \ \ \
\ \ \ \ \ \ \ \ \ \ \ \ \ \ \ \ \ \ \ \ \ \ \ \ \ \ \ \bigskip In present
article the original proposition is a generalization of \ the Einstein's
world tensor $g_{ij}$ by the introduction of pure inertial field tensor $%
\breve{g}_{ij}$ such that $R_{\mu \nu \lambda }^{\alpha }\left( \breve{g}%
_{ij}\right) \neq 0$. Bimetric\ \ theory of gravitational-inertial field\ is
considered\ for the\ case \bigskip when the gravitational-Inertial field is
governed by either a perfect magnetic fluid.

\bigskip

\bigskip\ \ \ \ \ \ \ \ \ \ \ \ \ \ \ \ \ \ \ \ \ \ \ \ \ \ \ \ \ \ \ \ \ \
\ \ \ \ \ \ \ \ \ \ \ \ \textbf{Contents\ \ \ }

\bigskip

\begin{itemize}
\item \textbf{I.Introduction.}

\item \textbf{I.1.GTR in Riemannian approximation.}

\item \textbf{I.2.GTR in Finsler-Lagrange approximation (GTRFL).}

\item \textbf{I.3.Theory of Gravitational-Inertial field with necessity}

\item \textbf{admit Einstein's "Strong Equivalence Principle" (SEP)}

\item \textbf{I.4.Bimetric theory of gravitational-inertial field in
Riemannian}

\item \textbf{\ approximation.}

\item \bigskip \textbf{II.1.Brief review of Rosen's bimetric theory of
gravitation.}

\item \textbf{II.2.Brief review of Davtyan's one-metric theory of \
gravitational-}

\item \textbf{inertial field.}

\item \textbf{II.2.1.Postulates of Davtyan's one-metric theory of
gravitational-}

\item \textbf{inertial field.}

\item \textbf{II.2.2.Lagrangian density and field equations in Davtyan's
theory }

\item \textbf{of gravitational-inertial field.}

\item \bigskip \textbf{II.2.3.Davtyan's field equations in Einstein
pproximation.}

\item \textbf{II.2.4.Weak field limit in Davtyan's theory of
gravitational-inertial }

\item \textbf{field.}

\item \textbf{III.Variational action principles in Rozen's bimetric theory
of }

\item \textbf{gravitation.}

\item \bigskip \textbf{III.1.Simple variational action principle in Rozen's
bimetric }

\item \textbf{theory.}

\item \textbf{III.2.General variational action principle in Rozen's bimetric 
}

\item \textbf{theory.}

\item \textbf{IV.Bimetric theory of gravitational-inertial field in
Riemannian}

\item \textbf{approximation.}

\item \textbf{IV.1.Simple variational action principle in bimetric theory of}

\item \textbf{\ gravitational-inertial field.}

\item \bigskip \textbf{IV.2.The weak field limit.}

\item \textbf{IV.3.General variational action principle in bimetric theory of%
}

\item \textbf{gravitational-inertial field.}

\item \textbf{IV.4.Linear post-Newtonian approximation to} \textbf{bimetric
theory }

\item \textbf{of gravitational-inertial field.Gravitational-Inertial }

\item \textbf{electromagnetism (GIEM).}

\item \textbf{IV.5. Bimetric theory of gravitational-inertial field in a
purely }

\textbf{inertial field approximation.}

\item \textbf{V}.\textbf{4.1.Linear Post-Newtonian Approximation to} \textbf{%
Bimetric }

\item \textbf{Theory of Gravitational-Inertial Field. Gravito-Inertial }

\item \textbf{Ectromagnetism in a purely inertial approximation.}

\item \textbf{IV.5. Bimetric theory of gravitational-inertial field in a
purely }

\textbf{inertial approximation.}

\item \textbf{IV.5.1.Bimetric theory of gravitational-inertial field in a
purely }

\textbf{inertial field approximation. Rosen type approximation.}

\item \textbf{IV.5.2.Bimetric theory of gravitational-inertial field in a
purely }

\textbf{inertial field approximation. Einstein type approximation.}

\item \textbf{IV.5.3. Gravitational-inertial black hole in a purely inertial
field approximation. Einstein type approximation.}

\item \textbf{V.Noninertial purely accelerated frame in bimetric theory of}

\textbf{gravitational-inertial \ field.}

\item \textbf{V.1.Holland's type comovin frame.}

\item \textbf{V.2.Bravais type relativistic comovin frame with curvature }

\textbf{and torsion. }

\item \bigskip \textbf{VI.Gauge theories} \textbf{of accelerated comovin
frame.}

\item \textbf{VI.1.Gravity type gauge theories} \textbf{of accelerated
comovin }

\item \textbf{frame.}

\item \textbf{VI.2. Charged Particles as defects In Bimetric Lorentzian }

\item \textbf{Manifold.}

\item \textbf{VI.3.Gravity type gauge theories} \textbf{of accelerated
comovin }

\item \textbf{frame formed by elastic media with defects.}

\item \textbf{VII.Bimetrical interpretation some exact solutions of the }

\textbf{Einstein field equations.}

\item \textbf{VII.1. General consideration.}

\item \textbf{VII.2. Bimetrical interpretation of the Shvartzshild solution.}
\end{itemize}

\bigskip

\bigskip 

\bigskip

\subsection{I.Introduction}

\bigskip

\bigskip \bigskip

\subsection{I.1.GTR in Riemannian Approximation.}

\bigskip

General theory of Relativity (GTR) in Riemannian approximation to proceed
from assumptions :

\begin{itemize}
\item (\textbf{I}) One-metric geometric structures of the space-time
continuum on the standard assumption of Lorentzianian geometry
\end{itemize}

\bigskip

$\ 
\begin{array}{cc}
\begin{array}{c}
\\ 
ds^{2}=g_{ik}dx^{i}dx^{k},g_{ik}=g_{ki},\det \left\Vert g_{ik}\right\Vert
\neq 0; \\ 
\end{array}
& \text{ \ }(1.1.1)%
\end{array}%
$

\bigskip

\begin{itemize}
\item (\textbf{II}) Equivalence of gravitational field and space-time metric
tenzor $g_{ik}.$
\end{itemize}

\bigskip

\bigskip

\subsection{I.2.GTR in Finsler-Lagrange Approximation (\textbf{GTRFL)}.}

\begin{itemize}
\item In contemporary literature pure formal a Finslerian-Lagrange extension
of general relativity was many developed [20]-[26]. Any extension of GTR
such that mentioned above based on a Finsler--Lagrange geometry [27]-[28].

Any Finsler geometry defined by a fundamental Finsler function $F(x,y)$,
being homogeneous of type $F(x,\lambda y)=|\lambda |F(x,y)$, for nonzero $%
\lambda \in $ $%
%TCIMACRO{\U{211d} }%
%BeginExpansion
\mathbb{R}
%EndExpansion
$, may be considered as a particular case of Lagrange space when $%
L(x,y)=F^{2}(x,y).$

A differentiable Lagrangian $L(x,y)$, i.e. a fundamental Lagrange function,

is defined by a map $L:(x,y)\in TM\rightarrow L(x,y)\in 
%TCIMACRO{\U{211d} }%
%BeginExpansion
\mathbb{R}
%EndExpansion
$ of class $C^{\infty }$ on $\widetilde{TM}=TM$ $\backslash \{0\}.$A

regular Lagrangian has non-degenerate Hessian

$%
\begin{array}{cc}
\begin{array}{c}
\\ 
^{L}g_{ik}\left( x,y\right) =\dfrac{\partial ^{2}L(x,y)}{\partial
x^{i}\partial y^{k}}, \\ 
\\ 
\text{rank}|^{L}g_{ik}|=n,\det \left\Vert ^{L}g_{ik}\right\Vert \neq 0. \\ 
\end{array}
& \text{ \ }(1.1.2)%
\end{array}%
$

A Lagrange space is a pair $L^{n}=[M,L(x,y)]$ with $^{L}g_{ij}$ being of
fixed signature over $V=$ $\widetilde{TM}$.\ \ The Euler--Lagrange equations

\ \ \ \ \ \ \ \ \ \ \ \ \ \ \ \ \ \ \ \ \ \ \ \ \ \ \ \ \ \ \ \ \ \ \ \ \ \
\ \ \ \ \ 
\end{itemize}

\bigskip\ $\ \ \ \ \ \ \ \ 
\begin{array}{cc}
\begin{array}{c}
\\ 
\dfrac{d}{d\tau }\left( \dfrac{\partial L}{\partial y^{i}}\right) -\dfrac{%
\partial L}{\partial x^{i}}=0 \\ 
\end{array}
& \text{ \ }(1.1.3)%
\end{array}%
$

where $y^{i}\equiv \dfrac{dx^{i}\left( \tau \right) }{d\tau }$ are
equivalent to the \textquotedblleft nonlinear\textquotedblright\ geodetic
equations\bigskip

$%
\begin{array}{cc}
\begin{array}{c}
\\ 
\dfrac{d^{2}x^{a}}{d\tau ^{2}}-2G^{a}\left( x^{k},\dfrac{dx^{b}\left( \tau
\right) }{d\tau }\right) =0 \\ 
\end{array}
& \text{ \ }(1.1.4)%
\end{array}%
$

\bigskip defining paths of a canonical semispray

\bigskip $%
\begin{array}{cc}
\begin{array}{c}
\\ 
S=y^{i}\dfrac{\partial }{\partial x^{i}}-2G^{a}\left( x,y\right) \dfrac{%
\partial }{\partial y^{a}} \\ 
\end{array}
& \text{ \ }(1.1.5)%
\end{array}%
$

\bigskip where

$\ 
\begin{array}{cc}
\begin{array}{c}
\\ 
G^{i}\left( x,y\right) =\dfrac{1}{2}\left( ^{L}g_{ij}\right) \left( \dfrac{%
\partial ^{2}L(x,y)}{\partial x^{k}\partial y^{i}}y^{k}-\dfrac{\partial L}{%
\partial x^{i}}\right) \\ 
\end{array}
& \text{ \ }(1.1.6)%
\end{array}%
$

\bigskip

There exists on $V\simeq \widetilde{TM}$ a canonical N--connection

\bigskip

\bigskip\ $%
\begin{array}{cc}
\begin{array}{c}
\\ 
^{L}N_{j}^{a}=\dfrac{\partial G^{a}\left( x,y\right) }{\partial y^{i}} \\ 
\end{array}
& \text{ \ }(1.1.7)%
\end{array}%
$\ \ 

\ \ \ \ \ 

defined by the fundamental Lagrange function $L(x,y)$, which prescribes
nonholonomic

frame structures [26]: $^{L}\mathbf{e}_{\nu }=\left( ^{L}\mathbf{e}%
_{i},e_{a}\right) $ and $^{L}\mathbf{e}^{\mu }=\left( e^{i},^{L}\mathbf{e}%
_{a}\right) .$One obtain the canonical metric structure

$%
\begin{array}{cc}
\begin{array}{c}
\\ 
^{L}\mathbf{g}=\text{ }\left[ ^{L}g_{ij}\left( x,y\right) \right] \left(
e^{i}\otimes e^{j}\right) +\left[ ^{L}g_{ij}\left( x,y\right) \right] \left[
\left( ^{L}e^{i}\right) \otimes \left( ^{L}e^{j}\right) \right]  \\ 
\end{array}
& \text{ \ }(1.1.8)%
\end{array}%
$

constructed as a Sasaki type lift from $M$ for $^{L}g_{ij}\left( x,y\right)
. $There is a unique canonical $\mathbf{d}$-connection $^{L}\mathbf{D=}(h\
^{L}\widehat{D},v\ ^{L}\widehat{D})$ with the coefficients $\ ^{L}\widehat{%
\Gamma }_{\ \beta \gamma }^{\alpha }=(\ ^{L}\widehat{L}_{\ jk}^{i},\ ^{L}%
\widehat{C}_{bc}^{a})$ computed by formulas

\bigskip

$%
\begin{array}{cc}
\begin{array}{c}
\\ 
\widehat{L}_{\ jk}^{i}=\frac{1}{2}g^{ih}(\mathbf{e}_{k}g_{jh}+\mathbf{e}%
_{j}g_{kh}-\mathbf{e}_{h}g_{jk}), \\ 
\\ 
\widehat{C}_{\ bc}^{a}=\frac{1}{2}%
g^{ae}(e_{b}g_{ec}+e_{c}g_{eb}-e_{e}g_{bc}), \\ 
\end{array}
& \text{ \ }(1.1.9)%
\end{array}%
$\ \ \ \ \ \ \ \ \ \ \ \ \ \ \ \ \ \ \ \ \ \ \ \ \ \ \ \ \ \ \ \ \ \ \ \ \ \
\ \ \ \ \ \ \ \ \ \ \ \ \ \ \ \ \ \ \ 

\bigskip

for the $\mathbf{d}$--metric (1.1.8) with respect to $^{L}\mathbf{e}_{\nu }$
and $^{L}\mathbf{e}^{\mu }.$ All such geometric objects, including the
corresponding to $^{L}\widehat{\Gamma }_{\ \beta \gamma }^{\alpha },\ ^{L}%
\mathbf{g}$ and $^{L}N_{j}^{a}\mathbf{d}$--curvatures

$\ $

\bigskip $\ 
\begin{array}{cc}
\begin{array}{c}
\\ 
\ ^{L}\widehat{\mathbf{\mathbf{R}}}\mathbf{_{\ \beta \gamma \delta }^{\alpha
}=}\left( \ ^{L}\widehat{R}_{\ hjk}^{i},\ ^{L}\widehat{P}_{\ jka}^{i},\ ^{L}%
\widehat{S}_{\ bcd}^{a}\right) , \\ 
\end{array}
& \text{ \ }(1.1.9^{\prime })%
\end{array}%
$

where

\bigskip 

\bigskip\ $%
\begin{array}{cc}
\begin{array}{c}
\\ 
\widehat{R}_{\ hjk}^{i}=\mathbf{e}_{k}\widehat{L}_{\ hj}^{i}-\mathbf{e}_{j}%
\widehat{L}_{\ hk}^{i}+\widehat{L}_{\ hj}^{m}\widehat{L}_{\ mk}^{i}-\widehat{%
L}_{\ hk}^{m}\widehat{L}_{\ mj}^{i}-\widehat{C}_{\ ha}^{i}\Omega _{\ kj}^{a},
\\ 
\\ 
\widehat{P}_{\ jka}^{i}=e_{a}\widehat{L}_{\ jk}^{i}-\widehat{\mathbf{D}}_{k}%
\widehat{C}_{\ ja}^{i},\ \widehat{S}_{\ bcd}^{a}=e_{d}\widehat{C}_{\
bc}^{a}-e_{c}\widehat{C}_{\ bd}^{a}+\widehat{C}_{\ bc}^{e}\widehat{C}_{\
ed}^{a}-\widehat{C}_{\ bd}^{e}\widehat{C}_{\ ec}^{a}, \\ 
\end{array}
& \text{ \ }(1.1.10)%
\end{array}%
$

\bigskip

where all indices $a,b,...,i,j,...$ run the same values and, for instance, $%
C_{\ bc}^{e}\rightarrow $ $C_{\ jk}^{i},...$ Thus any $\mathbf{d}$%
--curvatures$\ ^{L}\widehat{\mathbf{\mathbf{R}}}\mathbf{_{\ \beta \gamma
\delta }^{\alpha }}$ are completely defined by a Lagrange fundamental
function $L(x,y)$ for a nondegerate $^{L}g_{ij}.$Note that such locally
anisotropic configurations are not integrable if $\Omega _{\ kj}^{a}\neq 0,$
even the $\mathbf{d}$--torsion components $\widehat{T}_{\ jk}^{i}=0$ and $%
\widehat{T}_{\ bc}^{a}=0.$

General theory of Relativity (GTR) in Finsler-Lagrange approximation to
proceed from assumptions :

\begin{itemize}
\item \bigskip (\textbf{I}) One-metric Finsler-Lagrange geometric structures 
$^{L}\mathbf{g}$ of the space-time continuum on the standard assumption of \
Finsler-Lagrange geometry given by Eqs.(1.1.8)-(1.1.10). $\ \ \ \ \ \ \ \ \
\ \ \ \ \ \ \ \ \ \ \ \ \ \ \ \ $
\end{itemize}

\bigskip

\begin{itemize}
\item (\textbf{II}) Equivalence of gravitational field and space-time
structures:
\end{itemize}

\bigskip\ $\ 
\begin{array}{cc}
\begin{array}{c}
\\ 
^{L}\mathbf{g}=\text{ }\left[ ^{L}g_{ij}\left( x,y\right) \right] \left(
e^{i}\otimes e^{j}\right) +\left[ ^{L}g_{ij}\left( x,y\right) \right] \left[
\left( ^{L}e^{i}\right) \otimes \left( ^{L}e^{j}\right) \right] , \\ 
\\ 
\widehat{R}_{\ hjk}^{i}=\mathbf{e}_{k}\widehat{L}_{\ hj}^{i}-\mathbf{e}_{j}%
\widehat{L}_{\ hk}^{i}+\widehat{L}_{\ hj}^{m}\widehat{L}_{\ mk}^{i}-\widehat{%
L}_{\ hk}^{m}\widehat{L}_{\ mj}^{i}-\widehat{C}_{\ ha}^{i}\Omega _{\ kj}^{a},
\\ 
\\ 
\widehat{P}_{\ jka}^{i}=e_{a}\widehat{L}_{\ jk}^{i}-\widehat{\mathbf{D}}_{k}%
\widehat{C}_{\ ja}^{i},\ \widehat{S}_{\ bcd}^{a}=e_{d}\widehat{C}_{\
bc}^{a}-e_{c}\widehat{C}_{\ bd}^{a}+\widehat{C}_{\ bc}^{e}\widehat{C}_{\
ed}^{a}-\widehat{C}_{\ bd}^{e}\widehat{C}_{\ ec}^{a}. \\ 
\end{array}
& \text{ \ }(1.1.11)%
\end{array}%
$

\bigskip

\subsection{I.3.\textbf{Theory of Gravitational-Inertional field with
necessity admit Einstein's "Strong Equivalence Principle" (SEP)}}

\ 

\bigskip

When dealing with relativistic theories of gravity one is confronted with
three types of equivalence principles [36]:

\begin{itemize}
\item \bigskip \textbf{1.}The Weak Equivalence Principle (\textbf{WEP}),

\item \textbf{2.}The Einstein Equivalence Principle (\textbf{EEP}), and

\item \textbf{3.}The Strong Equivalence Principle (\textbf{SEP}).

\textbf{WEP: }In a pure geometrical view the WEP states that all test masses
move along geodesics in space-time $\tciLaplace _{\mathbf{Gr}.}=\left(
M,g_{ik}^{\mathbf{Gr}.}\right) $. Test masses are understood to be bodies
with negligible self-energy and therefore with negligible contribution to
space-time curvature \bigskip $R_{ik}^{l}\left( g_{ik}^{\mathbf{Gr}.}\right)
.$
\end{itemize}

\textbf{EEP:}The EEP demands, besides the validity of the WEP, that in local
Lorentz frames the non-gravitational laws of physics are those of special
relativity. The EEP implies that space-time has to be curved,i.e. $%
R_{ik}^{l}\left( g_{ik}^{\mathbf{Gr}.}\right) \neq 0$ and thus is the basic
ingredient of any metric theory of gravity.

\textbf{SEP: }The SEP states, besides the validity of the EEP, the
"universality of free fall for self-gravitating bodies".

Note that one has to be careful with the notion of a freely falling
self-gravitating bodies in an external gravitational field. There is no
rigorous definition for the SEP in relativistic theories of gravity. Because
of non-linearity the split of the metric field into an external and a local
part can only be approximate. For a discussion of the SEP within a
slow-motion weak-field approximation see [39],[40]. For metric theories of
gravity, other than general relativity, it has been found that they
typically introduce auxiliary gravitational fields (e.g. scalar fields) and
thus predict a violation of the SEP see [36],[37].

\bigskip \bigskip

\begin{definition}
2.3.1."Strong Equivalence Principle" (SEP) asserts that any gravitational
field $g_{ik}^{\mathbf{Gr}.}$ cannot be distinguished from a suitably chosen
accelerated reference frame $\tciFourier \left( g_{ik}^{\mathbf{ac.}}\right) 
$ - essentially because we cannot distinguish between the reciprocal cases
of spacetime $\tciLaplace _{\mathbf{Gr}.}=\left( M,g_{ik}^{\mathbf{Gr}%
.}\right) $ accelerating through us (gravity), or our own acceleration
through spacetime [4].
\end{definition}

\begin{itemize}
\item Hence SEP in fact asserts that the gravitational curvature cannot be
distinguished from a suitably chosen curved accelerated reference frame (as
curved Bravais frame or Hollands frame) - essentially because we cannot
distinguish between the reciprocal cases of curved space-time $\tciLaplace _{%
\mathbf{Gr}.}=\left( M,g_{ik}^{\mathbf{Gr}.}\right) ,R_{ik}^{l}\left(
g_{ik}^{\mathbf{Gr}.}\right) \neq 0$ accelerating through us (gravity), or
our own comoving curved space-time $\tciLaplace _{\mathbf{ac}.}=\left(
M,g_{ik}^{\mathbf{ac}.}\right) ,R_{ik}^{l}\left( g_{ik}^{\mathbf{ac}%
.}\right) \neq 0$ as curved Hollands frame or curved relativistic Bravais
frame.
\end{itemize}

\bigskip

However as shown by Fock [4] in fact SEP dos not was used by Einstein in
GTR, but only Einstein's\ "Weak Equivalence Principle" (\textbf{WEP}) was
used by Einstein in GTR.

Recall the \textbf{WEP}: all objects are observed to accelerate at the same
rate in a given gravitational field.Therefore, the inertial and
gravitational masses must be the same for any object.This has been verified
experimentally, with fractional difference in masses $<10^{-11}.$As a
consequence, the effects of gravity and of inertial forces (fictitious
forces associated with accelerated frames) cannot locally be
distinguished.\bigskip

\begin{itemize}
\item 
\begin{remark}
2.3.1.Recall the accelerational fields in canonical GTR (in contrast with
tensor gravitational fields $g_{ik}^{\mathbf{Gr}.}$) was introduced not as
objective physical field but only as fictive tensor fields $g_{ik}^{\mathbf{%
ac.}}$ which may be created by means of an arbitrary choice of coordinates
[1],[4],[35].
\end{remark}
\end{itemize}

But as shown by Fock [1],[4] the equivalence of gravitational fields $%
g_{ik}^{\mathbf{Gr}.}$ and accelerational fields $g_{ik}^{\mathbf{ac.}}$ 
\textit{such that mentioned above} is limited not only to sufficiently small
domaines of space and sufficiently short intervals of time, but generally to
weak and homogeneous fields and slow motions.

\begin{itemize}
\item 
\begin{remark}
2.3.2.Thus one can pointed-out with Fock [4] that SEP is inconsistent with
canonical GTR.
\end{remark}
\end{itemize}

By the way, here one should not confuse the law of equality of inertial and
gravitational masses with the mentioned principle of equivalence. The
mathematical expression of this principle is the possibility of introducing
the locally geodetic coordinate system such that

\bigskip

$%
\begin{array}{cc}
\begin{array}{c}
\\ 
g_{ik,l}=0. \\ 
\end{array}
& \text{ \ }(1.1.12)%
\end{array}%
$

However from this statement a not quite correct conclusion is drawn by Fock
namely since the possibility of introducing of locally-geodetic system is
contained in Riemannian geometry, therefore the pointed-out principle does
not constituate a separate physical hypothesis. Actually the availability of
such a possibility in the Riemannian space-time is not nessessary at all [1].

\bigskip

\begin{definition}
2.3.2.Every gravitational field theory which contained standard assumptions
of \ \textbf{GTR:} (\textbf{I)} and (\textbf{II}) and which is consistent
with \textbf{SEP }given by Def.(2.3.1), refers as \textbf{Theory of} \textbf{%
Gravitational-Inertional field in Riemannian Approximation} (\textbf{GIFTR}).
\end{definition}

\begin{notation}
\bigskip Thus every GIFTR in contrast with GTR to proceed with necessity
from additional assumption related to SEP:
\end{notation}

\bigskip\ 

\bigskip\ $\ \ \ \ \ \ \ \ \ \ \ \ \ \ \ \ \ \ 
\begin{array}{cc}
\begin{array}{c}
\\ 
R_{ikm}^{l}\left( g_{ik}^{\mathbf{ac.}}\right) \neq 0. \\ 
\end{array}
& \text{ \ }(1.1.13)%
\end{array}%
$

\begin{definition}
2.3.3.Every gravitational field theory which contained standard assumptions
of \textbf{GTRFL :} (\textbf{I)} and (\textbf{II}) and which is consistent
with \textbf{SEP}, refers as \textbf{Theory of} \textbf{%
Gravitational-Inertional field in Finsler-Lagrange Approximation }(\textbf{%
GIFTFL}).
\end{definition}

\bigskip

For the first time nontrivial gravitational-inertional field theory was
proposed by Davtyan

[1]-[3]. In contrast to GTR, in Davtyan's theory [1] of the
gravitational-inertial field tensor $\mathbf{g}_{ik}^{\mathbf{GI}}$ is not
related to the pure gravitational field $\mathbf{g}_{ik}^{\mathbf{Gr}}.$In
Davtyan's GIFT the real space-time metric tensor $g_{ik}$ of the
gravitational-inertial field is the metric tensor of the real world
(Universe), which formed by pure gravitational $g_{ik}^{\mathbf{Gr}}$ and $%
g_{ik}^{\mathbf{ac.}}$ pure inertial metric tensors. From the field of this
general metric tensor $g_{ik}^{\mathbf{U}}$ the Riemannian space is also
formed.

\bigskip Davtyan's field equations in general looks [1]:

$%
\begin{array}{cc}
\begin{array}{c}
\\ 
g^{lp}g^{mn}\nabla _{k}\left( \sqrt{-g}g^{ik}g_{np;i}\right) -\nabla
_{i}\left( \sqrt{-g}g^{ik}g_{np;k}\right) = \\ 
\\ 
=8\pi \varkappa k\sqrt{-g}g^{mn}\left( T_{n}^{\text{ }l}-\dfrac{1}{2}\delta
_{n}^{l}T\right) , \\ 
\\ 
k=\dfrac{2}{c^{4}}, \\ 
\end{array}
& \text{ \ }(1.1.14)%
\end{array}%
$

\bigskip

where

\bigskip\ $\ \ \ \ \ \ \ \ \ \ \ \ \ \ \ \ \ \ \ \ \ \ \ \ \ \ \ $

$\ 
\begin{array}{cc}
\begin{array}{c}
\\ 
\nabla _{l}g_{ik}=g_{ik\mathbf{;}l}=g_{ik,l}-\mathbf{\breve{\Gamma}}%
_{il}^{m}g_{mk}-\mathbf{\breve{\Gamma}}_{kl}^{m}g_{im}=0 \\ 
\end{array}
& \text{ \ }(1.1.15)%
\end{array}%
$

\bigskip 

\bigskip and

$\ 
\begin{array}{cc}
\begin{array}{c}
\\ 
\mathbf{\breve{\Gamma}}_{kl}^{i}=\dfrac{1}{2}\mathbf{\breve{g}}^{im}\left( 
\mathbf{\breve{g}}_{mk,l}+\mathbf{\breve{g}}_{ml,k}-\mathbf{\breve{g}}%
_{kl,m}\right) , \\ 
\\ 
\mathbf{\breve{g}}_{mk}\neq g_{mk}. \\ 
\end{array}
& \text{ \ }(1.1.16)%
\end{array}%
$

\bigskip

\begin{conclusion}
2.3.1.Thus in papers [1]-[3] the author proposes that Inertia, like
Gravitation, could be a curved spacetime phenomenon caused by accelerating
motion of matter in the full consent with SEP and in contrast with canonical
GTR.
\end{conclusion}

\bigskip

\begin{conclusion}
\bigskip 2.3.2.However one can pointed-out that:
\end{conclusion}

\begin{itemize}
\item \textbf{1.}The new theory proposed in [1] does not looks as true GIFT
but looks only as some kind of modified Einstein gravity with non metrical
connection $\mathbf{\breve{\Gamma}}_{il}^{m}$ and completely hidden of the
pure inertial field $g_{ik}^{\mathbf{ac.}}$ sector.

\item \textbf{2.}The new theory proposed in [1] explains the origin of the
pure inertial field $g_{ik}^{\mathbf{ac.}}$ as being a curved space-time
phenomenon, with the implication that accelerating matter might influence
the metrical tensor $g_{ik}$ of the real space-time by using of the
canonical energy-momentum tensor of matter and this influence completely
depend only from a small coupling constant $8\pi \varkappa k$ by using the
manner of the canonical GTR.
\end{itemize}

\bigskip\ 

\begin{itemize}
\item \textbf{3.} The gravitational-inertial field equations proposed in
paper [1] in a weak gravitational- inertional field limit admits only
equation of Newtonian gravity and weak pure gravitational wave equations:
\end{itemize}

(a) in a weak stationary field limit corresponds exactly to Newton's
equation of gravitation for continuous distribtion of \ masses i.e.

\bigskip $\ 
\begin{array}{cc}
\begin{array}{c}
\\ 
\Delta \Phi ^{\mathbf{Gr}}\left( x\right) =4\pi \varkappa \rho \left(
x\right) , \\ 
\\ 
\Phi ^{\mathbf{Gr}}=-\varkappa \dint \dfrac{\rho dV}{r}=-\dfrac{\varkappa m}{%
r}; \\ 
\end{array}
& (1.1.17)%
\end{array}%
$

(b) in a weak nonstationary field limit corresponds exactly to wave
equation, an equation of propagation of weak pure gravitational waves $%
h_{lm}^{\mathbf{Gr}}\simeq 0:$

\bigskip\ $%
\begin{array}{cc}
\begin{array}{c}
\\ 
\left( \Delta -\dfrac{1}{c^{2}}\dfrac{\partial ^{2}}{\partial t^{2}}\right)
h_{lm}^{\mathbf{Gr}}=0; \\ 
\end{array}
& (1.1.18)%
\end{array}%
$

\begin{itemize}
\item \textbf{4.} In GIFT proposed in paper [1] for a weak
gravitational-inertial field $g_{ik}\simeq 0$ space-time interval equals
\end{itemize}

\bigskip\ $\ 
\begin{array}{cc}
\begin{array}{c}
\\ 
ds^{2}=\left( 1+\dfrac{2\Phi ^{\mathbf{Gr}}}{c^{2}}\right)
dt^{2}-dx_{1}^{2}-dx_{2}^{2}-dx_{3}^{2}= \\ 
\\ 
\left( 1-\dfrac{2\varkappa m}{c^{2}}\right)
dt^{2}-dx_{1}^{2}-dx_{2}^{2}-dx_{3}^{2}. \\ 
\end{array}
& (1.1.19)%
\end{array}%
$

\bigskip

\begin{conclusion}
2.3.3.Thus one pointed-out that in a weak gravitational-inertial field limit 
$g_{ik}\simeq 0$ GIFT proposed in [1] \ does non admit any pure inertial
field phenomenon which corresponds directly with hidden pure inertial field
sector $g_{ik}^{\mathbf{ac.}}.$
\end{conclusion}

\ \ \ \ \ \ \ \ \ \ \ \ \ \ \ \ \ \ \ \ \ \ \ \ \ \ \ \ \ \ \ \ \ \ \ \ \ \
\ \ \ \ \ \ \ \ \ \ \ \ \ \ \ \ \ \ \ \ 

\bigskip

\begin{definition}
\bigskip 2.3.4. (1) $\Re \left( g_{ik}^{\mathbf{.}}\right) \triangleq
\sum_{l,i,k,m}\left\vert R_{ikm}^{l}\left( g_{ik}\right) \right\vert ,$ (2) $%
\Re ^{\mathbf{Gr}}\left( g_{ik}^{\mathbf{Gr}}\right) \triangleq
\sum_{l,i,k,m}\left\vert R_{ikm}^{l}\left( g_{ik}^{\mathbf{Gr}}\right)
\right\vert $ \ \ \ \ \ 

(3) $\Re ^{\mathbf{ac}}\left( g_{ik}^{\mathbf{ac.}}\right) \triangleq
\sum_{l,i,k,m}\left\vert R_{ikm}^{l}\left( g_{ik}^{\mathbf{ac.}}\right)
\right\vert .$
\end{definition}

\bigskip

\bigskip

\begin{claim}
One can pointed-out that the correct gravitational-inertial field equations
(GIFE) in a weak gravitational-inertional field limit with necessity admits:
\end{claim}

\bigskip

\ \ \ \ \ \ \ \ \ \ \ \ \ \ \ \ \ \ \ \ \ \ \ \ \ \ \ \ \ \ \ \ \ \ \ \ \ \
\ \ \ \ \ \ \ \ \ \ \ \ \ \ \ \ \ \ \ \ \ \ 

\begin{itemize}
\item \textbf{1.} In a weak pure gravitational field limit: $\Re \left(
g_{ik}^{\mathbf{.}}\right) \simeq \Re ^{\mathbf{Gr}}\left( g_{ik}^{\mathbf{Gr%
}}\right) ,g_{ik,l}^{\mathbf{Gr}}\simeq 0,$ GIFE admits equation of
Newtonian gravity and pure weak gravitational wave equations.

\ \ \ \ \ \ \ \ \ \ \ \ \ \ \ \ \ \ \ \ \ \ \ \ \ 

\item \textbf{2.}In a weak pure inertial field limit: $\Re \left( g_{ik}^{%
\mathbf{.}}\right) \simeq $ $\Re ^{\mathbf{ac}}\left( g_{ik}^{\mathbf{ac.}%
}\right) ,g_{ik,l}^{\mathbf{ac.}}\simeq 0,$ GIFE admits equation of
Newtonian inertia, which describe the Newtonian scalar potential of the
Newtonian inertial forces. If $\overrightarrow{F}$ denotes pure non
gravitational force acting on an sufficiently small object $O$ (particle), $%
\overrightarrow{r}$ denotes its position vector in an inertial frame, on can
obtain the Newtonian scalar potential $\Phi ^{\mathbf{ac.}}\left( x\right) $
directly from Newton's law of motion.

\item Let's consider Newtonian inertial forces $\overrightarrow{F}_{\mathbf{%
ac}}$ related by canonical manner to the some electric force $%
\overrightarrow{F}$ experienced by the charged particle in the external
stationary electric field $\overrightarrow{\mathbf{E}}.$ Thus

$%
\begin{array}{cc}
\begin{array}{c}
\\ 
\text{div}\overrightarrow{\mathbf{E}}=4\pi \rho \left( x\right) , \\ 
\\ 
\text{rot}\overrightarrow{\mathbf{E}}=0, \\ 
\\ 
\rho \left( x\right) =q\cdot \delta \left( x\right) . \\ 
\end{array}
& (1.1.20)%
\end{array}%
$

\ \ \ \ 

Substitution $\overrightarrow{\mathbf{E}}=-$grad$\Phi ^{\mathbf{ac.}}\left(
x\right) $ gives \ equation for the corresponding Newtonian scalar potential

\ \ \ \ \ \ \ \ \ \ \ \ \ \ \ \ \ \ \ \ \ \ \ \ \ \ \ \ \ \ \ \ \ \ \ \ \ \
\ \ \ \ \ \ \ \ \ \ \ \ \ \ \ \ \ \ \ \ \ 

$%
\begin{array}{cc}
\begin{array}{c}
\\ 
\Delta \Phi ^{\mathbf{ac}}\left( x\right) =4\pi q\rho \left( x\right) , \\ 
\\ 
\Phi ^{\mathbf{ac}}=-q\dint \dfrac{\rho dV}{r}; \\ 
\end{array}
& (1.1.21)%
\end{array}%
$ \ \ \ \ \ \ \ \ 

\ \ \ \ \ \ \ \ \ \ \ \ \ \ \ \ \ \ \ \ \ \ \ \ \ \ \ \ \ \ \ \ \ \ \textbf{%
\bigskip }

\item \textbf{3. }For any GIFTR in a weak pure inertial field limit: $\Re
\left( g_{ik}^{\mathbf{.}}\right) \simeq $ $\Re ^{\mathbf{ac}}\left( g_{ik}^{%
\mathbf{ac.}}\right) ,g_{ik}^{\mathbf{ac.}}\simeq 0,$ for the case of the
external stationary electric field $\overrightarrow{\mathbf{E}}$ given by
Eq.(1.1.20), space-time interval with necessity equals:
\end{itemize}

\bigskip\ $\ \ \ \ \ 
\begin{array}{cc}
\begin{array}{c}
\\ 
ds^{2}=\left( 1+\dfrac{2\Phi ^{\mathbf{ac}}}{c^{2}}\right)
dt^{2}-dx_{1}^{2}-dx_{2}^{2}-dx_{3}^{2}. \\ 
\end{array}
& (1.1.22)%
\end{array}%
$

\begin{itemize}
\item \textbf{4.}For any GIFT in a weak gravitational-inertial field limit: $%
g_{ik}^{\mathbf{Gr}}\simeq 0,g_{ik}^{\mathbf{ac.}}\simeq 0,$ for the case of
the external stationary electric field $\overrightarrow{\mathbf{E}}$ given
by Eq.(1.1.20), space-time interval with necessity equals:
\end{itemize}

\bigskip\ $\ 
\begin{array}{cc}
\begin{array}{c}
\\ 
ds^{2}=\left( 1+\dfrac{2\Phi ^{\mathbf{ac}}}{c^{2}}+\dfrac{2\Phi ^{\mathbf{Gr%
}}}{c^{2}}\right) dt^{2}-dx_{1}^{2}-dx_{2}^{2}-dx_{3}^{2}. \\ 
\end{array}
& (1.1.23)%
\end{array}%
$

\begin{itemize}
\item \textbf{5.}In a weak pure inertial field limit: $\Re \left( g_{ik}^{%
\mathbf{.}}\right) \simeq $ $\Re ^{\mathbf{ac}}\left( g_{ik}^{\mathbf{ac.}%
}\right) ,g_{ik,l}^{\mathbf{ac.}}\simeq 0,$ any GIFT (in a Finsler-Lagrange
approximation) with necessity admits equation of \ Post-Newtonian inertia,
which describe a sufficiently small relativistic inertial forces related to
external nonstationary electro-magnetic field $\left( \overrightarrow{%
\mathbf{E}},\overrightarrow{\mathbf{H}}\right) $.
\end{itemize}

\bigskip

\bigskip

\subsection{I.4. Bimetric Theory of Gravitational-Inertial Field in
Riemannian Approximation.}

\bigskip

Note that one can construct the Bimetric Theory of Gravitational-Inertial
Field in Riemannian Approximation by using Rosen type bimetric formalism.

Recall the Rosen bimetric gravitational field theory. Rosen [5]-[7] proposed
the bimetric gravitational field theory only with the purpose to remove some
of the unsatisfactory features of the Einstein gravity, in which there exist
two metric tensors at each point of bimetric Lorentzian space-time $\Re =\Re
\left( M,g_{ij},\gamma _{ij}\right) $ viz a physical metric tensor $g_{ij},$%
which described gravitation and the background flat metric $\gamma _{ij}$
which does not interact directly with matter fields and describes the
inertial forces associated with the acceleration of the frame of reference.
Note that in Rosen's theory of gravitation for the derivation\textbf{\ }
gravitational field equations one varies the quantities $g_{\mu \nu }$, not
the quantities $\gamma _{\mu \nu }$, in the variational principle.

\bigskip The gravitational field equations of Rozen reads

\bigskip\ $\ 
\begin{array}{cc}
\begin{array}{c}
\\ 
K_{\alpha \beta }-\dfrac{1}{2}Kg_{\alpha \beta }=-8\pi kT_{\alpha \beta },
\\ 
\\ 
k=\varkappa \sqrt{\dfrac{g}{\gamma }}, \\ 
\\ 
K_{\alpha \beta }=K_{\alpha }^{\beta }=\dfrac{1}{2}\breve{g}^{\mu \nu
}\left( g^{h\alpha }g_{h\beta |\mu }\right) _{|\nu }. \\ 
\end{array}
& \text{ \ }(1.1.24)%
\end{array}%
$

\bigskip Where vertical bar $|$ stands for covariant differentiation with
respect to $\gamma _{\mu \nu }.$

Let's consider a bimetric geometry $\Re _{2}=\Re \left( M,g_{ij},\breve{g}%
_{ij}\right) $ with metrics $g$ and $\breve{g}$ of Lorentzian signature that
define two different ways of measuring angles, distances and volumes on a
manifold $M.$In present article the original proposition is a generalization
of the real world tensor $g_{ij}$ by the introduction of a \textit{non flat}
inertial field tensor $\breve{g}_{ij}$ such that $K=K\left( g_{ij},\breve{g}%
_{ij}\right) \neq 0,$ $\nabla _{g}\breve{g}_{ij}\neq 0$ and $\ \ \breve{R}%
_{\mu \nu \lambda }^{\alpha }=R_{\mu \nu \lambda }^{\alpha }\left( \breve{g}%
_{ij}\right) \neq 0$. The first metric tensor $g_{ij}$ in \textbf{GIFTR},
refers to the curved Lorentzian space-time $\tciLaplace _{g}=\tciLaplace
\left( M,g_{\alpha \beta }\right) $ and describes Gravitational-Inertial
Field. The second metric tensor $\breve{g}_{ij}$ in \textbf{GIFTR}, refers
to the curved Lorentzian space-time $\tciLaplace _{\breve{g}}=\tciLaplace
\left( \breve{g}_{\alpha \beta }\right) $ and describes pure inertial
forces. The Rosen's tipe curvature tenzor $K_{\mu \nu }\left( g_{\alpha
\beta },\breve{g}_{\mu \nu }\right) $\ refers to the curved Lorentzian
space-time $\tciLaplace \left( M,g_{\alpha \beta }\right) $ and describes
pure gravitational field. \ \ \ \ \ \ \ \ \ \ \ \ \ \ \ \ \ \ \ \ \ \ \ \ \
\ \ \ \ \ \ \ \ \ \ \ \ \ 

This demands to use as a Action of the gravitational-inertial \ field the
expression

\bigskip $%
\begin{array}{cc}
\begin{array}{c}
\\ 
\mathbf{S}\left( K,\breve{R}\right) =\left[ \mathbf{S}_{1}\left( K,\breve{R}%
\right) \right] _{\breve{g}}+\left[ \mathbf{S}_{2}\left( \breve{R}\right) %
\right] _{\breve{g}}= \\ 
\\ 
\int \tciLaplace _{1}\left( K\left( g_{\alpha \beta },\breve{g}_{\mu \nu
}\right) \right) \sqrt{-\breve{g}}d^{4}x+\ \int \tciLaplace _{2}\left( 
\breve{R}\left( \breve{g}_{\mu \nu }\right) \right) \sqrt{-\breve{g}}d^{4}x.
\\ 
\end{array}
& \text{ }(1.1.25)%
\end{array}%
$

On the basis of variational principle a system of more general Rozen's tipe
covariant equations of the gravitational-inertial field is obtained:\ \ \ \
\ \ \ \ \ \ \ \ \ \ \ \ \ \ 

\bigskip\ \ \ \ \ \ \ \ \ \ \ \ \ \ \ \ \ \ \ \ \ \ \ \ \ \ \ \ \ \ \ $\ \ \
\ \ 
\begin{array}{cc}
\begin{array}{c}
\\ 
\dfrac{\delta \mathbf{S}\left( K,\breve{R}\right) }{\delta g_{\alpha \beta }}%
=K_{\alpha \beta }-\dfrac{1}{2}Kg_{\alpha \beta }=-8\pi \varkappa \sqrt{%
\dfrac{g}{\breve{g}}}\left[ T_{\alpha \beta }\right] _{g}, \\ 
\\ 
K_{\alpha \beta }=K_{\alpha }^{\beta }=\dfrac{1}{2}\breve{g}^{\mu \nu
}\left( g^{h\alpha }g_{h\beta ||\mu }\right) _{||\nu } \\ 
\\ 
\dfrac{\delta \mathbf{S}_{1}\left( K,\breve{R}\right) }{\delta \breve{g}%
_{\mu \nu }}+\dfrac{\delta \mathbf{S}_{2}\left( K,\breve{R}\right) }{\delta 
\breve{g}_{\mu \nu }}=\breve{\Theta}_{\mu \nu }+\breve{E}_{\mu \nu }=k_{1}%
\left[ \breve{T}^{\mu \nu }\right] _{\breve{g}}, \\ 
\\ 
\breve{\Theta}_{\mu \nu }=\dfrac{\delta \mathbf{S}\left( K,\breve{R}\right) 
}{\delta \breve{g}_{\mu \nu }}, \\ 
\\ 
\breve{E}_{\mu \nu }=\breve{R}_{\mu \nu }-\dfrac{1}{2}\breve{R}g_{\mu \nu },
\\ 
\\ 
\left[ \breve{T}_{;\nu }^{\mu \nu }\right] _{\breve{g}}=\breve{F}^{\mu }. \\ 
\end{array}
& \text{ }(1.1.26)%
\end{array}%
$

\bigskip

Here, a subscripts $g,\breve{g}$ stands for specifying that the labelled
quantity is defined by curved space-time metrics $ds_{1}^{2}=g_{\alpha \beta
}dx^{\alpha }dx^{\alpha }$ and $ds_{2}^{2}=\breve{g}_{\mu \nu }dx^{\mu
}dx^{\nu }$ respectively and $\breve{F}^{\mu }$ denote 4-vector of a pure
nongravitational force and vertical duble-bar $||$ stands for covariant
differentiation with respect to $\breve{g}_{\mu \nu }.$

\bigskip

In the Rosen approximation ($\breve{R}_{\mu \nu \lambda }^{\alpha }\approx
0, $ $k_{1}\left[ \breve{T}^{\mu \nu }\right] _{\breve{g}}\approx 0$) $%
\Theta ^{\alpha \beta }\approx K_{\mu \nu }$ and\ these equations reduce to
the field equations of Rosen: $K_{\mu \nu }-\tfrac{1}{2}Kg_{\mu \nu }\approx
-8\pi kT_{\mu \nu }.$

In the general theory of relativity by means of the new equations gives the
same results as the solution by means of Rosen's equations only in the Rosen
approximation.

However, application of these equations to the standard astrophysical and
cosmologic models coupled with a sufficiently strong electromagnetic field
or another nongravitational fields, gives a result different from that
obtained by Einstein's or Rosen's equations.\ In particular, the solution
gives Kantowski-Sachs model [8],[9] with source cosmic cloud strings coupled
with strong electromagnetic field in contrast with corresponding solution
gives in Rosen's bimetric theory [5]-[7].

In this paper we also propose an nontrivial extension of General Relativity
with noninertial frames $\tciFourier \left[ g\right] $ that experience
space-time to have a metric $g$ different from usual metric of noninertial
frames given in canonical General Relativity.

\bigskip

\subsection{II.1.\textbf{Brief review} of Rosen's Bimetric Theory}

\bigskip

Rosen [1] proposed some simplest type the bimetric gravitational field
theory such that at each point of Lorentzian space-time $\left( \tciLaplace
,g_{ij}\right) $ a flat Lorentzian metric tensor $\gamma _{ij}$ in addition
to the curved Lorentzian metric tensor $g_{ij}.$ Thus at each point of \
Rosen's space-time $\Re =\Re \left( g_{ij},\gamma _{ij}\right) $ there are
two metrics:

\bigskip\ $%
\begin{array}{cc}
\begin{array}{c}
\\ 
ds_{1}^{2}=g_{ij}dx^{i}dx^{j}, \\ 
\end{array}
& (2.1.1)%
\end{array}%
$

and

\bigskip $%
\begin{array}{cc}
\begin{array}{c}
\\ 
ds_{2}^{2}=\gamma _{ij}dx^{i}dx^{j}. \\ 
\end{array}
& \text{ }(2.1.2)%
\end{array}%
$

The first metric tensor $g_{ij}$ in Rosen's theory, refers to the curved
space-time and thus the gravitational field. The second metric tensor $%
\gamma _{ij}$ in Rosen's theory, refers to the flat space-time or space-time
of constant curvature, for example such that: Eq.(2.1.3)

$%
\begin{array}{cc}
\begin{array}{c}
\\ 
ds_{2}^{2}=\left( 1-\dfrac{r^{2}}{a^{2}}\right) dt^{2}- \\ 
\\ 
\dfrac{dr^{2}}{1-r^{2}/a^{2}}-r^{2}\left( d\theta ^{2}+\sin \theta d\phi
^{2}\right) . \\ 
\end{array}
& \text{ \ \ \ }(2.1.3)%
\end{array}%
$

\ 

\begin{remark}
2.1.1.\textbf{\ }Note that in Rosen's theory\textbf{\ }element (2.1.3)
corresponds only to a background space-time of constant curvature, to which
the physical metric reduces in the absence of any kind of energy.\ 
\end{remark}

\ \ \ \ \ \ \ \ \ \ \ \ \ \ \ \ \ \ \ \ \ \ \ \ \ \ \ \ \ \ \ \ \ \ \ \ \ \
\ \ \ \ \ \ \ \ \ \ \ \ \ \ \ \ \ \ \ \ \ \ \ \ \ \ \ \ \ \ \ \ \ \ \ \ \ \ 

The Christoffel symbols formed from $g_{ij}$ and $\gamma _{ij}$ are denoted
by $\left\{ \QDATOP{i}{jk}\right\} $ and $\Gamma _{jk}^{i}$ respectively.
The quantities $\Delta _{jk}^{i}$ are defined via formulae

\bigskip\ $%
\begin{array}{cc}
\begin{array}{c}
\\ 
\Delta _{jk}^{i}=\left\{ \QATOP{i}{jk}\right\} -\Gamma _{jk}^{i}. \\ 
\end{array}
& \text{ \ }(2.1.4)%
\end{array}%
$ \ \ \ \ 

\ \ \ \ \ \ \ \ \ \ \ \ 

\begin{remark}
2.1.2.\textbf{\ }Let $R_{\mu \nu \lambda }^{\alpha }$\textbf{\ }and $S_{\mu
\nu \lambda }^{\text{ }\alpha }$\textbf{\ }be the curvature tensors
calculated from $g_{\mu \nu }$ and $\gamma _{\mu \nu }$ respectively. Note
that In the Rosen's approach as $ds_{2}^{2}=\gamma _{\mu \nu }dx^{i}dx^{j}$
is the flat metric, the curvature tensor is zero $S_{\mu \nu \lambda }^{%
\text{ }\alpha }=0$.
\end{remark}

Now there arise two kinds of covariant differentiation:

(1) $g$-differentiation based on $g_{\mu \nu }$ (denoted by semicolon $(;)$)

$\ \ \ 
\begin{array}{cc}
\begin{array}{c}
\\ 
A_{\mu \nu ;\lambda }=\left( A_{\mu \nu ,\lambda }-\left\{ \QATOP{\alpha }{%
\mu \text{ }\lambda }\right\} A_{\alpha \nu }-\left\{ \QATOP{\alpha }{\nu 
\text{ }\lambda }\right\} A_{\alpha \mu }\right)  \\ 
\end{array}
& \text{ }(2.1.5)%
\end{array}%
$

\bigskip

(2) differentiation based on $\gamma _{ij}$ (denoted by a slash $(|)$)

\bigskip 

$%
\begin{array}{cc}
\begin{array}{c}
\\ 
A_{\mu \nu |\lambda }=\left( A_{\mu \nu ,\lambda }-\Gamma _{\mu \text{ }%
\lambda }^{\alpha }A_{\alpha \nu }-\Gamma _{\nu \text{ }\lambda }^{\alpha
}A_{\alpha \mu }\right) , \\ 
\end{array}
& \text{ \ }(2.1.6)%
\end{array}%
$

\bigskip

where ordinary partial derivatives are denoted by comma $(,)$.

\bigskip The straightforward calculations gives\bigskip\ 

$\ 
\begin{array}{cc}
\begin{array}{c}
\\ 
R_{\mu \nu \lambda }^{\alpha }=-\Delta _{\mu \nu |\lambda }^{\alpha }+\Delta
_{\mu \lambda |\nu }^{\alpha }-\Delta _{\beta \nu }^{\alpha }\Delta _{\mu
\lambda }^{\beta }-\Delta _{\beta \lambda }^{\alpha }\Delta _{\mu \nu
}^{\beta }. \\ 
\end{array}
& \text{ }(2.1.7)%
\end{array}%
$\ \ \ \ 

\ \ 

\bigskip

Hence

$%
\begin{array}{cc}
\begin{array}{c}
\\ 
R_{\mu \nu }=-\Delta _{\mu \nu |\alpha }^{\alpha }+\Delta _{\alpha \mu |\nu
}^{\alpha }-\Delta _{\alpha \beta }^{\alpha }\Delta _{\mu \nu }^{\beta
}-\Delta _{\beta \mu }^{\alpha }\Delta _{\alpha \nu }^{\beta }. \\ 
\end{array}
& \text{ }(2.1.8)%
\end{array}%
$\ \ 

\ \ 

This is the curvature tensor $R_{\mu \nu }$ associated with the curvature
effects of pure gravitation acting in the spacetime.

\bigskip

The geodesic equation in Rosen's bimetric relativity takes the form

\bigskip 

\bigskip\ $%
\begin{array}{cc}
\begin{array}{c}
\\ 
\dfrac{d^{2}x^{i}}{ds}+\Delta _{jk}^{i}\dfrac{dx^{j}}{ds}\dfrac{dx^{k}}{ds}=
\\ 
\\ 
\dfrac{d^{2}x^{i}}{ds}+\left\{ \QDATOP{i}{jk}\right\} \dfrac{dx^{j}}{ds}%
\dfrac{dx^{k}}{ds}-\Gamma _{jk}^{i}\dfrac{dx^{j}}{ds}\dfrac{dx^{k}}{ds}=0.
\\ 
\end{array}
& \text{ }(2.1.9)%
\end{array}%
$

\bigskip

It is seen from Eqs. (2.1.4) and (2.1.9) that $\Gamma _{jk}^{i}$ can be
regarded as describing the flat inertial field because it vanishes by a
suitable coordinate transformation.

\bigskip \bigskip \bigskip \bigskip \bigskip

\subsection{II.2.\textbf{Brief review of} Davtyan's One-Metric Theory of
Gravitational-Inertial Field.}

\bigskip \bigskip \bigskip

Einstein theory of General Relativity (GTR) and Einstein Gravitational field
theory to proceed from assumptions:

\begin{itemize}
\item (\textbf{I}) One-metric geometric structures of the space-time
continuum on the standard assumption of Lorentzianian geometry
\end{itemize}

\bigskip

$\ \ \ \ \ \ 
\begin{array}{cc}
\begin{array}{c}
\\ 
ds^{2}=g_{ik}dx^{i}dx^{k},g_{ik}=g_{ki},\det \left\Vert g_{ik}\right\Vert
\neq 0; \\ 
\end{array}
& \text{ \ }(2.2.1)%
\end{array}%
$

\bigskip

\begin{itemize}
\item (\textbf{II}) From equivalence of gravitational field and spacc-tirne
metric tenzor

$g_{ik}\left( M\right) =g_{ik}\left( x_{1},x_{2},x_{3},,x_{4}\right) ;$

\item (\textbf{III}) From equivalence of \ accelerational field $\gamma
_{ik} $ and flat space-time metric tenzor

$\gamma _{ik}\left( M\right) =\gamma _{ik}\left(
x_{1},x_{2},x_{3},,x_{4}\right) ,$ i.e. $R_{klm}^{i}\left[ \gamma _{ik}%
\right] =0$.
\end{itemize}

\begin{axiom}
\bigskip 2.2.1.\textbf{GTR is based on the following postulates:}
\end{axiom}

\begin{itemize}
\item (\textbf{1}) In nonrelativistic approximation and very far from the
localized masses the metric
\end{itemize}

\bigskip\ \ \ \ \ \ tensor describes a flat space-time, i.e. $R_{klm}^{i}=0$.

\begin{itemize}
\item (\textbf{2}) A sufficiently small domain of Lorentzian space-time is
flat, i.e. $R_{klm}^{i}\approx 0$.

\item \bigskip (\textbf{3}) Homogenious gravitational field and
accelerational field are equivalent.

\item (\textbf{4}) Every accelerational field $g_{ik}^{\mathbf{ac.}}$ is
flat, i.e. $R_{klm}^{i}\left( g_{ik}^{\mathbf{ac.}}\right) =0.$
\end{itemize}

\bigskip

\bigskip Let's consider this postulates in more detail [1].

\textbf{1.}The exsistence of remote masses of this Universe, including the
field masses, will undoubtedly influence metric of the real world and create
a general metrical background in the Universe different from Galileo's.
Therefore everywhere in the Universe in difference to the ideas of \ GTR $%
g_{ik,l}\neq 0.$This metrical background will be called henceforth an
inertial field.In order to find the way out of this situation the authors of
the well-known scalar theory Brans and Dicke [36]-[37] proceeding from
Mach's principle suggested an idea within that theory according to which
tlere exists a scalar field, besides the usual tensor field, with a
long-range radius of action and caused by universal mass - the
\textquotedblleft mass of fixed stars\textquotedblright .

\bigskip

\textbf{2. }A small domain of mathematical Riemannian or Lorentzian space
cannot be flat, it should be approximately similar to the flat space, to be
more exact : for each of Rieniannian space a tangent flat space may be
constructed. Hence it follows that though locally $R_{klm}^{i}$ are
sufficiently small quantities $R_{klm}^{i}\approx 0$ of higher order,
nevertheless $R_{klm}^{i}\neq 0.$

\textbf{3. }From the mentioned principle of equivalence it follows that the
theory automatically permits arbitrary \textit{holonomic transformations of
coordinates}, under which at linear conditions the gravitational field
vanishes or at other conditions new physical fields originate.This is
evidently not correct, because the true gravitational field, which is
equivalent to the geometric structure of Riemannian spacc-time, cannot be
eliminated by means of choosing coordinates. On the other hand no objectivc
physical field (in contrast to fictive fields) may be created by means of an
arbitrary choice of coordinates. Moreover it is well known that, the
principle of equivalencc in thc mentioned sense has only a local and
approximate character.

By the way, here one should not confuse the law of equality of inertial and
gravitational masses with the nientioned principle of equivalence. The
mathematical expression of this principle is the possibility of introducing
the locally geodetic coordinatc system such that $g_{ik,l}=0.$

However from this statement a not quite correct conclusion is drawn by Fock
[4], namely since the possibility of introducting of locally-geodetic system
is contained in Riemannian geometry, therefore the pointed-out principle
does not constitute a separate physical hypothesis. Actually the
availability of such a possibility in the Riemannian space-time is not
nessessary at all. On the contrary, as we have seen, everywhere in this
space $g_{ik,l}\neq 0.$

\bigskip

Therefore the principle of equivalence of Einstein may be exprcssed
mathematically in terms of any abstract tensor $\mathbf{\breve{g}}_{ik}:$ $%
g_{ik,l}\simeq $ $\mathbf{\breve{g}}_{ik,l}=0$ in locally-geodetic
coordinate system or,

$\ \ 
\begin{array}{cc}
\begin{array}{c}
\\ 
g_{ik,l}\simeq \mathbf{\breve{g}}_{ik,l}=0 \\ 
\end{array}
& \text{ \ }(2.2.2)%
\end{array}%
$

in all coordinate systems. Without this assumption one cannot construct the
EGTR. Thus since in the locally-geodetic system $g_{ik;l}$ may be equal to
zero only approximately: $g_{ik,l}\simeq 0$ the approximate character of the
principle of equivalence follows.

Actually one more very important fact (usually unnoticed) follows from the
principle of equivalencc, namely that the geodetic line is identified with a
trajectory of motion of a free material particle. Indeed it is well known
that the notion of affine connection $\mathbf{\breve{\Gamma}}_{kl}^{i}.$On
the basis of parallel transfer the whole tensor analysis may be Constructed,
the expression for the tensor of curvature $R_{klm}^{i}$ obtained, and
geodetic lines be constructed, i.e., the curves of parallel transfer of
vector or tensor with their equations, without introducing the notion of
metrical tensor $g_{ik}$.

Indeed let an arbitrary scalar parameter $t$ be taken as a parameter
changing along the curve of parallel transfer of vector $u^{i}$, i.e. the
curve is parametrically defined by the equation $x_{i}=x_{i}(t)$ and $%
u^{i}=dx_{i}/dt$ is a unit tangent vector to the curve. Then the variations
of vector components as a result of parallel transfcr from point $\mathbf{M}$
to point $\mathbf{M}^{\prime }$ along \ the curve will be equal to

\bigskip 

$\ 
\begin{array}{cc}
\begin{array}{c}
\\ 
\dfrac{dx_{i}(\mathbf{M}^{\prime })-dx_{i}(\mathbf{M})}{dt}= \\ 
\\ 
\dfrac{dx_{i}(\mathbf{M}^{\prime })}{dt}-\dfrac{dx_{i}(\mathbf{M})}{dt}= \\ 
\\ 
=-\mathbf{\breve{\Gamma}}_{kl}^{i}dx_{k}dt\times \Delta x_{l}. \\ 
\end{array}
& \text{ \ }(2.2.3)%
\end{array}%
\ \ \ \ \ \ \ \ \ \ \ \ \ \ $

\bigskip

Dividing these equations by the value of transfer $\Delta t$ from $\mathbf{M}%
^{\prime }$ to $\mathbf{M}^{\prime }$ and taking the limit when $\Delta
t\rightarrow 0$ we obtain the equations of the geodetic line

\ \ \ \ 

\bigskip $%
\begin{array}{cc}
\begin{array}{c}
\\ 
\dfrac{d^{2}x_{i}}{dt^{2}}+\mathbf{\breve{\Gamma}}_{kl}^{i}\dfrac{dx_{k}}{dt}%
\dfrac{dx_{l}}{dt}=0. \\ 
\end{array}
& \text{ \ }(2.2.4)%
\end{array}%
$

\bigskip

The affine coefficients $\mathbf{\breve{\Gamma}}_{kl}^{i}$ and the scalar
parameter $t$ in these equations are not related

to the metric of space-time. However $\mathbf{\breve{\Gamma}}_{kl}^{i}$ may
be expressed in terms of any abstract

tensor $\mathbf{\breve{g}}_{ik}$ (and its first derivatives) satisfying $%
\mathbf{\breve{g}}_{ik;l}=0$ on the basis of expressions for

covariant derivatives :

\bigskip

$%
\begin{array}{cc}
\begin{array}{c}
\\ 
\mathbf{\breve{g}}_{ik\mathbf{;}l}=\mathbf{\breve{g}}_{ik,l}-\mathbf{\breve{%
\Gamma}}_{il}^{m}\mathbf{\breve{g}}_{mk}-\mathbf{\breve{\Gamma}}_{kl}^{m}%
\mathbf{\breve{g}}_{im}=0 \\ 
\end{array}
& \text{ \ }(2.2.5)%
\end{array}%
$

\bigskip

and consequently one obtain

\bigskip

$%
\begin{array}{cc}
\begin{array}{c}
\\ 
\mathbf{\breve{\Gamma}}_{kl}^{i}=\dfrac{1}{2}\mathbf{\breve{g}}^{im}\left( 
\mathbf{\breve{g}}_{mk,l}+\mathbf{\breve{g}}_{ml,k}-\mathbf{\breve{g}}%
_{kl,m}\right)  \\ 
\end{array}
& \text{ \ }(2.2.6)%
\end{array}%
$

\bigskip

Thus it follows that in the EGTR the real space-time metric tensor $g_{ik}$
is identified with an

abstract tensor $\mathbf{\breve{g}}_{ik.}.$

\bigskip

Meanwhile the trajectory of motion of a free particle, in contrast to thc
geodetic

line, can be obtained only on the basis of real space-time metric tensor $%
g_{ik}$ from the

principle of least action

\bigskip 

\bigskip $%
\begin{array}{cc}
\begin{array}{c}
\\ 
\delta S=-mc\delta \dint ds=0,ds^{2}=g_{ik}dx^{i}dx^{k} \\ 
\end{array}
& \text{ \ }(2.2.7)%
\end{array}%
$

\bigskip

and consequently from Eq.(2.2.7) one obtain

\bigskip $%
\begin{array}{cc}
\begin{array}{c}
\\ 
\dfrac{d^{2}x_{i}}{dt^{2}}+\Gamma _{kl}^{i}\dfrac{dx_{k}}{dt}\dfrac{dx_{l}}{%
dt}=0, \\ 
\\ 
\Gamma _{kl}^{i}=\dfrac{1}{2}g^{im}\left( g_{mk,l}+g_{ml,k}-g_{kl,m}\right) .
\\ 
\end{array}
& \text{ \ }(2.2.8)%
\end{array}%
$

\bigskip

The curvature tensor$\ \ R_{klm}^{i}\ \ $is also defined through parallel
transfer and one can expressed curvature tensor by the affine coefficients $%
\mathbf{\breve{\Gamma}}_{kl}^{i}$ and their derivatives. In particular

$\bigskip $

$%
\begin{array}{cc}
\begin{array}{c}
\\ 
\mathbf{\breve{R}}_{ik}=\mathbf{\breve{\Gamma}}_{i,kl}^{l}-\mathbf{\breve{%
\Gamma}}_{il,k}^{l}+\mathbf{\breve{\Gamma}}_{ik}^{l}\mathbf{\breve{\Gamma}}%
_{lm}^{m}-\mathbf{\breve{\Gamma}}_{il}^{m}\mathbf{\breve{\Gamma}}_{km}^{l},
\\ 
\\ 
\mathbf{\breve{R}}=\mathbf{\breve{g}}^{ik}\mathbf{\breve{R}}_{ik}. \\ 
\end{array}
& \text{ \ }(2.2.9)%
\end{array}%
$

\bigskip

Hence Einstein's field equations reads

\bigskip $%
\begin{array}{cc}
\begin{array}{c}
\\ 
\mathbf{\breve{R}}_{ik}-\dfrac{1}{2}\mathbf{\breve{g}}_{ik}\mathbf{\breve{R}=%
}\dfrac{8\pi \varkappa }{c^{4}}T_{ik}. \\ 
\end{array}
& \text{ \ }(2.2.10)%
\end{array}%
$

\bigskip \bigskip

\bigskip

\subsection{II.2.1.Postulates of Davtyan's One-Metric Theory of
Gravitational-Inertial Field.}

\bigskip

The fundamental principle of \ Davtyan's gravitational-inertial theory [1],
like that of EGTR, is the assumption of equivalence of the
gravitational-inertial field with the geometric structure of space-time on
the basis of statement and conditions (2.2.1) of Rienimanian geometry and
its \textquotedblleft extension\textquotedblright\ all over the Universe.The
essence of such an \textquotedblleft extension\textquotedblright\ lies in
the point that because of the existence of the inertial field (besides the
gravitational fields) far from the masses and also locally the metric tensor
of the world is everywhere distinct from Calileio's metric, i.e.,

\bigskip\ $\ \ \ \ \ \ \ \ \ \ \ \ \ \ \ \ \ \ \ \ \ \ \ \ \ \ \ \ \ \ \ \ \
\ \ \ \ \ \ \ \ \ \ \ \ \ \ \ \ \ 
\begin{array}{cc}
\begin{array}{c}
\\ 
g_{ik,l}\neq 0 \\ 
\end{array}
& \text{ \ }(2.2.11)%
\end{array}%
$

\bigskip

The space is permanently related to a weak metrical background $h_{\left(
J\right) }^{lm}$. The field formed by this tensor background $h_{\left(
J\right) }^{lm}\left( \mathbf{M}\right) $, as already noted, will be called
inertial field. The cause of formation of such tensor background, as we
shall see in the later development of the theory, is the world
energy-momentum tensor $T_{\left( J\right) }^{lm}$ related to a field mass
of gravitational electromagnetic radiation surrounding the metagalaxy.
Further, as already shown, there are no theoretical are experimental reasons
to considering $g_{ik}$ that represents a physical field as identical with $%
\mathbf{\breve{g}}_{ik}$ entering into the coefficients of affine connection 
$\mathbf{\breve{\Gamma}}_{km}^{l}$, which is being abstractly constructed
for operations in tensor analysis. Independently of gravitational fields, in
the inertial frame of reference and in the locally geodetic coordinate system

\bigskip

\bigskip\ $\ \ \ \ \ \ \ \ \ \ \ \ \ \ \ \ \ \ \ \ \ \ \ \ \ \ \ \ \ \ \ \ \
\ \ \ \ \ \ \ \ \ \ \ 
\begin{array}{cc}
\begin{array}{c}
\\ 
\mathbf{\breve{g}}_{ik,l}=0. \\ 
\end{array}
& \text{ \ }(2.2.12)%
\end{array}%
$ \ \ \ 

\ \ \ \ \ 

\begin{remark}
\ \textbf{2.2.1.1. }It should be recalled that according to GTR the
gravitational field is defined by the tensor $\mathbf{\breve{g}}_{ik}$.
Therefore in the presence of a gravitational field in the locally-geodctic
coordinate system though $\mathbf{\breve{g}}_{ik,l}=0$ nevertheless:
\end{remark}

\bigskip\ $\ \ \ \ \ \ \ \ \ \ \ \ \ \ \ \ \ \ \ \ \ \ \ \ \ \ \ \ \ \ \ \ \
\ \ \ \ \ \ \ \ $

$\ \ \ \ \ \ \ \ \ \ \ \ \ \ \ \ \ \ \ 
\begin{array}{cc}
\begin{array}{c}
\\ 
\mathbf{\breve{g}}_{ik,lm}=\dfrac{\partial ^{2}\mathbf{\breve{g}}_{ik}}{%
\partial x_{l}\partial x_{m}}\neq 0. \\ 
\end{array}
& \text{ \ }(2.2.13)%
\end{array}%
$ \ \ \ \ \ \ \ \ \ \ \ \ \ \ \ \ \ \ \ \ \ \ \ \ \ \ \ \ \ \ \ \ \ \ \ \ \
\ \ \ \ \ \ \ 

\bigskip

\ In contrast to this, in Davtyan's theory [1] of the gravitational inertial
field $\mathbf{\breve{g}}_{ik}$ is not related to the gravitational field,
so that even at the presence of such a field we have in the locally-geodetic
coordinate system\bigskip

\bigskip\ $\ \ \ \ \ \ \ \ \ \ \ \ \ \ \ \ \ \ \ \ \ \ \ \ \ \ \ \ \ \ \ \ \
\ \ \ \ 
\begin{array}{cc}
\begin{array}{c}
\\ 
\mathbf{\breve{g}}_{ik,lm}=\dfrac{\partial ^{2}\mathbf{\breve{g}}_{ik}}{%
\partial x_{l}\partial x_{m}}=0. \\ 
\end{array}
& \text{ \ }(2.2.14)%
\end{array}%
$ \ \ \ 

\bigskip

Thus on the basis of listed propositions one may conclude that the universal
metric tensor $g_{ik}$ of the gravitational-inertial field is the metric
tensor of the real world, formed by gravitational $g_{ik}^{\mathbf{Gr}}$ and 
$g_{ik}^{\mathbf{In}}$ inertial metric tensors. From the field of this
general metric tensor $g_{ik}(\mathbf{M})$ the Riemannian space is also
formed.

On the basis of original proposition given by Eq. (2.2.11) the following
quite obvious theorem may be proved [1]: If first derivatives of the metric
tensor gir are nonzero in the locally geodetic coordinate system then also
nonzero will be its covariant derivatives in all coordinate systems
i.e.\bigskip

\bigskip\ $\ \ \ \ \ \ \ \ \ \ \ \ \ \ \ \ \ \ \ \ \ \ \ \ \ \ \ \ \ \ \ \ \ 
\begin{array}{cc}
\begin{array}{c}
\\ 
g_{ik;l}=g_{ik,l}-\mathbf{\breve{\Gamma}}_{il}^{m}g_{mk}-\mathbf{\breve{%
\Gamma}}_{ki}^{m}g_{im}\neq 0, \\ 
\end{array}
& \text{ \ }(2.2.15)%
\end{array}%
$ \ \ 

\ \ \ \ \ \ \ \ \ \ \ \ \ \ \ \ \ \ \ \ \ \ \ \ 

where

\bigskip

\bigskip\ $\ \ \ \ \ \ \ \ \ \ \ \ \ \ \ \ \ \ \ \ \ \ \ \ \ \ \ \ \ \ \ \ 
\begin{array}{cc}
\begin{array}{c}
\\ 
\mathbf{\breve{\Gamma}}_{kl}^{i}=\dfrac{1}{2}\mathbf{\breve{g}}^{im}\left( 
\mathbf{\breve{g}}_{mk,l}+\mathbf{\breve{g}}_{ml,k}-\mathbf{\breve{g}}%
_{kl,m}\right) . \\ 
\end{array}
& \text{ \ }(2.2.16)%
\end{array}%
$ \ \ 

\bigskip

Since, according to Eq.(2.2.12) in the locally-geodetic coordinate system $%
\breve{g}_{ik,l}=0$, then in Eq.(2.2.15) $\mathbf{\breve{\Gamma}}_{il}^{m}$
and $\mathbf{\breve{\Gamma}}_{ki}^{m}$ should be equal to zero and according
to Eq. (2.2.11) actually $\mathbf{\breve{g}}_{ik;l}\neq 0.$ But since the
quantity $\mathbf{\breve{g}}_{ik;l}$ is a tensor, it will be nonzero in all
other coordinate systems if it is nonzero in one of them.

\begin{itemize}
\item 
\begin{remark}
2.2.1.2. Note that the expression given by Eqs.(2.2.15)-(2.2.16) is another
important original proposition of Davtyan's gravitational-inertial field
theory of the Universe. This proposition actually means that the
gravitational-inertia1 field $g_{ik}$ and Riemannian space $\tciLaplace
\left( M,g_{ik}\right) $ formed by a general metric tensor of real world
represents a truely physical field, that may not be eliminated by means of
transformation of coordinates.\bigskip
\end{remark}
\end{itemize}

\bigskip

Meanwhile the affine field $\mathbf{\breve{\Gamma}}_{il}^{m}$ containing the
tensor $\mathbf{\breve{g}}_{ik}$ may be eliminated by the choice of a
special coordinate system, $\mathbf{\breve{\Gamma}}_{kl}^{i}=0$. Further if
the world is considered as pseudoeuclidean, i.e. as a world without
gravitational and inertial fields, then in curvilinear coordinates or
generally in non-inertial frames of reference $\mathbf{\breve{\Gamma}}%
_{kl}^{i}\neq 0.$ Thus whereas the real nietric tensor $g_{ik}$ forms a
truely physical field $g_{ik}(\mathbf{M})$, the tensor $\mathbf{\breve{g}}%
_{ik}$ causes various kineinntical-dynamical effects due to origination of
fictitious fields in noninertial frames of reference.

\begin{remark}
2.2.1.3. From the introduced propositions of \ Davtyan's theory it follows
apart from the Riemannian space tensor $g_{ik}(\mathbf{M})$ representing the
gravitationel-inertial field (of real world) with the quadratic form of
space-time interval element
\end{remark}

\bigskip

\bigskip\ $\ \ \ \ \ \ \ \ \ \ \ \ \ \ \ \ \ \ \ \ \ \ \ \ \ \ \ \ \ \ \ \ \
\ \ \ \ \ \ \ \ \ $

$\ \ \ \ \ \ \ \ \ \ \ \ \ \ \ \ \ \ \ \ \ \ \ \ \ \ \ \ \ \ \ \ \ \ \ \ \ \
\ \ \ \ \ \ 
\begin{array}{cc}
\begin{array}{c}
\\ 
ds^{2}=g_{ik}dx^{i}dx^{k} \\ 
\end{array}
& \text{ \ }(2.2.17)%
\end{array}%
$

no other physical space or new tensor or scalar is being introduced, as done
in bimetrical and scalar theories.

\bigskip

\begin{remark}
\textbf{\ }2.2.1.4.\textbf{\ }Though it is formally supposed in Davtyan's
theory that $g_{ik}$ may be represented \ sum of a gravitational tensor $%
g_{ik}^{\mathbf{Gr}}$ and the metrical tensor background $h_{ik}^{\left(
J\right) }$, this assumption is not used in the Davtyan's field equations
(see [1] section 3.).
\end{remark}

\bigskip

\begin{remark}
2.2.1.5.The potentials of the gravitational-inertial field obtained in [1]
represent themselves only the components of the universal metric tensor $%
g_{ik}$, entering in (2.2.17). Further, as we have seen, in the
locally-geodetic coordinate system
\end{remark}

\bigskip

\bigskip\ $\ \ \ \ \ \ \ \ \ \ \ \ \ \ \ \ \ \ \ \ \ \ \ \ \ \ \ \ \ \ \ \ \
\ \ \ \ \ \ \ \ \ $

$\ \ \ \ \ \ \ \ \ \ \ \ \ \ \ \ \ \ \ \ \ \ \ \ \ \ \ \ \ \ \ \ \ \ \ \ \ \
\ \ \ \ \ \ \ 
\begin{array}{cc}
\begin{array}{c}
\\ 
g_{ik,l}=g_{ik;l}\simeq 0. \\ 
\end{array}
& \text{ \ }(2.2.18)%
\end{array}%
$

\bigskip

On the same basis the scalar quantity

\bigskip

\bigskip\ $\ \ \ \ \ \ \ \ \ \ \ \ \ \ \ \ \ \ \ \ \ \ \ \ \ \ \ \ \ \ \ \ \
\ \ \ \ \ \ \ \ \ \ \ \ \ \ \ \ \ \ 
\begin{array}{cc}
\begin{array}{c}
\\ 
g^{_{ik}}g_{ik}\simeq 1 \\ 
\end{array}
& \text{ \ }(2.2.19)%
\end{array}%
$

\bigskip

in all coordinate systems.

\bigskip \bigskip

\subsection{II.2.2.Lagrangian Density and Field Equations in Davtyan's
Theory of Gravitational-Inertial Field.}

\bigskip

The expression (2.2.15) allows us to use the following variational action
for obtaining field equations [5]:

\bigskip\ $\ \ \ \ \ \ \ \ \ \ \ \ \ \ \ \ \ \ \ \ \ \ \ \ \ \ \ \ \ \ 
\begin{array}{cc}
\begin{array}{c}
\\ 
\mathbf{S}=\mathbf{S}_{g}+\mathbf{S}_{m}=\dfrac{1}{c}\dint \left( \Lambda
_{g}+\Lambda _{m}\right) \sqrt{-g}d^{4}x, \\ 
\end{array}
& \text{ \ }(2.2.20)%
\end{array}%
$

\bigskip

where

\bigskip\ $\ \ \ \ \ \ \ \ \ \ \ \ \ \ \ \ \ \ \ \ \ \ \ \ \ \ \ \ \ \ \ \ \
\ \ \ \ \ \ \ $

$\ \ \ \ \ \ \ \ \ \ \ \ \ \ \ \ \ \ \ \ \ \ \ \ \ \ \ \ \ \ \ \ \ \ \ \ \ 
\begin{array}{cc}
\begin{array}{c}
\\ 
\Lambda _{g}=\dfrac{1}{8\pi \varkappa }g_{lm;i}\cdot g_{;k}^{lm}g^{ik}. \\ 
\end{array}
& \text{ \ }(2.2.21)%
\end{array}%
$

\bigskip

This corresponds to the fact, that in Euclidean space the Lagrange density
is defined as a square of gradient of the potential $\Phi :$

\bigskip\ $\ \ \ \ \ \ \ \ \ \ \ \ \ \ \ \ \ \ \ \ \ \ \ \ \ \ \ \ \ \ \ \ \
\ \ \ \ \ \ \ \ 
\begin{array}{cc}
\begin{array}{c}
\\ 
\Lambda _{g}=\dfrac{1}{8\pi \varkappa }\left[ \mathbf{grad}\left( \Phi
\right) \right] ^{2} \\ 
\end{array}
& \text{ \ }(2.2.22)%
\end{array}%
$

\bigskip

In Riemann space this quantity should be generalized to an inner product of
covariant derivatives of the metric tensor as in Eq.(2.2.21).

\bigskip By analogy with Eq.(2.2.21) the Lagrangian density for matter will
be defined as$\bigskip $

$\ \ \ \ \ \ \ \ \ \ \ \ \ \ \ \ \ \ \ \ \ \ \ \ \ \ \ \ \ \ \ \ \ 
\begin{array}{cc}
\begin{array}{c}
\\ 
\Lambda _{m}=k\cdot f\left( g^{ik},g_{i}^{lm}\right) \left( \sqrt{-g}\right)
^{-1} \\ 
\end{array}
& \text{ \ }(2.2.23)%
\end{array}%
$

\bigskip

\bigskip

and the action will be equal to

$\ \ \ \ \ \ \ \ \ \ \ \ \ \ \ \ \ \ \ \ \ \ \ \ 
\begin{array}{cc}
\begin{array}{c}
\\ 
\mathbf{S}_{m}=\dfrac{1}{c}\dint \Lambda _{m}\sqrt{-g}d^{4}x=\dfrac{k}{c}%
\dint f\left( g^{ik},g_{i}^{lm}\right) d^{4}x \\ 
\end{array}
& \text{ \ }(2.2.24)%
\end{array}%
$

\bigskip where $k$ is a constant. As was shown by Davtyan's in [1] that

$\bigskip \ \ \ \ \ \ \ \ \ \ \ \ \ \ \ \ \ \ \ \ \ \ \ \ \ \ \ \ \ \ \ \ \
\ \ \ \ \ \ \ \ \ \ \ \ \ \ 
\begin{array}{cc}
\begin{array}{c}
\\ 
k=\dfrac{2}{c^{4}} \\ 
\end{array}
& \text{ \ }(2.2.25)%
\end{array}%
$

or $k=\dfrac{4}{c^{4}}.$

\bigskip According to Eqs.(2.2.20)-(2.2.24) by using variational principle

\bigskip\ $\ \ \ \ \ \ \ \ \ \ \ \ \ \ \ \ \ \ \ \ \ \ \ \ \ \ \ \ \ \ \ \ \
\ \ \ \ 
\begin{array}{cc}
\begin{array}{c}
\\ 
\delta \mathbf{S}=\delta \left( \mathbf{S}_{g}+\mathbf{S}_{m}\right) =0 \\ 
\end{array}
& \text{ \ }(2.2.26)%
\end{array}%
$

\bigskip

one obtain [1]:

\bigskip\ $\ \ \ \ \ \ \ \ \ \ \ \ \ \ \ \ \ \ \ \ \ \ \ 
\begin{array}{cc}
\begin{array}{c}
\\ 
\delta \mathbf{S}=\delta \left( \mathbf{S}_{g}+\mathbf{S}_{m}\right) = \\ 
\\ 
\dint \left\{ \dfrac{1}{8\pi \varkappa }\left[ g^{lp}g^{mn}\nabla _{k}\left( 
\sqrt{-g}g^{ik}g_{np;i}\right) \right. \right. \\ 
\\ 
\left. \left. -\nabla _{i}\left( \sqrt{-g}g^{ik}g_{np;k}\right) \right]
-kg^{lp}g^{mn}\sqrt{-g}T_{np}\right\} , \\ 
\\ 
\sqrt{-g}T_{lm}=\dfrac{\partial \Lambda _{m}\sqrt{-g}}{\partial g^{lm}}-%
\dfrac{\partial }{\partial x_{i}}\left( \dfrac{\partial \Lambda _{m}\sqrt{-g}%
}{\partial g_{,i}^{lm}}\right) . \\ 
\\ 
\end{array}
& \text{ \ }(2.2.27)%
\end{array}%
$

\bigskip

Finally Davtyan's Gravitational-Inertial field equations sees\ [5]:

\bigskip

\bigskip\ $\ \ \ \ \ \ \ \ \ \ \ \ \ \ \ \ \ 
\begin{array}{cc}
\begin{array}{c}
\\ 
g^{lp}g^{mn}\nabla _{k}\left( \sqrt{-g}g^{ik}g_{np;i}\right) -\nabla
_{i}\left( \sqrt{-g}g^{ik}g_{np;k}\right) = \\ 
\\ 
=8\pi \varkappa k\sqrt{-g}g^{lp}g^{mn}T_{np}= \\ 
\\ 
8\pi \varkappa k\sqrt{-g}g^{mn}T_{n}^{l}. \\ 
\end{array}
& \text{ \ }(2.2.28)%
\end{array}%
$

\bigskip

In the cquations (2.2.28) the components of $g_{ik}$ represent the
potentials of the gravitational-inertial field. The components of tensor $%
\breve{g}_{ik}$ entering only into Christoffel symbols $\Gamma
_{kl}^{i}\left( \breve{g}_{ik},\breve{g}_{ik,l}\right) $ may be eliminated.
The possibility of eliminating $\breve{g}_{ik}$ may be explained in virtue
of the fact that, was pointed out, the space of affine connection $\mathbf{%
\breve{\Gamma}}_{kl}^{i}\left( \mathbf{M}\right) $ auxiliary, abstract
mathematical space while $g_{ik}\left( \mathbf{M}\right) $ ) defines a
Riemannian world and the gravitational-inertia1 field. Therefore \ $\mathbf{%
\breve{\Gamma}}_{kl}^{i}\left( \mathbf{M}\right) $ may be chosen
arbitrarily\ [1]. Thus the $\mathbf{\breve{\Gamma}}_{kl}^{i}$ may be defined
in such a way that $\breve{g}_{ik,l}=0$ and hence $\mathbf{\breve{\Gamma}}%
_{kl}^{i}=0.$It sllould be noted that the elimination of Christoffel symbols
in (2.2.28) actually means the elimination of various fictive fields. In
order that no misunderstanding of these items will arise (due to existing
traditional habits) we consider it necessary to specify the essence of one
of the major differences between the GTR and present theory. As already
pointed out, in GTR it is assumed that $g_{ik}\equiv \breve{g}_{ik}.$%
Therefore if $g_{ik}$ is taken in some coordinate system, then $\mathbf{%
\breve{\Gamma}}_{kl}^{i}$ should be taken in the same coordinate system. In
contrast to this in the present theory [1] the Riemann space $\left(
M,g_{ik}\right) $ is everywere and always curved, therefore $g_{ik}$ is
always taken in curvilinear coordinates. Since $g_{ik}$ is not physically
related to $\breve{g}_{ik}$ the coordinate system for $\breve{g}_{ik}$ may
be chosen independently of that for $g_{ik}$.

\bigskip

For example, in the locally geodetic coordinate system $\breve{g}_{ik,l}=0$
while $g_{ik,l}\neq 0.$This means that $\breve{g}_{ik}$ is taken in Carhian
coordinates, while $g_{ik}$ is still related to the curved space and may be
taken in arbitrary curvilinear coordinates. This also means that, as
mentioned above, the parallel transfer operation and its curve do not depend
on the curvature of space-time and hence on the trajectory of motion of a
free particle. Thus according to conditions (2.2.12)-(2.2.14) for $\mathbf{%
\breve{\Gamma}}_{kl}^{i}=0$ in (2.2.28).Hence all Christoffel symbols are
being eliminated and (2.2.28) are transformed into the equations

\bigskip\ $\ \ \ \ \ \ \ \ \ \ \ \ \ \ \ \ \ \ \ \ \ \ \ \ \ \ \ \ \ \ \ \ \
\ \ \ \ \ \ \ \ \ \ \ \ \ \ 
\begin{array}{cc}
\begin{array}{c}
\\ 
\nabla _{k}g^{lm}=g_{,k}^{lm}, \\ 
\\ 
\nabla _{i}\nabla _{k}g^{lm}=g_{,ki}^{lm}. \\ 
\end{array}
& \text{ \ }(2.2.29)%
\end{array}%
$

\bigskip

Hence all Christoffe1 symbols are being eliminated and (2.2.28) are
transformed into the equations [1]:

\bigskip

$\ \ \ \ \ \ \ \ \ \ \ \ \ \ \ \ \ \ \ \ \ 
\begin{array}{cc}
\begin{array}{c}
\\ 
g^{lp}g^{mn}\dfrac{\partial }{\partial x_{k}}\left( \sqrt{-g}%
g^{ik}g_{np;i}\right) -\dfrac{\partial }{\partial x_{i}}\left( \sqrt{-g}%
g^{ik}g_{np;k}\right) = \\ 
\\ 
=8\pi \varkappa k\sqrt{-g}g^{mn}\left( T_{n}^{\text{ }l}-\dfrac{1}{2}\delta
_{n}^{l}T\right) . \\ 
\end{array}
& \text{ \ }(2.2.30)%
\end{array}%
$

$\ \ \ \ \ \ \ \ \ \ \ \ \ \ \ \ \ \ \ \ \ \ \ \ \ \ \ $

\bigskip

These equations are very attractive not only because Christoffel symbols are
absent in them (and hence they are extremely simple), but mainly because the
solution of fundamental problems of GTR by means of these equations gives
the same results as the solution by means of the Einstein equations [1].

\bigskip $\ \ \ \ \ \ \ \ \ \ \ \ \ $

\bigskip

\subsection{ II.2.3. Davtyan's Field Equations in Einstein approximation.}

\bigskip

It is necessary to observe that the original variational equations make it
possible to obtain another version of field equations, somewhat different
from Eq.(2.2.28) Indeed in the proccss of variation of Lagrangian $\Lambda
_{m}$, for the matter (2.2.23) a tensor density is obtained in the form
(2.2.27). This expression with some coefficient may also be considered as
the tensor density\bigskip

\bigskip\ $\ \ \ \ \ \ \ \ \ \ \ \ \ \ \ \ \ \ \ \ \ \ \ \ \ \ \ \ \ $

$\ \ \ \ \ \ \ \ \ \ \ \ \ \ \ \ \ \ \ \ \ \ \ \ \ \ \ \ \ \ \ \ \ \ 
\begin{array}{cc}
\begin{array}{c}
\\ 
\sqrt{-g}\breve{T}_{lm}=\sqrt{-g}\left( T_{lm}-\dfrac{1}{2}g_{lm}T\right) \\ 
\end{array}
& \text{ \ }(2.2.31)%
\end{array}%
$ \ \ \ \ \ \ \ \ \ \ \ \ \ \ \ \ \ \ \ \ \ \ \ \ \ \ \ \ \ \ \ \ 

\bigskip

Anyway one may always (with equal hasis) proceed from the fact. That for the
gravitational- inertial field the following formula holds

\bigskip

\bigskip\ $\ \ \ \ \ \ \ \ \ \ \ \ \ \ \ \ \ \ \ \ \ \ \ \ \ \ \ \ \ 
\begin{array}{cc}
\begin{array}{c}
\\ 
\delta \mathbf{S}_{m}=\dfrac{k}{c}\dint \breve{T}\sqrt{-g}\delta
g^{lm}d^{4}x= \\ 
\\ 
\dfrac{k}{c}\dint \left( T_{lm}-\frac{1}{2}g_{lm}T\right) \sqrt{-g}%
g^{lm}d^{4}x. \\ 
\end{array}
& \text{ \ }(2.2.32)%
\end{array}%
$

\bigskip

Therefore the choice of tensor $T_{lm}$, is mathematically equivalent to the
choice of $\ \breve{T}_{lm}$ However the analysis of equations (2.2.28)
shows that in the Einstein approximation, i.e. in the limit $g_{ik}\simeq 
\breve{g}_{ik}$ they reduce to

\bigskip

\bigskip\ $\ \ \ \ \ \ \ \ \ \ \ \ \ \ \ \ \ \ \ \ \ \ \ \ \ \ \ \ \ \ \ \ \
\ \ \ \ \ \ \ \ \ \ \ 
\begin{array}{cc}
\begin{array}{c}
\\ 
R^{lm}=4\pi k\varkappa T^{lm}. \\ 
\end{array}
& \text{ \ }(2.2.33)%
\end{array}%
$

\bigskip

Where $R^{lm}$ is the Ricci tensor. Therefore, in order to satisfy the
continuity law as well as the Bianchi identity the choice of the second
tensor version-the source of gravitational- inertional field as

\bigskip\ $\ \ \ \ \ \ \ \ \ \ \ \ \ \ \ \ \ \ \ \ \ \ \ \ \ \ \ \ \ \ \ \ \
\ \ \ \ \ \ \ \ \ 
\begin{array}{cc}
\begin{array}{c}
\\ 
\breve{T}^{lm}=T^{lm}-\dfrac{1}{2}g^{lm}T \\ 
\end{array}
& \text{ \ }(2.2.34)%
\end{array}%
$

\bigskip

is more expedient. Then the gravitional-inertial field equations take the
following form

\bigskip

\bigskip\ $\ \ \ \ \ \ \ \ \ \ \ \ \ \ \ \ \ \ \ \ \ \ \ 
\begin{array}{cc}
\begin{array}{c}
\\ 
g^{lp}g^{mn}\nabla _{k}\left( \sqrt{-g}g^{ik}g_{np;i}\right) -\nabla
_{i}\left( \sqrt{-g}g^{ik}g_{np;k}\right) = \\ 
\\ 
=8\pi \varkappa k\sqrt{-g}g^{mn}\left( T_{n}^{\text{ }l}-\dfrac{1}{2}\delta
_{n}^{l}T\right) . \\ 
\end{array}
& \text{ \ }(2.2.35)%
\end{array}%
$

\bigskip In the above mentioned limit one get

\bigskip\ $\ \ \ \ \ \ \ \ \ \ \ \ \ \ \ \ \ \ \ \ \ \ \ \ \ \ \ \ \ \ \ \ \
\ 
\begin{array}{cc}
\begin{array}{c}
\\ 
R^{lm}=4\pi k\varkappa \left( T^{lm}-\dfrac{1}{2}g^{lm}T\right) = \\ 
\\ 
\dfrac{8\pi \varkappa }{c^{4}}\left( T^{lm}-\dfrac{1}{2}g^{lm}T\right) . \\ 
\end{array}
& \text{ \ }(2.2.36)%
\end{array}%
$

\ \ \ \ \ \ \ \ \ \ \ \ \ \ \ \ \ \ \ \ \ \ \ \ \ \ \ \ \ \ \ \ \ \ \ \ \ \
\ \ \ \ \ \ \ \ \ \ \ \ \ \ \ \ \ \ \ \ \ \ 

\bigskip

Thus in the Einstein approximation the gravitational-inertial field
equations reduce to the usual Einstein equations of the gravitational field.

\bigskip

\bigskip \bigskip

\subsection{ II.2.4. Weak Field limit in Davtyan's Theory of
Gravitational-Inertial Field.}

\bigskip

\bigskip Under these conditions the metric of space-time is close to the
Galileo metric:

\bigskip\ $\ \ \ \ \ \ \ \ \ \ \ \ \ \ \ \ \ \ \ \ \ \ \ \ \ \ \ \ 
\begin{array}{cc}
\begin{array}{c}
\\ 
g_{11}^{0}=g_{22}^{0}=g_{33}^{0}=-1,g_{44}^{0}=1, \\ 
\\ 
g_{ik}^{0}=0,i\neq k. \\ 
\end{array}
& \text{ \ }(2.2.37)%
\end{array}%
$

with a certain background of an inertial field $h_{\left( J\right) }^{jk}$.
A weak perturbation, caused by gratvitational field (plus the inertial field 
$h_{\left( J\right) }^{il}$)) may be represented by a tensor $h_{ik}$, which
is a first order small quantity:

\bigskip\ $\ \ \ \ \ \ \ \ \ \ \ \ \ \ \ \ \ \ \ \ \ \ \ \ \ \ \ \ \ \ \ \ \
\ \ \ \ 
\begin{array}{cc}
\begin{array}{c}
\\ 
g_{lm}=g_{lm}^{0}+h_{lm}. \\ 
\end{array}
& \text{ \ }(2.2.38)%
\end{array}%
$

\bigskip

With the same accuracy one obtain the expression for the determinant of the
inetricsl tensor :

\bigskip

\bigskip\ $\ \ \ \ \ \ \ \ \ \ \ \ \ \ \ \ \ \ \ \ \ \ \ \ \ \ \ 
\begin{array}{cc}
\begin{array}{c}
\\ 
g=\det \left\Vert g_{lm}\right\Vert =-\left( 1+g^{lm}h_{lm}\right) . \\ 
\end{array}
& \text{ \ }(2.2.39)%
\end{array}%
$

\bigskip

On the basis of these simplifications one may proceed to the solution of the
gravitational field equations (2.2.30). According to Eq.(2.2.38) one obtain

\bigskip\ $\ \ \ \ \ \ \ \ \ \ \ \ \ \ \ \ \ \ \ \ \ \ \ \ \ \ \ \ $

$\ \ \ 
\begin{array}{cc}
\begin{array}{c}
\\ 
\left( g^{lp\left( 0\right) }-h^{lp}\right) \left( g^{nm\left( 0\right)
}-h^{mn}\right) \dfrac{\partial }{\partial x_{k}}\left[ \left( g^{ik\left(
0\right) }-h^{ik}\right) \dfrac{\partial }{\partial x_{i}}\left(
g_{np}^{\left( 0\right) }+h_{np}\right) \right] - \\ 
\\ 
-\dfrac{\partial }{\partial x_{i}}\left[ \left( g^{ik\left( 0\right)
}-h^{ik}\right) \dfrac{\partial }{\partial x_{k}}\left( g^{lm\left( 0\right)
}-h^{lm}\right) \right] = \\ 
\\ 
=8\pi k\varkappa g^{mn\left( 0\right) }\left( T_{n}^{i}-\dfrac{1}{2}\delta
_{n}^{i}T\right) . \\ 
\end{array}
& \text{ \ }(2.2.40)%
\end{array}%
$

\bigskip

\bigskip From Eq.(2.2.40) one obtain

\bigskip\ $\ \ \ \ \ \ \ \ \ \ \ \ \ \ \ \ \ \ \ \ \ \ \ \ \ \ \ \ \ \ \ \ 
\begin{array}{cc}
\begin{array}{c}
\\ 
\mathbf{\square }h_{ml}=4\pi k\varkappa g^{nm\left( 0\right) }\left(
T_{n}^{i}-\dfrac{1}{2}\delta _{n}^{i}T\right) \\ 
\end{array}
& \text{ \ }(2.2.41)%
\end{array}%
$

\bigskip

The solution of this equation is

\bigskip

\bigskip\ $\ \ \ \ \ \ \ \ \ \ \ \ \ \ \ \ \ \ \ \ \ \ \ \ \ \ \ \ \ \ 
\begin{array}{cc}
\begin{array}{c}
\\ 
h_{ml}=-k\varkappa \dint \dfrac{\left( g^{nm\left( 0\right) }\left(
T_{n}^{i}-\dfrac{1}{2}\delta _{n}^{i}T\right) \right) }{r}d^{3}x. \\ 
\end{array}
& \text{ \ }(2.2.42)%
\end{array}%
$ \ \ \ \ \ \ \ \ \ \ \ \ \ \ \ \ \ \ \ \ \ \ \ 

\bigskip

The energy-momentum tensor may be taken in the form

\bigskip

\bigskip\ $\ \ \ \ \ \ \ \ \ \ \ \ \ \ \ \ \ \ \ \ \ \ \ \ \ \ \ \ \ \ \ \ \
\ \ \ \ \ \ 
\begin{array}{cc}
\begin{array}{c}
\\ 
T_{l}^{n}=\rho v_{l}v^{n}, \\ 
\\ 
v_{l}=\dfrac{dx_{l}}{dt},v^{n}=\dfrac{dx^{n}}{dt}, \\ 
\\ 
l,n=1,2,3,4 \\ 
\end{array}
& \text{ \ }(2.2.43)%
\end{array}%
$

\bigskip

where $\rho $ is the density of mass.

Since according to condition the field is weak the components of
thrce-dimensional velocity should be very small with respect to the
fudamental velocity $c$. Therefore all space components of velocity in
Eq.(2.2.43) inay he neglected. Hence only the time coinponent $c^{2}$
remains. \ Hence $T_{4}^{4}=c^{2}\rho $ and system (2.2.41) consisting of 10
equations turns into a single equation

\bigskip\ $\ \ \ \ \ \ \ \ \ \ \ \ \ \ \ \ \ \ \ \ \ \ \ \ \ \ \ \ \ \ \ \ \
\ \ \ \ \ \ \ 
\begin{array}{cc}
\begin{array}{c}
\\ 
\square h_{44}=2\pi c^{2}k\varkappa \rho \\ 
\end{array}
& \text{ \ }(2.2.44)%
\end{array}%
$

\bigskip

If the field is stationary we have $\dfrac{\partial h_{44}}{\partial x_{4}}$
and hence from Eq.(2.2.44) one obtain

\bigskip

$\ \ \ \ \ \ \ \ \ \ \ \ \ \ \ \ \ \ \ \ \ \ \ \ \ \ \ \ \ \ \ \ \ \ \ \ \ \
\ \ \ 
\begin{array}{cc}
\begin{array}{c}
\\ 
\Delta h_{44}=2\pi c^{2}k\varkappa \rho . \\ 
\end{array}
& \text{ \ }(2.2.45)%
\end{array}%
$

\bigskip

Thc solution of this equation is

\bigskip

\bigskip\ $\ \ \ \ \ \ \ \ \ \ \ \ \ \ \ \ \ \ \ \ \ \ \ \ \ \ \ \ \ \ \ \ \
\ \ \ \ \ \ 
\begin{array}{cc}
\begin{array}{c}
\\ 
h_{44}=\dfrac{k}{2}c^{2}\varkappa \dint \dfrac{\rho }{r}d^{3}x. \\ 
\end{array}
& \text{ \ }(2.2.46)%
\end{array}%
$

\bigskip Thus

\bigskip\ $\ \ \ \ \ \ \ \ \ \ \ \ \ \ \ \ \ \ \ \ \ \ \ \ \ \ 
\begin{array}{cc}
\begin{array}{c}
\\ 
h^{44}=g_{44}^{\left( 0\right) }-\dfrac{k}{2}c^{2}\varkappa \dint \dfrac{%
\rho }{r}d^{3}x, \\ 
\\ 
h_{44}=\dfrac{k}{2}c^{2}\Phi ,g_{44}=g_{44}^{\left( 0\right) }+\dfrac{k}{2}%
c^{2}\Phi , \\ 
\\ 
\Phi =-\varkappa \dint \dfrac{\rho }{r}d^{3}x \\ 
\end{array}
& \text{ \ }(2.2.47)%
\end{array}%
$

\bigskip

and according to Eq.(2.2.45)

\bigskip

\bigskip\ $\ \ \ \ \ \ \ \ \ \ \ \ \ \ \ \ \ \ \ \ \ \ \ \ \ \ \ \ \ \ \ \ \
\ \ \ \ \ \ \ \ \ \ 
\begin{array}{cc}
\begin{array}{c}
\\ 
\Delta \Phi =4\pi \varkappa \rho . \\ 
\end{array}
& \text{ \ }(2.2.48)%
\end{array}%
$

This expression corresponds exactly to Newton's equation of gravitation for
continuous

distribution of \ masses.

\bigskip $\ \ \ \ \ \ \ \ \ \ \ \ \ $

\bigskip\ \ \ \ \ \ \ \ \ \ \ \ \ \ \ \ \ \ \ \ \ \ \ \ \ \ \ \ \ \ \ \ \ \
\ \ \ \ \ \ \ \ \ \ \ \ \ \ \ \ 

\bigskip

\bigskip

\subsection{III. Variational Action Principles in Rozen's Bimetric Theory.}

\bigskip

\bigskip

General variational Action Principle [38] was introduced, in the Rosen's
Theory of Gravitation, in view of deriving field equations, motion
equations,canonical energy tensor. and conservative principles. Using the
constraint of metric invariance during the variational process along the
trajectory, a certain relationship between the canonical tensor and the
motion equations is established as a test for selfconsistency. As was
pointed out in the paper [38] the Equivalence Principle between
gravitational mass and inertial mass does hold in a weak version, i.e.
equality of masses but not also of their space distributions.

\bigskip

\subsection{III.1.Simple Variational Action Principle in Rozen's Bimetric \
Theory. Field Equations.\ \ \ \ \ \ \ \ \ \ \ \ \ \ \ \ \ \ \ \ \ \ \ \ \ \
\ \ \ \ \ \ \ \ \ \ \ \ \ \ \ \ \ \ \ \ \ \ \ \ \ \ \ \ \ \ \ \ \ \ \ \ \ \
\ \ \ \ \ \ \ \ \ }

\bigskip

The field and motion equations, as well as the canonical energy tensor, may
by derived, in the case of Rosen's theory of gravitation [5]-[7], from a
certain Action Principle adopting the perfect magnetic fluid scheme for the
matter tensor and specifying the field part of the Action as depending on
Minkowskian quantities of definite variance not exceeding the first order
derivatives, the Action integral takes the form

\bigskip\ $\ \ \ \ \ \ \ \ \ \ \ \ \ \ \ \ \ \ \ \ \ \ \ \ \ \ 
\begin{array}{cc}
\begin{array}{c}
\\ 
\mathbf{S=}\dfrac{1}{64\pi \varkappa }\dint \tciLaplace \left( g_{\mu \nu },%
\mathbf{\gamma }_{\mu \nu }\right) \sqrt{-\mathbf{\gamma }}d^{4}x+\mathbf{S}%
_{m} \\ 
\end{array}
& \text{ \ \ \ \ \ \ \ \ }(3.1.1)%
\end{array}%
$

where

\bigskip\ $\ \ \ \ \ \ \ \ \ \ \ \ \ \ \ \ \ \ \ \ \ \ \ \ \ \ \ 
\begin{array}{cc}
\begin{array}{c}
\\ 
\tciLaplace \left( g_{\mu \nu },\mathbf{\gamma }_{\mu \nu }\right) = \\ 
\\ 
\mathbf{\gamma }^{\mu \nu }g^{\alpha \beta }g^{\gamma \delta }\left(
g_{\alpha \gamma |\mu }g_{\alpha \delta |\nu }-\dfrac{1}{2}g_{\alpha \beta
|\mu }g_{\gamma \delta |\nu }\right) , \\ 
\end{array}
& \text{ \ \ \ \ }(3.1.2)%
\end{array}%
$

\bigskip

where the bar $\left( "|"\right) $ denotes covariant derivative with respect
to $\mathbf{\gamma }_{\mu \nu }$. The corresponding field equations may be
written in the form:

\bigskip\ $\ \ \ \ \ \ \ \ \ \ \ \ \ \ \ \ \ \ 
\begin{array}{cc}
\begin{array}{c}
\\ 
\square _{\gamma }g_{\mu \nu }-g^{\alpha \beta }\mathbf{\gamma }^{\gamma
\delta }=-16\pi \varkappa \sqrt{g/\mathbf{\gamma }}\left( T_{\mu \nu }-%
\dfrac{1}{2}g_{\mu \nu }T\right) \\ 
\end{array}
& \text{ \ \ \ \ \ \ }(3.1.3)%
\end{array}%
$

\bigskip

or in the form

\bigskip

$\ \ \ \ \ \ \ \ \ \ \ \ \ \ \ \ \ \ 
\begin{array}{cc}
\begin{array}{c}
\\ 
\gamma ^{\alpha \beta }g_{\mu \nu |\alpha \beta }-g^{\alpha \beta }\mathbf{%
\gamma }^{\gamma \delta }=-16\pi \varkappa \sqrt{g/\mathbf{\gamma }}\left(
T_{\mu \nu }-\dfrac{1}{2}g_{\mu \nu }T\right) \\ 
\end{array}
& \text{ \ \ }(3.1.4)%
\end{array}%
$

\bigskip

\bigskip \bigskip\ 

\begin{remark}
3.1.1.\textbf{\ }Note that in Rosen's theory of gravitation for the
derivetion\textbf{\ } gravitational field equations by simple variational
principle one varies the quantities $g_{\mu \nu }$, not the quantities $%
\gamma _{\mu \nu }$, in the simple variational principle.
\end{remark}

Suppose that the second metric tensor $\gamma _{ij}$ in Rosen's theory,
refers to space-time of constant curvature, given by Eq.(2.1.3).The
gravitational field equations reads

\bigskip\ $\ \ \ \ \ \ \ \ \ \ \ \ \ \ \ \ \ \ \ \ \ \ \ \ \ \ \ \ \ \ \ \ \
\ \ \ \ \ \ 
\begin{array}{cc}
\begin{array}{c}
\\ 
K_{\mu \nu }-\dfrac{1}{2}Kg_{\mu \nu }=-8\pi kT_{\mu \nu } \\ 
\end{array}
& \text{ \ \ \ \ \ \ }(3.1.5)%
\end{array}%
$

or\bigskip

\ 

$\bigskip \ \ \ \ \ \ \ \ \ \ \ \ \ \ \ \ \ 
\begin{array}{cc}
\begin{array}{c}
\\ 
R_{\mu \nu }-\dfrac{1}{2}Rg_{\mu \nu }-\dfrac{3}{a^{2}}\left( \gamma _{\mu
\nu }-\dfrac{1}{2}g^{\alpha \underline{\alpha }}g_{\mu \nu }\right) =-8\pi
kT_{\mu \nu }, \\ 
\\ 
k=\dfrac{\varkappa }{c^{4}}. \\ 
\end{array}
& \text{ \ \ \ \ \ \ \ \ }(3.1.6)%
\end{array}%
$

\bigskip

One sees that (3.1.6) differs from the Einstein field equations by an
additional term on the left hand side.\bigskip

\bigskip Suppose that the background space-time metric $\gamma _{\mu \nu }$
corresponding to the metric $g_{\mu \nu }$ is

\bigskip\ $\ \ \ \ \ \ \ \ \ \ \ \ \ \ \ \ \ \ \ \ \ \ \ \ \ \ \ \ \ \ \ \ \
\ 
\begin{array}{cc}
\begin{array}{c}
\\ 
ds_{2}^{2}=dt^{2}-dx^{2}-dy^{2}-dz^{2}. \\ 
\end{array}
& \text{ \ \ \ \ \ \ }(3.1.7)%
\end{array}%
$

The energy momentum tensor for perfect fluid is given by

\bigskip\ $\ \ \ \ \ \ \ \ \ \ \ \ \ \ \ \ \ \ \ \ \ \ \ \ \ \ \ \ \ \ \ \ \
\ \ 
\begin{array}{cc}
\begin{array}{c}
\\ 
T_{\alpha \beta }^{FL}=\left( P+\rho \right) u_{\alpha }u_{\beta }-g_{\alpha
\beta }P. \\ 
\end{array}
& \text{ \ \ \ \ \ }(3.1.8)%
\end{array}%
$

\bigskip The Rozen-Maxwell equations for ideal MHD flow are

\bigskip\ $\ \ \ \ \ \ \ \ \ \ \ \ \ \ \ \ \ \ \ \ \ \ \ \ \ \ \ \ \ \ 
\begin{array}{cc}
\begin{array}{c}
\\ 
K_{\alpha \beta }-\dfrac{1}{2}Kg_{\alpha \beta }=-8\pi k\left( T_{\alpha
\beta }^{FL}+T_{\alpha \beta }^{EM}\right) , \\ 
\\ 
k=\varkappa \sqrt{\dfrac{g}{\gamma }}, \\ 
\\ 
K_{\alpha \beta }=K_{\alpha }^{\beta }=\dfrac{1}{2}\breve{g}^{\mu \nu
}\left( g^{h\alpha }g_{h\beta |\mu }\right) _{|\nu }, \\ 
\end{array}
& \text{ \ \ \ \ \ }(3.1.9)%
\end{array}%
$

\bigskip

where the energy momentum tensor of the electromagnetic field denoted by $%
T_{\alpha \beta }^{EM}.\ \ \ \ \ \ \ \ \ \ \ \ \ \ \ \ \ \ \ \ \ \ \ \ \ \ \
\ \ \ \ \ \ \ \ \ \ \ \ \ \ \ \ \ \ \ \ \ $

\bigskip

The electromagnetic field tensor for the MHD fluid is given by the covariant
expression\bigskip\ \ \ \ \ \ \ \ \ \ \ \ \ \ \ \ \ \ \ \ \ 

\bigskip\ $\ \ \ \ \ \ \ \ \ \ \ \ \ \ \ \ \ \ \ \ \ \ \ \ \ \ \ \ \ 
\begin{array}{cc}
\begin{array}{c}
\\ 
\ F_{\alpha \beta }=u_{\alpha }E_{\beta }-u_{\beta }E_{\alpha }+\eta
_{\alpha \beta \gamma \delta }u^{\gamma }B^{\delta } \\ 
\end{array}
& \text{ \ \ \ }(3.1.10)%
\end{array}%
$

\bigskip and similarly in a contravariant form

\bigskip\ $\ \ \ \ \ \ \ \ \ \ \ \ \ \ \ \ \ \ \ \ \ \ \ \ \ \ \ 
\begin{array}{cc}
\begin{array}{c}
\\ 
\ F^{\alpha \beta }=u^{\alpha }E^{\beta }-u^{\beta }E^{\beta }+\eta ^{\alpha
\beta \gamma \delta }u_{\gamma }B_{\delta }, \\ 
\end{array}
& \text{ \ \ \ }(3.1.11)%
\end{array}%
$

where the four vectors $E_{\alpha }$ and $B_{\alpha },$ denoting the
electric and magnetic field components in the four dimensional spacetime,
are orthogonal to the velocity four vector $u^{\alpha }.$ Here the volume
element 4-form of $V_{4}$ namely $\eta _{\alpha \beta \gamma \delta }$ and
its dual $\eta ^{\alpha \beta \gamma \delta }$ is defined

\bigskip

$\ \ \ \ \ \ \ \ \ \ \ \ \ \ \ \ \ \ \ \ \ \ \ \ \ \ \ \ \ \ \ \ \ \ \ \ \ \
\ \ \ 
\begin{array}{cc}
\begin{array}{c}
\\ 
\eta _{\alpha \beta \gamma \delta }=\sqrt{-g}\epsilon _{\alpha \beta \gamma
\delta }, \\ 
\\ 
\eta ^{\alpha \beta \gamma \delta }=\dfrac{1}{\sqrt{-g}}\epsilon ^{\alpha
\beta \gamma \delta }, \\ 
\end{array}
& \text{ \ \ \ }(3.1.13)%
\end{array}%
$

\bigskip

where $g$ represents the determinant of the metric tensor g$_{\alpha \beta }$
and $\epsilon _{\alpha \beta \gamma \delta }$ is the Levi -Civita symbol,
which is $+1,-1,$ and $0$ for a cyclic, anti-cyclic, and noncyclic
permutation of $\alpha \beta \gamma \delta $ respectively. It should be
noticed that the choice of spacetime $V_{4}$ is quite general here, namely
of a four dimensional vector space, or more generally of a differentiable
manifold of four dimensions. Also the signature of an axially symmetric
metric defined on this manifold must be either $(+---)$ or $(-+++).$

\bigskip \bigskip

\subsection{III.2.General Variational Action Principle in Rozen's Bimetric
Theory.}

\bigskip

Let's consider bimetric Rosen's space-time $\Re \left( M,g_{\mu \nu },\breve{%
g}_{\mu \nu }\right) $ with $R\left( g_{\mu \nu }\right) \neq 0,$ $R\left( 
\breve{g}_{\mu \nu }\right) =0.$

\bigskip Action integral takes the form

\bigskip\ $\ \ \ \ \ \ \ \ \ \ \ \ \ \ \ \ \ \ \ \ \ \ \ 
\begin{array}{cc}
\begin{array}{c}
\\ 
\mathbf{S=}\dfrac{1}{c}\dint \tciLaplace \sqrt{-\breve{g}}d^{4}x, \\ 
\\ 
\tciLaplace \left( g_{\mu \nu },\breve{g}_{\mu \nu }\right) =\tciLaplace
_{m}\left( g_{\mu \nu },\breve{g}_{\mu \nu }\right) +\tciLaplace _{f}\left(
\chi ^{\mu \nu };g_{\mu \nu |\lambda },\breve{h}^{\mu \nu }\right) , \\ 
\end{array}
& \text{ \ \ \ \ }(3.2.1)%
\end{array}%
$

where

\bigskip\ $\ \ \ \ \ \ \ \ \ \ \ \ \ \ \ \ \ \ \ \ \ \ \ \ \ \ 
\begin{array}{cc}
\begin{array}{c}
\\ 
\tciLaplace _{m}=\left[ \left( c^{2}+\Xi \right) \rho -p\right] _{g}\cdot
K,K=\sqrt{\dfrac{-g}{-\breve{g}}}, \\ 
\\ 
\tciLaplace _{f}=\dfrac{c^{4}}{16\pi G}f\left( \chi ^{\alpha \beta
},g_{\alpha \beta |\lambda };\breve{h}^{\alpha \beta }\right) . \\ 
\\ 
g_{\mu \lambda }\chi ^{\lambda \nu }=\delta _{\mu }^{\nu };\text{ }\breve{g}%
_{\mu \lambda }\breve{h}^{\lambda \nu }=\delta _{\mu }^{\nu }, \\ 
\\ 
g^{\alpha \beta }=g_{\mu \lambda }\breve{h}^{\mu \alpha }\breve{h}^{\nu
\beta }\neq \chi ^{\alpha \beta }, \\ 
\\ 
\chi _{\alpha \beta }=\chi ^{\mu \nu }\breve{g}_{\mu \alpha }\breve{g}_{\nu
\beta }\neq g_{\alpha \beta }. \\ 
\\ 
\Xi =\int_{0}^{p\left( \rho \right) }\dfrac{dp}{p\left( \rho \right) }, \\ 
\\ 
p=p\left( \rho \right) ,\rho =\rho \left( p\right) , \\ 
\\ 
g=\det \left\Vert g_{\mu \lambda }\right\Vert <0,\breve{g}=\det \left\Vert 
\breve{g}_{\mu \lambda }\right\Vert <0. \\ 
\end{array}
& \text{ \ \ \ \ \ \ \ \ }(3.2.2)%
\end{array}%
$

Here, a subscript $g$ stands for specifying that the labelled quantity is
defined by curved space-time metric $ds_{1}^{2}=g_{\alpha \beta }dx^{\alpha
}dx^{\alpha }.$ Besid this, a flat space-time metric is $ds_{2}^{2}=\breve{g}%
_{\mu \nu }dx^{\mu }dx^{\nu }.$

\bigskip

The quantity $K$ in (3.2.2)) is a Minkowskian scalar, while the quantities $%
\rho $ (mass density) $p$ (pressure) and $\Xi $ (Helmoltz potential) are
scalar in both $M_{4}$ and $\tciLaplace \left( M,g_{\alpha \beta }\right) $.

\bigskip

For ensuring coherence of the whole variational process, not only the
coordinates $x^{\alpha }$ and the signatures (time-like) should be the same
for the two metrics, but also the metric tensor $g_{\alpha \beta }$ of the
curved Lorentzian space-time $\tciLaplace _{g}=\tciLaplace \left(
M,g_{\alpha \beta }\right) $ must be considered as a tensor in $M_{4}$
(Minkowskian space time). Arbitrary coordinates in $M_{4}$ are adopted
(necessary for carrying out the variational calculations) and $x^{0}=ct$ is
taken as the time coordinate (with physical dimension of a length). So, in
Rosen's theory the role of the two metrics is strongly dissymmetrized, $%
g_{\alpha \beta }$ are some quantities preserving only the meaning of
gravitational potentials, and the metric $ds_{1}^{2}=g_{\alpha \beta
}dx^{\alpha }dx^{\alpha }$ turns out to be a simple mathematical artifact
necessary to formulate the specific coupling of gravitational field to its
sources This bimetric philosophy (which restricts the main role of
Lorentzian metric $ds_{1}^{2}$ to the motion equations) entitles us to treat
the quantities $g_{\mu \nu }$, as true gravitational potentials, distinct
from the metric functions $\breve{g}_{\mu \nu }$. Now, performing the
variations against $g_{\mu \nu },$ in the action integral, we come to field
equations; variation against $\breve{g}_{\mu \nu }$ delivers a canonical
energy tensor, while the variation against the coordinates of a fluid
particle delivers motion equations.

The Minkowskian covariant derivatives are denoted by a vertical bar ($|$)
followed by a certain (Greek) subscript, or (equivalently) by a derivative
symbol ($D_{\alpha }$) followed by the same subscript. For example

\bigskip

\bigskip\ $\ \ \ \ \ \ \ \ \ \ \ \ \ \ \ \ \ \ \ \ \ \ \ 
\begin{array}{cc}
\begin{array}{c}
\\ 
g_{\alpha \beta |\lambda }=\breve{D}_{\lambda }g_{\alpha \beta }=g_{\alpha
\beta ,\lambda }-\breve{G}_{\alpha \lambda }^{\text{ }\sigma }g_{\alpha
\beta }-\breve{G}_{\beta \lambda }^{\text{ }\sigma }g_{\sigma \alpha }, \\ 
\\ 
\breve{G}_{\mu \nu }^{\lambda }=\dfrac{1}{2}\breve{h}^{\lambda \sigma
}\left( \breve{g}_{\mu \sigma ,\eta }+\breve{g}_{\nu \sigma ,\rho }-\breve{g}%
_{\mu \nu ,\sigma }\right) . \\ 
\end{array}
& \text{ \ \ \ \ \ \ \ \ \ \ \ \ \ \ }(3.2.3)%
\end{array}%
$

\bigskip

For obtaining the field equations and the energy canonical tensor one obtain
to the following identity

\bigskip\ $\ \ 
\begin{array}{cc}
\begin{array}{c}
\\ 
\dfrac{1}{\sqrt{-\breve{g}}}\delta \left( \sqrt{-\breve{g}}L\right) =\breve{D%
}_{\lambda }q^{\lambda }- \\ 
\\ 
\dfrac{1}{2}K\left( \left[ T_{\alpha \beta }\right] _{g}+\dfrac{c^{4}}{8\pi G%
}E_{\alpha \beta }\right) \delta \chi ^{\alpha \beta }-\dfrac{1}{2}\Im
_{\alpha \beta }\delta \breve{h}^{\alpha \beta }, \\ 
\end{array}
& \text{ \ }(3.2.4)%
\end{array}%
$

\bigskip

where

\bigskip 

\bigskip $%
\begin{array}{cc}
\begin{array}{c}
\\ 
q^{\lambda }=\mathbf{P}^{\alpha \beta ||\lambda }\delta g_{\alpha \beta
}+2\gamma _{\sigma \nu }\mathbf{P}^{\mu \nu ||\eta }\Omega _{\nu \mu
||\alpha \beta }\delta h^{\alpha \beta }, \\ 
\\ 
\mathbf{P}^{\alpha \beta ||\lambda }=\dfrac{\partial L_{f}}{\partial
g_{\alpha \beta |\lambda }}, \\ 
\\ 
\breve{\Omega}_{\nu \mu ||\alpha \beta }^{\lambda \sigma }=\dfrac{1}{4}%
\left( \delta _{\beta }^{\lambda }\delta _{\nu }^{\sigma }\breve{g}_{\mu
\alpha }+\delta _{\lambda }^{\alpha }\delta _{\nu }^{\sigma }\breve{g}_{\mu
\beta }+\delta _{\beta }^{\lambda }\delta _{\mu }^{\sigma }\breve{g}_{\nu
\alpha }+\right.  \\ 
\\ 
+\delta _{\alpha }^{\lambda }\delta _{\mu }^{\sigma }\breve{g}_{\nu \beta }-%
\breve{h}^{\lambda \sigma }\breve{g}_{\mu \alpha }\breve{g}_{\nu \beta }-%
\breve{h}^{\lambda \sigma }\breve{g}_{\mu \beta }\breve{g}_{\nu \alpha }),
\\ 
\\ 
\left[ T_{\alpha \beta }\right] _{g}=\left\{ \left( c^{2}+\Xi \right) \rho
U_{\alpha }U_{\beta }-pg_{\alpha \beta }\right\} _{g}, \\ 
\\ 
E_{\alpha \beta }=\widetilde{R}_{\alpha \beta }-\dfrac{1}{2}g_{\alpha \beta
}\left( \chi ^{\mu \nu }\widetilde{R}_{\mu \nu }\right)  \\ 
\\ 
\widetilde{R}_{\alpha \beta }=-\dfrac{1}{K}\left\{ \left( \dfrac{\partial f}{%
\partial \chi _{\alpha \beta }}-\dfrac{1}{2}g_{\alpha \beta }\chi ^{\mu \nu }%
\dfrac{\partial f}{\partial \chi ^{\mu \nu }}\right) \right. + \\ 
\\ 
\left. \left( g_{\mu \alpha }g_{\nu \beta }-\dfrac{1}{2}g_{\alpha \beta
}g_{\mu \nu }\right) D_{\lambda }p^{\mu \nu ||\lambda }\right\} , \\ 
\\ 
p^{\alpha \beta ||\lambda }=\dfrac{16\pi G}{c^{4}}\mathbf{P}^{\alpha \beta
||\lambda }, \\ 
\\ 
\Im _{\alpha \beta }=\breve{D}_{\sigma }Q_{\alpha \beta }^{\sigma }-2\left( 
\dfrac{\partial L_{f}}{\partial \breve{h}^{\alpha \beta }}-\dfrac{1}{2}%
\breve{g}_{\alpha \beta }L_{f}\right)  \\ 
\\ 
Q_{\alpha \beta }^{\sigma }=4g_{\lambda \zeta }\mathbf{P}^{\mu \zeta ||\nu }%
\breve{\Omega}_{\mu \nu ||\alpha \beta }^{\lambda \sigma }. \\ 
\end{array}
& \text{ \ }(3.2.5)%
\end{array}%
$

Putting the variation of the action integral against $\chi ^{\mu \nu }$ to
vanish, one obtains the field equations

\bigskip\ $%
\begin{array}{cc}
\begin{array}{c}
\\ 
\widetilde{R}_{\alpha \beta }-\dfrac{1}{2}g_{\alpha \beta }\left( \chi ^{\mu
\nu }\widetilde{R}_{\mu \nu }\right) =-\dfrac{8\pi G}{c^{4}}\left[ T_{\alpha
\beta }\right] _{g} \\ 
\end{array}
& \text{ \ }(3.2.6)%
\end{array}%
$

Rosen's gravitation field equations may be obtained out of the general
bimetric theory so far presented by specifying the function $f$

\bigskip

$%
\begin{array}{cc}
\begin{array}{c}
\\ 
f=-\dfrac{1}{8}\breve{h}^{\alpha \beta }W^{\lambda \sigma \rho \tau }\left(
g_{\lambda \sigma |\alpha }\cdot g_{\rho \tau |\beta }\right) , \\ 
\\ 
W^{\lambda \sigma \rho \tau }=\chi ^{\lambda \rho }\chi ^{\sigma \tau }+\chi
^{\sigma \rho }\chi ^{\lambda \tau }-\chi ^{\lambda \sigma }\chi ^{\rho \tau
}. \\ 
\end{array}
& \text{ \ }(3.2.7)%
\end{array}%
$

\bigskip

Hence

\bigskip 

$%
\begin{array}{cc}
\begin{array}{c}
\\ 
\widetilde{R}_{\mu \nu }=\dfrac{1}{2K}\left( \breve{h}^{\alpha \beta }\breve{%
D}_{\alpha }\breve{D}_{\beta }g_{\mu \nu }-\chi ^{\lambda \sigma }g_{\mu
\lambda |\alpha }g_{\nu \sigma }^{\text{ }|\alpha }\right) , \\ 
\end{array}
& \text{ \ }(3.2.8)%
\end{array}%
$

\bigskip

and

\bigskip 

$\ 
\begin{array}{cc}
\begin{array}{c}
\\ 
\Im _{\alpha \beta }=\dfrac{c^{4}}{32\pi G}\left\{ \chi ^{\mu \lambda }\chi
^{\nu \sigma }\left( g_{\mu \nu |\alpha }g_{\lambda \sigma |\beta }-\dfrac{1%
}{2}\breve{g}_{\alpha \beta }g_{\mu \nu |\eta }g_{\lambda \sigma }^{\text{ }%
|\eta }\right) \right. - \\ 
\\ 
2\left[ \left( \ln K\right) _{,\alpha }\left( \ln K\right) _{,\beta }-\dfrac{%
1}{2}\breve{g}_{\alpha \beta }\breve{h}^{\mu \nu }\left( \ln K\right) _{,\mu
}\left( \ln K\right) _{,\nu }\right] -\mathbf{P}_{\alpha \beta }, \\ 
\end{array}
& \text{ \ }(3.2.9)%
\end{array}%
$

where

\bigskip 

\bigskip $%
\begin{array}{cc}
\begin{array}{c}
\\ 
\mathbf{P}_{\alpha \beta }=\left( g_{\mu \beta |\alpha }\chi ^{\mu \nu
}\right) _{|\nu }+\left( g_{\mu \alpha |\beta }\chi ^{\mu \nu }\right)
_{|\nu }+\left( g_{\mu \beta }^{\text{ }|\nu }\chi _{\alpha }^{\mu }\right)
_{|\nu }+ \\ 
\\ 
\left( g_{\mu \alpha }^{\text{ }|\nu }\chi _{\beta }^{\mu }\right) _{|\nu
}-\left( g_{\mu |\alpha }^{\text{ }\nu }\chi _{\beta }^{\mu }\right) _{|\nu
}-\left( g_{\mu |\beta }^{\text{ }\nu }\chi _{\alpha }^{\mu }\right) _{|\nu
}-2\breve{g}_{\alpha \beta }\breve{h}^{\alpha \beta }\breve{D}_{\alpha }%
\breve{D}_{\beta }\ln K, \\ 
\end{array}
& \text{ \ }(3.2.10)%
\end{array}%
$

\bigskip

where the covariant derevative $\left( ...^{|\alpha }\right) $ is a lift via 
$\breve{h}^{\alpha \beta }.$\bigskip \bigskip \bigskip \bigskip

\subsection{IV.Bimetric Theory of Gravitational-Inertial Field \textbf{in
Riemannian approximation.}}

\bigskip

Bimetric Theory of Gravitational-Inertial Field in Riemannian approximation
to proceed from assumptions:

\begin{itemize}
\item \textbf{I}. Bimetric geometrical structures $\Re _{2}\left( M,g_{ij},%
\breve{g}_{ij}\right) $ of the space-time continuum on the standard
assumption of Bimetric Lorentzianian geometry:
\end{itemize}

\bigskip

$\ \ 
\begin{array}{cc}
\begin{array}{c}
\\ 
ds_{1}^{2}=g_{ik}dx^{i}dx^{k},g_{ik}=g_{ki},\det \left\Vert
g_{ik}\right\Vert \neq 0; \\ 
\\ 
ds_{2}^{2}=\mathbf{\breve{g}}_{ik}dx^{i}dx^{k},\mathbf{\breve{g}}_{ik}=%
\mathbf{\breve{g}}_{ki},\det \left\Vert \breve{g}_{ik}\right\Vert \neq 0; \\ 
\end{array}
& \text{ \ }(4.1)%
\end{array}%
$

\bigskip

\begin{itemize}
\item \textbf{II.} Equivalence of gravitational-inertial field and
spacc-time metric tenzor
\end{itemize}

$\bigskip $

$g_{ik}\left( M\right) =g_{ik}\left( x_{1},x_{2},x_{3},,x_{4}\right) ;$

\bigskip

\begin{itemize}
\item \bigskip \textbf{III}. Equivalence of pure\ accelerational field $%
\mathbf{\breve{g}}_{ik}$ and space-time metric tenzor
\end{itemize}

$\mathbf{\breve{g}}_{ik}\left( M\right) =\mathbf{\breve{g}}_{ik}\left(
x_{1},x_{2},x_{3},,x_{4}\right) .$

\begin{axiom}
Bimetric Theory of Gravitational-Inertial Field in Riemannian approximation
is based on the following postulates:
\end{axiom}

\textbf{1}.In nonrelativistic approximation, i.e. if reference frame (body)
is accelerate by a sufficiently small external nongravitational force and
very far from the localized masses, then the metric tensors $g_{ik}$ and $%
\mathbf{\breve{g}}_{ik}$ describes a flat space-time, i.e. $%
R_{klm}^{i}\left( g_{ik}\right) =0$ and $\mathbf{\breve{R}}_{klm}^{i}\left( 
\mathbf{\breve{g}}_{ik}\right) =0.$

\textbf{2}. A sufficiently small domain of \ bimetric space-time $\Re
_{2}\left( M,g_{ij},\mathbf{\breve{g}}_{ij}\right) $ is flat, i.e. $%
R_{klm}^{i}\approx 0$ and $\mathbf{\breve{R}}_{klm}^{i}\approx 0.$

\textbf{3.} If reference frame (body) is accelerate by an arbitrary external
nongravitational force but very far avay from the localized masses, then $%
g_{ik||l}\simeq 0.$

\textbf{4. }SEP is satisfied. In particular:\textbf{\ } there is no
experiment observers can perform to distinguish whether an acceleration
arises because of a gravitational force or because their reference frame is
accelerating by an external nonravitational force.

\bigskip

\bigskip

\bigskip

\subsection{IV.1.Simple Variational Action Principle in Bimetric Theory of
Gravitational-Inertial Field.}

\bigskip

Let's consider Gravitational-Inertial field theory (GIFTR) such that at each
point of Lorentzian space-time $\left( \tciLaplace ,g_{ij}\right) $ a curved
Lorentzian metric tensor $\breve{g}_{ij}$ in addition to the curved
Lorentzian metric tensor $g_{ij}.$ Thus at each point of \ generalized\
Rosen's space-time $\Re _{2}=\Re _{2}\left( M,g_{ij},\breve{g}_{ij}\right) $
there are two curved metrics:

\bigskip

\bigskip\ $\ 
\begin{array}{cc}
\begin{array}{c}
\\ 
ds_{1}^{2}=g_{ij}dx^{i}dx^{j}, \\ 
\\ 
ds_{2}^{2}=\breve{g}_{ij}dx^{i}dx^{j}. \\ 
\end{array}
& \text{ \ \ }(4.1.1)%
\end{array}%
$

\begin{notation}
The first metric tensor $g_{ij}$ in GIFTR, refers to the curved space-time
and describes Gravitational-Inertial Field. The second metric tensor $\breve{%
g}_{ij}\triangleq g_{ij}^{\mathbf{ac.}}$ in GIFTR, space-time and describes
pure inertial forces. The Christoffel symbols formed from $g_{ij}$ and $%
\breve{g}_{ij}$ are denoted by $\Gamma _{jk}^{i}$ and $\breve{\Gamma}%
_{jk}^{i}$ respectively.
\end{notation}

\bigskip The quantities $\tilde{\Delta}_{jk}^{i}$ are defined via formulae

\bigskip\ $\ \ 
\begin{array}{cc}
\begin{array}{c}
\\ 
\tilde{\Delta}_{jk}^{i}=\Gamma _{jk}^{i}-\breve{\Gamma}_{jk}^{i}. \\ 
\end{array}
& \text{ }(4.1.2)%
\end{array}%
$

\begin{remark}
4.1.1.\textbf{\ }Let $R_{\mu \nu \lambda }^{\alpha }$\textbf{\ }and $\breve{R%
}_{\mu \nu \lambda }^{\alpha }$\textbf{\ }be the curvature tensors
calculated from $g_{\mu \nu }$ and $\breve{g}_{\mu \nu }$ respectively. We
set $R_{\mu \nu \lambda }^{\alpha }\neq 0,\breve{R}_{\mu \nu \lambda
}^{\alpha }\neq 0.$
\end{remark}

\bigskip Now there arise two kinds of covariant differentiation:

\begin{itemize}
\item (\textbf{1}) $g$-differentiation based on $g_{\mu \nu }$ (denoted by a
semicolon $(;)$)
\end{itemize}

$\ \ \ 
\begin{array}{cc}
\begin{array}{c}
\\ 
A_{\mu \nu ;\lambda }=\left( A_{\mu \nu ,\lambda }-\Gamma _{\mu \text{ }%
\lambda }^{\alpha }A_{\alpha \nu }-\Gamma _{\nu \text{ }\lambda }^{\alpha
}A_{\alpha \mu }\right) \\ 
\end{array}
& \text{ }(4.1.3)%
\end{array}%
$

\bigskip

\begin{itemize}
\item (\textbf{2}) $\breve{g}$-differentiation based on $\breve{g}_{\mu \nu
} $ (denoted by a bislash $(||)$)
\end{itemize}

\bigskip

$\ \ 
\begin{array}{cc}
\begin{array}{c}
\\ 
A_{\mu \nu ||\lambda }=\left( A_{\mu \nu ,\lambda }-\breve{\Gamma}_{\mu 
\text{ }\lambda }^{\alpha }A_{\alpha \nu }-\breve{\Gamma}_{\nu \text{ }%
\lambda }^{\alpha }A_{\alpha \mu }\right) , \\ 
\end{array}
& (4.1.4)%
\end{array}%
$

\bigskip

\bigskip

where ordinary partial derivatives are denoted by comma $(,)$.

\bigskip The straightforward calculations gives

\bigskip $%
\begin{array}{cc}
\begin{array}{c}
\\ 
\tilde{R}_{\mu \nu \lambda }^{\alpha }=-\tilde{\Delta}_{\mu \nu ||\lambda
}^{\alpha }+\tilde{\Delta}_{\mu \lambda ||\nu }^{\alpha }-\tilde{\Delta}%
_{\beta \nu }^{\alpha }\tilde{\Delta}_{\mu \lambda }^{\beta }-\tilde{\Delta}%
_{\beta \lambda }^{\alpha }\tilde{\Delta}_{\mu \nu }^{\beta }. \\ 
\end{array}
& \text{ \ }(4.1.5)%
\end{array}%
$

\bigskip

Hence

\bigskip 

$%
\begin{array}{cc}
\begin{array}{c}
\\ 
\tilde{R}_{\mu \nu }=-\tilde{\Delta}_{\mu \nu ||\alpha }^{\alpha }+\tilde{%
\Delta}_{\alpha \mu ||\nu }^{\alpha }-\tilde{\Delta}_{\alpha \beta }^{\alpha
}\Delta _{\mu \nu }^{\beta }-\tilde{\Delta}_{\beta \mu }^{\alpha }\tilde{%
\Delta}_{\alpha \nu }^{\beta }. \\ 
\end{array}
& \text{ \ \ }(4.1.6)%
\end{array}%
$\ \ 

\ \ 

\bigskip

This is the curvature tensor $\tilde{R}_{\mu \nu }$ associated with the
curvature effects of pure gravitation acting in the bimetric spacetime $\Re
_{2}=\Re _{2}\left( M,g_{ij},\breve{g}_{ij}\right) .$

\bigskip The geodesic equation in bimetric spacetime $\Re _{2}$ takes the
form:

\bigskip $%
\begin{array}{cc}
\begin{array}{c}
\\ 
\dfrac{d^{2}x^{i}}{ds_{1}}+\tilde{\Delta}_{jk}^{i}\dfrac{dx^{j}}{ds_{1}}%
\dfrac{dx^{k}}{ds_{1}}= \\ 
\\ 
\dfrac{d^{2}x^{i}}{ds_{1}}+\Gamma _{jk}^{i}\dfrac{dx^{j}}{ds_{1}}\dfrac{%
dx^{k}}{ds_{1}}-\breve{\Gamma}_{jk}^{i}\dfrac{dx^{j}}{ds_{1}}\dfrac{dx^{k}}{%
ds_{1}}=0. \\ 
\end{array}
& \text{ }(4.1.7)%
\end{array}%
$

\bigskip

Action integral takes the form

\bigskip

$%
\begin{array}{cc}
\begin{array}{c}
\\ 
\mathbf{S=}\left[ \mathbf{S}_{1}\left( K,\breve{R}\right) \right] _{\breve{g}%
}+\left[ \mathbf{S}_{2}\left( \breve{R}\right) \right] _{\breve{g}}+\mathbf{S%
}_{m}= \\ 
\\ 
\dfrac{1}{64\pi \varkappa }\dint \tciLaplace _{1}\left( g_{\mu \nu },\mathbf{%
\breve{g}}_{\mu \nu }\right) \sqrt{-\mathbf{\breve{g}}}d^{4}x+\dfrac{1}{8\pi
\varkappa }\dint \tciLaplace _{2}\left( \mathbf{\breve{g}}_{\mu \nu }\right) 
\sqrt{-\mathbf{\breve{g}}}d^{4}x+\mathbf{S}_{m}, \\ 
\end{array}
& \text{ \ }(4.1.8)%
\end{array}%
$

\bigskip

where

\bigskip 

\bigskip $%
\begin{array}{cc}
\begin{array}{c}
\\ 
\tciLaplace _{1}\left( g_{\mu \nu },\mathbf{\breve{g}}_{\mu \nu }\right) =
\\ 
\\ 
\mathbf{\breve{g}}^{\mu \nu }g^{\alpha \beta }g^{\gamma \delta }\left(
g_{\alpha \gamma ||\mu }g_{\alpha \delta ||\nu }-\dfrac{1}{2}g_{\alpha \beta
||\mu }g_{\gamma \delta ||\nu }\right) , \\ 
\end{array}
& \text{ }(4.1.9)%
\end{array}%
$

\bigskip

where the duble bar $\left( "||"\right) $ denotes covariant derivative with
respect to $\mathbf{\breve{g}}_{\mu \nu }$. The corresponding field
equations may be written in the form:

\bigskip 

\bigskip $%
\begin{array}{cc}
\begin{array}{c}
\\ 
\square _{\mathbf{\breve{g}}}g_{\mu \nu }-g^{\alpha \beta }\mathbf{\breve{g}}%
^{\gamma \delta }=-16\pi \varkappa \sqrt{g/\mathbf{\breve{g}}}\left( \left[
T_{\mu \nu }\right] _{g}-\dfrac{1}{2}g_{\mu \nu }\left[ T\right] _{g}\right)
, \\ 
\\ 
\breve{\Theta}_{\mu \nu }+\breve{E}_{\mu \nu }=k_{1}\left[ \breve{T}^{\mu
\nu }\right] _{\breve{g}}, \\ 
\\ 
\breve{\Theta}_{\mu \nu }=\dfrac{\delta \mathbf{S}\left( R,\breve{R}\right) 
}{\delta \breve{g}_{\mu \nu }}, \\ 
\\ 
\breve{E}_{\mu \nu }=\breve{R}_{\mu \nu }-\dfrac{1}{2}\breve{R}g_{\mu \nu },
\\ 
\\ 
\left[ \breve{T}_{;\nu }^{\mu \nu }\right] _{\breve{g}}=\breve{F}^{\mu }. \\ 
\end{array}
& \text{ \ \ }(4.1.10)%
\end{array}%
$

\bigskip

or in the form

\bigskip 

\bigskip $%
\begin{array}{cc}
\begin{array}{c}
\\ 
\mathbf{\breve{g}}^{\alpha \beta }g_{\mu \nu ||\alpha \beta }-g^{\alpha
\beta }\mathbf{\breve{g}}^{\gamma \delta }=-16\pi \varkappa \sqrt{g/\mathbf{%
\breve{g}}}\left( \left[ T_{\mu \nu }\right] _{g}-\dfrac{1}{2}g_{\mu \nu }%
\left[ T\right] _{g}\right) , \\ 
\\ 
\breve{\Theta}_{\mu \nu }+\breve{E}_{\mu \nu }=k_{1}\left[ \breve{T}^{\mu
\nu }\right] _{\breve{g}}. \\ 
\end{array}
& \text{ \ }(4.1.11)%
\end{array}%
$

\bigskip

Here, a subscripts $g,\breve{g}$ stands for specifying that the labelled
quantity is defined by curved space-time metrics $ds_{1}^{2}=g_{\alpha \beta
}dx^{\alpha }dx^{\alpha }$ and $ds_{2}^{2}=\breve{g}_{\mu \nu }dx^{\mu
}dx^{\nu }$ respectively and $\breve{F}^{\mu }$ denote 4-vector of a pure
nongravitational force and vertical duble bar $||$ stands for covariant
differentiation with respect to $\breve{g}_{\mu \nu }.$

\bigskip

\subsection{IV.2.The Weak Field Limit of the Bimetric Theory of
Gravitational-Inertial Field in Riemannian Approximation.}

\bigskip

\bigskip

\subsection{IV.2.1.}

\bigskip

Let us recall the weak field limit procedure of the GTR.

Using Cartesian space coordinates, the metric determined from the radar
method is,

\bigskip

\bigskip $%
\begin{array}{cc}
\begin{array}{c}
\\ 
g_{ab}=g^{ab}=\left\Vert 
\begin{array}{cccc}
k^{-2} & 0 & 0 & 0 \\ 
0 & -k^{2} & 0 & 0 \\ 
0 & 0 & -k^{2} & 0 \\ 
0 & 0 & 0 & -k^{2}%
\end{array}%
\right\Vert  \\ 
\end{array}
& \text{\ }(4.2.1.1)%
\end{array}%
$

\bigskip $\ \ \ \ \ \ \ \ \ \ \ \ \ \ \ \ \ \ \ \ \ \ \ \ \ \ \ \ \ \ \ \ \
\ $

The partial derivatives of the metric are

\bigskip

\bigskip\ $%
\begin{array}{cc}
\begin{array}{c}
\\ 
g_{ab,c}=\left\Vert 
\begin{array}{cccc}
-2k^{-3}k_{,c} & 0 & 0 & 0 \\ 
0 & -2kk_{,c} & 0 & 0 \\ 
0 & 0 & -2kk_{,c} & 0 \\ 
0 & 0 & 0 & -2kk_{,c}%
\end{array}%
\right\Vert  \\ 
\end{array}
& \text{ \ }(4.2.1.2)%
\end{array}%
$

\bigskip

The geodesic equation is $\dot{x}^{a}=-\Gamma _{bc}^{a}\dot{x}^{b}\dot{x}%
^{c} $ where the Christoffel symbols are

$\Gamma _{abc}=\left( g_{ab,c}+g_{ac,b}-g_{cb,a}\right) /2.$ For the
Christoffel symbols to be non-zero, two indices must be the same, and, in a
constant field, the other must not be zero. For non-relativistic velocities,
terms in the order of velocity squared can be ignored, and we have $\dot{x}%
^{0}\approx 1$. Then, 3-acceleration is given by, for $a\neq 0,\ddot{x}%
^{a}\approx -\Gamma _{00}^{a}=-g^{ab}\Gamma _{b00}=-k\left( -k^{2}\right)
\left( -2kk_{,c}/2\right) =-k^{-1}k_{,c}.$ Writing $k=1+\varphi ,$ where $%
\varphi $ is small, we have that acceleration is minus the gradient of $%
\varphi $,i.e. $\ddot{x}^{a}\approx -\varphi _{,a}.$So, gravitational
redshift can be identified with the scalar potential in Newtonian gravity.
Hence the time component of the metric is $g_{00}=k^{-2}\approx 1+2\varphi
\left( x,y,z\right) .$\ $\ \ \ \ \ \ \ \ \ \ \ \ \ \ \ \ \ \ \ \ \ \ \ \ \ \
\ \ \ \ \ \ \ \ \ $

This is called the \textit{weak gravitational field limit. }$m\varphi $ has
units of kinetic energy, $\frac{1}{2}mv^{2}.$ To convert to conventional
units, we must divide by $c^{2},$ i.e.

$%
\begin{array}{cc}
\begin{array}{c}
\\ 
g_{00}=k^{-2}\approx 1+\dfrac{2\varphi \left( x,y,z\right) }{c^{2}}, \\ 
\\ 
k=1-\dfrac{\varphi \left( x,y,z\right) }{c^{2}}. \\ 
\end{array}
& \text{ \ }(4.2.1.3)%
\end{array}%
$

\bigskip Note that Lagrangian of a relativistic particle in a weak
gravitational field is:

\bigskip $%
\begin{array}{cc}
\begin{array}{c}
\\ 
L\left( t\right) =-mc^{2}\sqrt{1-\dfrac{\mathbf{v}^{2}\left( t\right) }{c^{2}%
}}-m\varphi \left( x,y,z\right) . \\ 
\end{array}
& \text{ \ }(4.2.1.4)%
\end{array}%
$

\bigskip In nonrelativistic limit $\mathbf{v}^{2}/c^{2}\rightarrow 0$ from
Eq.(4.2.1.4) one obtain

$%
\begin{array}{cc}
\begin{array}{c}
\\ 
\mathbf{S}=\int L\left( t\right) dt=-mc\dint \left( c-\dfrac{\mathbf{v}%
^{2}\left( t\right) }{2c}+\dfrac{\varphi \left( x,y,z\right) }{c}\right) dt.
\\ 
\end{array}
& \text{ \ \ }(4.2.1.5)%
\end{array}%
$

\bigskip

Take into account that $\mathbf{S}=-mc\int ds$ from Eq.(4.2.1.5) one obtain

\bigskip

\bigskip $%
\begin{array}{cc}
\begin{array}{c}
\\ 
ds=\left( c-\dfrac{\mathbf{v}^{2}\left( t\right) }{2c}+\dfrac{\varphi \left(
x,y,z\right) }{c}\right) dt. \\ 
\end{array}
& \text{ }(4.2.1.6)%
\end{array}%
$

\bigskip Thus (take into account that $d\mathbf{r}=\mathbf{v}dt$ ) in
nonrelativistic limit $\mathbf{v}^{2}/c^{2}\rightarrow 0$ we obtain

$%
\begin{array}{cc}
\begin{array}{c}
\\ 
ds^{2}=\left( c-\dfrac{\mathbf{v}^{2}\left( t\right) }{2c}+\dfrac{\varphi
\left( x,y,z\right) }{c}\right) ^{2}dt^{2}-dr^{2}\approx  \\ 
\\ 
\approx c^{2}\left( 1+\dfrac{2\varphi \left( x,y,z\right) }{c^{2}}\right)
dt^{2}-dr^{2}. \\ 
\end{array}
& \text{ \ }(4.2.1.7)%
\end{array}%
$

\bigskip

Finally Eq.(4.2.1.7) gives again the same \textit{weak gravitational field
limit:}

$%
\begin{array}{cc}
\begin{array}{c}
\\ 
g_{00}\simeq -\left[ 1+\dfrac{2\varphi \left( x,y,z\right) }{c^{2}}\right] .
\\ 
\end{array}
& \text{ \ }(4.2.1.8)%
\end{array}%
$

\bigskip Einstein's field equation, written in terms of the Ricci tensor is

\bigskip $%
\begin{array}{cc}
\begin{array}{c}
\\ 
R^{ab}-\dfrac{1}{2}g^{ab}R=\kappa T^{ab}. \\ 
\end{array}
& \text{ }(4.2.1.9)%
\end{array}%
$

In the case of a static body of uniform density, $\rho ,$ $T^{ab}$ is

\bigskip\ $%
\begin{array}{cc}
\begin{array}{c}
\\ 
T^{ab}=\rho 
\begin{bmatrix}
1 & 0 & 0 & 0 \\ 
0 & 0 & 0 & 0 \\ 
0 & 0 & 0 & 0 \\ 
0 & 0 & 0 & 0%
\end{bmatrix}
\\ 
\end{array}
& \text{ }(4.2.1.10)%
\end{array}%
$

Contract the indices $a$ and $b,$ noting from the standard summation
convention that

\bigskip $g_{ab}g^{ab}=\delta _{a}^{a}=4$ one obtain

$%
\begin{array}{cc}
\begin{array}{c}
\\ 
g_{ab}R^{ab}-\dfrac{1}{2}g_{ab}g^{ab}R=\kappa g_{ab}T^{ab}, \\ 
\\ 
R-2R=\kappa \rho , \\ 
\\ 
R=-\kappa \rho . \\ 
\end{array}
& \text{ }(4.2.1.11)%
\end{array}%
$

\bigskip

Thus

\bigskip 

\bigskip $%
\begin{array}{cc}
\begin{array}{c}
\\ 
R^{ab}=\kappa T^{ab}+\dfrac{1}{2}g^{ab}R=\kappa T^{ab}-\dfrac{1}{2}%
g^{ab}\kappa \rho , \\ 
\\ 
R^{00}=\dfrac{1}{2}\kappa \rho . \\ 
\end{array}
& \text{ }(4.2.1.12)%
\end{array}%
$

\bigskip 

The Ricci tensor is

$%
\begin{array}{cc}
\begin{array}{c}
\\ 
R_{ab}=R_{acb}^{c}=\Gamma _{ab,c}^{c}-\Gamma _{ac,b}^{c}+\Gamma
_{ab}^{e}\Gamma _{ec}^{c}-\Gamma _{ac}^{e}\Gamma _{eb}^{c}, \\ 
\end{array}
& \text{ }(4.2.1.13)%
\end{array}%
$

In the Newtonian approximation, the metric $g$, is slowly varying in space
and constant in time. We may neglect terms of second order in derivatives of
the metric, and set time derivatives to zero. Then from Eq.(4.2.1.12) one
obtain

\bigskip $%
\begin{array}{cc}
\begin{array}{c}
\\ 
R_{00}\approx \Gamma _{00,c}^{c}\approx \left( g^{cd}\Gamma _{d00}\right)
_{c}=-\dfrac{1}{2}g^{cd}g_{00,dc} \\ 
\end{array}
& \text{ }(4.2.1.14)%
\end{array}%
$

\bigskip In Cartesian, $x,y,z$ -coordinates one obtain

\bigskip\ $\ 
\begin{array}{cc}
\begin{array}{c}
\\ 
R_{00}\approx \dfrac{1}{2}\left( \dfrac{\partial ^{2}}{\partial x^{2}}+%
\dfrac{\partial ^{2}}{\partial y^{2}}+\dfrac{\partial ^{2}}{\partial z^{2}}%
\right) k=\Delta k, \\ 
\\ 
\Delta k=\dfrac{1}{2}\kappa \rho \left( x,y,z\right) . \\ 
\end{array}
& \text{\ }(4.2.1.15)%
\end{array}%
$

\bigskip

Hence Poisson's equation for a Newtonian gravitational potential, $k,$ due
to a mass distribution of density $\rho $ is

$%
\begin{array}{cc}
\begin{array}{c}
\\ 
\Delta k=\kappa \rho \left( x,y,z\right) , \\ 
\\ 
\kappa =8\pi G. \\ 
\end{array}
& \text{ }(4.2.1.16)%
\end{array}%
$

\bigskip

\bigskip

\bigskip

\subsection{IV.2.2.}

\bigskip

Let us consider the motion of a charged particle with a charge $\pm e$ and
masses $m$ in any external electric field $\mathbf{E}^{Ext}\left(
x,y,z,t\right) $. Note that Lagrangian of a relativistic particle in
electric field is [65]:

\bigskip 

$%
\begin{array}{cc}
\begin{array}{c}
\\ 
L\left( t\right) =-mc^{2}\sqrt{1-\dfrac{\mathbf{v}^{2}\left( t\right) }{c^{2}%
}}-e\varphi \left( x,y,z,t\right) . \\ 
\end{array}
& \text{ \ }(4.2.2.1)%
\end{array}%
$

\bigskip 

In nonrelativistic approximation, i.e. $v/c\simeq 0$ from Eq.(4.2.2.1) one
obtain

\bigskip 

$%
\begin{array}{cc}
\begin{array}{c}
\\ 
L\left( t\right) =-mc^{2}+\dfrac{m\mathbf{v}^{2}\left( t\right) }{2}%
-e\varphi \left( x,y,z,t\right) . \\ 
\end{array}
& \text{ \ }(4.2.2.2)%
\end{array}%
$

\bigskip 

\bigskip The action for relativistic charged particle in electric field is:

\bigskip $%
\begin{array}{cc}
\begin{array}{c}
\\ 
\mathbf{S}=\dint\nolimits_{t_{1}}^{t_{2}}L\left( t\right) dt= \\ 
\\ 
\dint\nolimits_{t_{1}}^{t_{2}}\left[ -mc^{2}\sqrt{1-\dfrac{\mathbf{v}%
^{2}\left( t\right) }{c^{2}}}-e\cdot \varphi \left( x,y,z,t\right) \right]
dt. \\ 
\end{array}
& \text{ \ }(4.2.2.3)%
\end{array}%
$

In nonrelativistic approximation, i.e. $v/c\simeq 0$ from
Eqs.(4.2.2.2)-(4.2.2.3) one obtain

\bigskip

$%
\begin{array}{cc}
\begin{array}{c}
\\ 
\mathbf{S}=\dint\nolimits_{t_{1}}^{t_{2}}L\left( t\right) dt= \\ 
\\ 
-mc\dint\nolimits_{t_{1}}^{t_{2}}\left( c-\dfrac{\mathbf{v}^{2}\left(
t\right) }{2c}+\dfrac{e\cdot \varphi \left( x,y,z,t\right) }{m\cdot c}%
\right) dt. \\ 
\end{array}
& \text{ \ }(4.2.2.4)%
\end{array}%
$

\bigskip

But from other side we have [65]:

$%
\begin{array}{cc}
\begin{array}{c}
\\ 
\mathbf{S}=-mc\int ds. \\ 
\end{array}
& \text{ \ \ }(4.2.2.5)%
\end{array}%
$

\bigskip 

Thus from Eq.(4.2.2.4) and Eq.(4.2.2.4) one obtain\bigskip 

\bigskip $%
\begin{array}{cc}
\begin{array}{c}
\\ 
ds=\left[ c-\dfrac{\mathbf{v}^{2}\left( t\right) }{2c}+\left( \dfrac{e}{mc}%
\right) \cdot \varphi \left( x,y,z,t\right) \right] dt. \\ 
\end{array}
& \text{ }(4.2.2.6)%
\end{array}%
$

\bigskip 

\bigskip Thus (take into account that $d\mathbf{r}=\mathbf{v}dt$ ) in
nonrelativistic limit $\mathbf{v}^{2}/c^{2}\rightarrow 0$ we obtain

\bigskip 

\bigskip $%
\begin{array}{cc}
\begin{array}{c}
\\ 
ds^{2}=\left( c-\dfrac{\mathbf{v}^{2}\left( t\right) }{2c}+\left( \dfrac{e}{%
mc}\right) \cdot \varphi \left( x,y,z,t\right) \right) ^{2}dt^{2}= \\ 
\\ 
c^{2}dt^{2}+\dfrac{\mathbf{v}^{4}\left( t\right) }{4c^{2}}dt^{2}+\left( 
\dfrac{e}{mc}\right) ^{2}\cdot \varphi ^{2}\left( x,y,z,t\right) dt^{2}-%
\mathbf{v}^{2}\left( t\right) dt^{2}+ \\ 
\\ 
\left( \dfrac{2e}{m}\right) \cdot \varphi \left( x,y,z,t\right) dt^{2}-%
\dfrac{\mathbf{v}^{2}\left( t\right) }{c}\left( \dfrac{e}{mc}\right) \cdot
\varphi \left( x,y,z,t\right) dt^{2}= \\ 
\\ 
c^{2}\left( 1+\dfrac{2e\cdot \varphi \left( x,y,z,t\right) }{m\cdot c^{2}}%
+\left( \dfrac{e}{mc^{2}}\right) ^{2}\cdot \varphi ^{2}\left( x,y,z,t\right)
dt^{2}+\dfrac{\mathbf{v}^{4}\left( t\right) }{4c^{4}}\right) dt^{2}-d\mathbf{%
r}^{2}\simeq  \\ 
\\ 
c^{2}\left( 1+\dfrac{2e\cdot \varphi \left( x,y,z,t\right) }{m\cdot c^{2}}%
+\left( \dfrac{e}{mc^{2}}\right) ^{2}\cdot \varphi ^{2}\left( x,y,z,t\right)
dt^{2}\right) dt^{2}-d\mathbf{r}^{2} \\ 
\end{array}
& \text{ \ }(4.2.2.7.a)%
\end{array}%
$

\bigskip and

$%
\begin{array}{cc}
\begin{array}{c}
\\ 
ds^{2}\simeq c^{2}\left( 1+\dfrac{2e\cdot \varphi \left( x,y,z,t\right) }{%
m\cdot c^{2}}\right) dt^{2}-d\mathbf{r}^{2}. \\ 
\end{array}
& \text{ \ }(4.2.2.7.b)%
\end{array}%
$

\bigskip

Let us consider now a system of charges located the electric field generated
by a charge distribution is

\bigskip $%
\begin{array}{cc}
\begin{array}{c}
\\ 
\varphi =\dsum \dfrac{e_{a}}{\left\vert \mathbf{R}_{0}-\mathbf{r}%
_{a}\right\vert } \\ 
\end{array}
& \text{ }(4.2.2.8)%
\end{array}%
$

\bigskip $%
\begin{array}{cc}
\begin{array}{c}
\\ 
\varphi \approx \dfrac{\sum e_{a}}{\left\vert \mathbf{R}_{0}\right\vert }%
-\left( \sum e_{a}\mathbf{r}_{a}\right) \cdot \mathbf{grad}\left( \dfrac{1}{%
\left\vert \mathbf{R}_{0}\right\vert }\right) = \\ 
\\ 
\dfrac{\sum e_{a}}{\left\vert \mathbf{R}_{0}\right\vert }-\mathbf{d\cdot grad%
}\left( \dfrac{1}{\left\vert \mathbf{R}_{0}\right\vert }\right) , \\ 
\end{array}
& \text{ }(4.2.2.9)%
\end{array}%
$

where sum $\sum e_{a}\mathbf{r}_{a}\triangleq \mathbf{d}$ is colled dipole
moment [65]. In the complete expansion of the $\varphi $ in powers $%
\left\vert \mathbf{R}_{0}\right\vert ^{-1}$

\bigskip 

$%
\begin{array}{cc}
\begin{array}{c}
\\ 
\varphi =\varphi _{0}+\varphi _{1}+...+\varphi _{n}+..., \\ 
\\ 
\varphi _{n}\text{ }\symbol{126}\text{ }\left\vert \mathbf{R}_{0}\right\vert
^{-\left( n+1\right) }. \\ 
\end{array}
& \text{ }(4.2.2.10)%
\end{array}%
$

\bigskip

We saw that:

\bigskip 

$%
\begin{array}{cc}
\begin{array}{c}
\\ 
\varphi _{0}=\left\vert \mathbf{R}_{0}\right\vert ^{-1}\sum e_{a}, \\ 
\\ 
\varphi _{1}=-\mathbf{d\cdot grad}\left( \dfrac{1}{\left\vert \mathbf{R}%
_{0}\right\vert }\right) . \\ 
\end{array}
& \text{ }(4.2.2.11)%
\end{array}%
$

\bigskip 

If the total charge is thero, term $\varphi _{0}$ is vanishes, and

\bigskip 

$%
\begin{array}{cc}
\begin{array}{c}
\\ 
\varphi =-\mathbf{d\cdot grad}\left( \dfrac{1}{\left\vert \mathbf{R}%
_{0}\right\vert }\right) +O\left( \left\vert \mathbf{R}_{0}\right\vert
^{-3}\right) . \\ 
\end{array}
& \text{ }(4.2.2.12)%
\end{array}%
$

\bigskip 

The second term $\varphi _{1}$ is colled dipole potential of the system.

\bigskip

Let us consider now a system of charges located in an external electric
field $\mathbf{E}^{Ext}\left( x,y,z,t\right) $. We denote the potential of
this external electric field by $\varphi ^{Ext}\left( \mathbf{r,}t\right) =$ 
$\varphi ^{Ext}\left( x,y,z,t\right) .$Total potential energy of the system
is [65]:

\bigskip 

\bigskip\ $%
\begin{array}{cc}
\begin{array}{c}
\\ 
U=\dint j_{0}\varphi ^{Ext}\left( x,y,z,t\right) dV= \\ 
\\ 
\dsum\nolimits_{i=1}^{n}\dint e_{i}\cdot \delta \left( \mathbf{r-r}%
_{i}\right) \varphi ^{Ext}\left( \mathbf{r,}t\right) dV= \\ 
\\ 
\dsum\nolimits_{i=1}^{n}e_{i}\cdot \varphi ^{Ext}\left( \mathbf{r}%
_{i},t\right) . \\ 
\end{array}
& \text{ }(4.2.2.13)%
\end{array}%
$

We introduce another coordinate system with its origin anywhere within the
system of charges: $\mathbf{r}_{i}$ is the radius vector of the charge $%
e_{i} $ in these coordinate.We assume that the external field $\mathbf{E}%
^{Ext}\left( \mathbf{r}_{i},t\right) $ changes slowly: (1) over region of
the system of charges and (2) over region of the time $t\in \left[ 0,\infty %
\right] ,$i.e $\mathbf{E}^{Ext}\left( \mathbf{r}_{i},t\right) \simeq 
\widetilde{\mathbf{E}}^{Ext}\left( \mathbf{r}_{i}\right) .$ Then one can
expand the energy $U$ is powers of $\mathbf{r}_{i}$

\bigskip

\bigskip $%
\begin{array}{cc}
\begin{array}{c}
\\ 
U=U_{0}+U_{1}+... \\ 
\end{array}
& \text{ }(4.2.2.14)%
\end{array}%
$

\bigskip In this expansion the first term is

\bigskip $%
\begin{array}{cc}
\begin{array}{c}
\\ 
U_{0}=\varphi _{0}^{Ext}\cdot \dsum\nolimits_{i=1}^{n}e_{i}, \\ 
\end{array}
& \text{ }(4.2.2.15)%
\end{array}%
$

where $\varphi _{0}^{Ext}$ is the walue of the potential at the origin. In
this approximation, the energy of the system is the same as it would be if
all the charges were located at one point: the origin. The second term in
the expansion is

\bigskip 

\bigskip $%
\begin{array}{cc}
\begin{array}{c}
\\ 
U_{1}=\left( \mathbf{grad}\varphi ^{Ext}\right) _{0}\cdot
\dsum\nolimits_{i=1}^{n}e_{i}\cdot \mathbf{r}_{i}. \\ 
\end{array}
& \text{ }(4.2.2.16)%
\end{array}%
$

\bigskip Hence

\bigskip $%
\begin{array}{cc}
\begin{array}{c}
\\ 
U_{1}=-\mathbf{d}\cdot \mathbf{E}_{0}^{Ext}, \\ 
\\ 
\mathbf{E}_{0}=\left( \mathbf{grad}\varphi ^{Ext}\right) _{0}. \\ 
\end{array}
& \text{ }(4.2.2.17)%
\end{array}%
$

The total force acting on the system in the external quasiuniform field is
(to the order $O\left( \left\vert \mathbf{R}_{0}\right\vert ^{-3}\right) $
we are considering above)\bigskip 

$\ 
\begin{array}{cc}
\begin{array}{c}
\\ 
\mathbf{F}=\mathbf{E}_{0}\cdot \dsum\nolimits_{i=1}^{n}e_{i}+\left[ \nabla
\left( \mathbf{d}\cdot \mathbf{E}\right) \right] _{0}. \\ 
\end{array}
& \text{ }(4.2.2.18)%
\end{array}%
$

\bigskip 

\bigskip If the total charge is thero, the first term in Eq.(4.2.2.18) is
vanishes, and we obtain

\bigskip $%
\begin{array}{cc}
\begin{array}{c}
\\ 
\mathbf{F}=\left( \nabla \cdot \mathbf{d}\right) \mathbf{E}_{0}, \\ 
\end{array}
& \text{ }(4.2.2.19)%
\end{array}%
$

where the derivatives of the field intensity taken at the origin $\mathbf{R}%
_{0}.$

\bigskip Eq.(4.2.2.7) gives \textit{weak inertional field limit:}

\bigskip $%
\begin{array}{cc}
\begin{array}{c}
\\ 
\breve{g}_{00}\simeq -1-\dfrac{2e\varphi \left( x,y,z\right) }{mc^{2}}. \\ 
\end{array}
& \text{ \ }(4.2.2.20)%
\end{array}%
$

Let us consider Bimetric theory of gravitational-inertial field in pure
inertial field approximation\textbf{\ }(see section IV.5.2). Field equations
we take in the form:

\ 

$%
\begin{array}{cc}
\begin{array}{c}
\\ 
\breve{R}_{i}^{k}\simeq \breve{\kappa}\left( \check{T}_{i}^{k}-\dfrac{1}{2}%
\delta _{i}^{k}\breve{T}\right) . \\ 
\end{array}
& \text{ \ }(4.2.2.21)%
\end{array}%
$

Now consider any discrete distribution of charged matter, with a
four-current charge density

\bigskip 

\bigskip $%
\begin{array}{cc}
\begin{array}{c}
\\ 
s_{i}=\left( \dfrac{\rho \mathbf{u}}{c},i\rho \right) , \\ 
\\ 
\rho =\dfrac{\rho ^{0}}{\sqrt{1-\dfrac{u^{2}}{c^{2}}}}, \\ 
\\ 
\rho ^{0}=\dsum\nolimits_{j=1}^{n}\delta \left( \mathbf{r-r}_{j}\right)
\cdot e_{j} \\ 
\end{array}
& \text{ \ }(4.2.2.22)%
\end{array}%
$ \ \ \ \ \ \ \ \ \ \ \ \ \ \ \ \ \ \ \ \ \ \ \ \ \ \ \ \ \ \ \ \ \ \ \ \ \
\ 

\bigskip 

in a given external electromagnetic field $F_{ik}^{Ext}$, \bigskip where $%
\rho ^{0}$ is the charge density in the rest system $\mathbf{S}$.Consider a
definite point in space at a definite time; the charged

matter at this point is moving with a certain velocity.Now, let $\mathbf{S}%
^{0}$ be the momentary rest system of the matter at this point. The
components of $s_{i}$ in this system are then

\bigskip\ $\ \ \ \ \ \ \ \ \ \ \ \ \ \ \ \ \ \ \ \ \ \ \ \ \ \ \ \ \ \ \ \ \
\ \ \ \ \ \ \ \ \ \ \ \ \ \ \ \ \ \ \ \ $

$%
\begin{array}{cc}
\begin{array}{c}
\\ 
s_{i}^{0}=\left( 0,0,0,i\rho ^{0}\right)  \\ 
\end{array}
& \text{ \ }(4.2.2.23)%
\end{array}%
$

\bigskip 

The action of the electromagnetic field $F_{ik}$ with charged particles is $%
S=\dint\nolimits_{\Omega }LdV$ with

\bigskip 

$%
\begin{array}{cc}
\begin{array}{c}
\\ 
L=-\dfrac{1}{4}F_{ik}^{Ext}F_{ik}^{Ext}+A_{i}^{Ext}s_{i}-\mu ^{0}c^{2}= \\ 
\\ 
-\dfrac{1}{4}\left( \dfrac{\partial A_{k}^{Ext}}{\partial x_{i}}-\dfrac{%
\partial A_{i}^{Ext}}{\partial x_{k}}\right) \left( \dfrac{\partial
A_{k}^{Ext}}{\partial x_{i}}-\dfrac{\partial A_{i}^{Ext}}{\partial x_{k}}%
\right) +A_{i}^{Ext}s_{i}-\mu ^{0}c^{2}, \\ 
\\ 
\mu ^{0}=\dsum\nolimits_{j=1}^{n}\delta \left( \mathbf{r-r}_{j}\right) \cdot
\mu _{j}. \\ 
\end{array}
& \text{ }(4.2.2.24)%
\end{array}%
$

\bigskip

Note that Lagrangian $L$ does not contain any derivatives of the metric $%
\breve{g}_{ik}.$ Hence the energy-momentum tensor of the electromagnetic
field $F_{ik}$ with charged particles is

$%
\begin{array}{cc}
\begin{array}{c}
\\ 
\breve{T}_{ik}=-2\dfrac{\partial L}{\partial \breve{g}_{ik}}+\breve{g}_{ik}L.
\\ 
\end{array}
& \text{ \ }(4.2.2.25)%
\end{array}%
$

In the weak (pure inertional) field\textit{\ }approximation, the metric $%
\breve{g},$ is slowly varying in space and constant in time. We may neglect
terms of second order in derivatives of the metric, and set time derivatives
to zero.

\bigskip

\bigskip Substitution Eq.(4.2.2.20) into \ Eq.(4.2.2.)\ gives\ \ 

\bigskip $%
\begin{array}{cc}
\begin{array}{c}
\\ 
\breve{R}_{00}=-\breve{R}_{0}^{0}=\dfrac{\partial \breve{\Gamma}%
_{00}^{\alpha }}{\partial x^{\alpha }}, \\ 
\\ 
\breve{\Gamma}_{00}^{\alpha }\simeq -\dfrac{1}{2}\breve{g}^{\alpha \alpha }%
\dfrac{\partial \breve{g}_{00}}{\partial x^{\alpha }}=\dfrac{e}{mc^{2}}%
\dfrac{\partial \varphi }{\partial x^{\alpha }}. \\ 
\end{array}
& \text{ \ \ }(4.2.2.26)%
\end{array}%
$

Hence

\bigskip $%
\begin{array}{cc}
\begin{array}{c}
\\ 
\breve{R}_{0}^{0}\simeq -\dfrac{e}{mc^{2}}\dfrac{\partial ^{2}\varphi }{%
\partial x^{\alpha 2}}=-\dfrac{e}{mc^{2}}\Delta \varphi . \\ 
\end{array}
& \text{ \ }(4.2.2.27)%
\end{array}%
$

\textbf{Suppose that: }

\begin{itemize}
\item (1) $\dsum\nolimits_{i=1}^{n}e_{i}\neq 0,$

\item (2) $\left( \mathbf{grad}\varphi ^{Ext}\right) _{0}\ll 1,$i.e. $%
U_{1}\simeq 0,$

\item (3) $u^{\alpha }\simeq 0,\alpha =1,2,3;u^{0}=-u_{0}=1.$
\end{itemize}

\bigskip

\bigskip $%
\begin{array}{cc}
\begin{array}{c}
\\ 
\breve{T}_{i}^{k}\simeq 0,i\neq k, \\ 
\\ 
\breve{T}_{0}^{0}\simeq -\dfrac{\rho ^{0}}{\sqrt{-\breve{g}}}\varphi
_{0}^{Ext}=-\dfrac{1}{\sqrt{-\breve{g}}}\varphi _{0}^{Ext}\cdot
\dsum\nolimits_{j=1}^{n}\delta \left( \mathbf{r-r}_{j}\right) \cdot e_{j}.
\\ 
\end{array}
& \text{ \ }(4.2.2.28)%
\end{array}%
$

\bigskip From the field equations (4.2.2.21) for the case $i=k=0$ one obtain

\bigskip $%
\begin{array}{cc}
\begin{array}{c}
\\ 
\breve{R}_{0}^{0}\simeq -\breve{\kappa}\rho ^{0}\varphi _{0}^{Ext}= \\ 
\\ 
-\breve{\kappa}\varphi _{0}^{Ext}\dsum\nolimits_{j=1}^{n}\delta \left( 
\mathbf{r-r}_{j}\right) \cdot e_{j}. \\ 
\end{array}
& \text{ }(4.2.2.29)%
\end{array}%
$

\bigskip Substitution Eq.(4.2.2.29) into Eq.(4.2.2.27) gives

\bigskip $\ 
\begin{array}{cc}
\begin{array}{c}
\\ 
\dfrac{e}{mc^{2}}\Delta \varphi =\breve{\kappa}\rho ^{0}\left( x,y,z\right)
\varphi _{0}^{Ext}. \\ 
\end{array}
& \text{ }(4.2.2.30)%
\end{array}%
$

Poisson's equation for a Newtonian gravitational-inertional potential, $%
\varphi $, due to a distribution of charge density $\rho ^{0}$ is

\bigskip

$%
\begin{array}{cc}
\begin{array}{c}
\\ 
\Delta \varphi =\left( \dfrac{e}{mc^{2}}\right) ^{-1}\breve{\kappa}\rho
^{0}\left( x,y,z\right) \varphi _{0}^{Ext}. \\ 
\end{array}
& \text{ \ }(4.2.2.31)%
\end{array}%
$

\bigskip Thus, 

\ \ \ \ \ \ 

$%
\begin{array}{cc}
\begin{array}{c}
\\ 
\breve{\kappa}=\dfrac{e}{\varphi _{0}^{Ext}mc^{2}}, \\ 
\end{array}
& \text{ \ }(4.2.2.32)%
\end{array}%
$

\bigskip

and field equation is

\bigskip 

\bigskip $%
\begin{array}{cc}
\begin{array}{c}
\\ 
\breve{R}_{i}^{k}\simeq \breve{\kappa}\left( \check{T}_{i}^{k}-\dfrac{1}{2}%
\delta _{i}^{k}\breve{T}\right) . \\ 
\end{array}
& \text{ \ }(4.2.2.21)%
\end{array}%
$

\bigskip

\bigskip

\bigskip

\subsection{IV.2.3. Weak field limit of the geodesic equation for the motion
of a free test particle.}

\bigskip

\bigskip In the linear approximation, pure inertial field tensor $\breve{g}$
can be written as

$%
\begin{array}{cc}
\begin{array}{c}
\\ 
\breve{g}_{\mu \nu }=\eta _{\mu \nu }+\breve{h}_{\mu \nu }, \\ 
\end{array}
& \text{ \ }(4.2.3.1)%
\end{array}%
$

where $\eta _{\mu \nu }$ is the Minkowski metric tensor with signature $+2$
and $\breve{h}_{\mu \nu }$ is a first-order perturbation.Under
transformation of the background coordinates $x^{\mu }$\bigskip $=(ct,%
\mathbf{\vec{x}}),$ $x^{\mu }\rightarrow $ $x^{\mu }-\varepsilon ^{\mu },$%
the pure inertial field potentials $\breve{h}_{\mu \nu }$ transform as

$%
\begin{array}{cc}
\begin{array}{c}
\\ 
\breve{h}_{\mu \nu }\rightarrow \breve{h}_{\mu \nu }+\varepsilon _{\mu ,\nu
}+\varepsilon _{\nu ,\mu }. \\ 
\end{array}
& \text{ \ }(4.2.3.2)%
\end{array}%
$

Henceforth, the nertial field potentials are considered to be gauge
dependent, while the background global inertial coordinate system is in
effect fixed. The accelerated spacetime curvature $\breve{R}\left( \breve{g}%
\right) $ is, however, gauge invariant. It is useful to introduce the

trace-reversed pure inertial potentials

\bigskip $%
\begin{array}{cc}
\begin{array}{c}
\\ 
\widetilde{h}_{\mu \nu }\triangleq \breve{h}_{\mu \nu }-\dfrac{1}{2}h\eta
_{\mu \nu }, \\ 
\\ 
h=tr\left( h_{\mu \nu }\right) . \\ 
\end{array}
& \text{ \ }(4.2.3.3)%
\end{array}%
$

By imposing the transverse gauge condition $\widetilde{h}_{,\nu }^{\mu \nu
}=0,$the gravitational-inertial field equations take the form\bigskip 

\bigskip $%
\begin{array}{cc}
\begin{array}{c}
\\ 
\square \widetilde{h}_{\mu \nu }=-\breve{k}T_{\mu \nu }^{\mathbf{EM}}. \\ 
\end{array}
& \text{ \ }(4.2.3.4)%
\end{array}%
$

Where $T_{\mu \nu }^{\mathbf{EM}}$ is the corresponding electromagnetic
stress-energy tensor

\bigskip The general solution of (4.2.3.4) is given by the special retarded
solution

$%
\begin{array}{cc}
\begin{array}{c}
\\ 
\widetilde{h}_{\mu \nu }=\widetilde{h}_{\mu \nu }^{\mathbf{a.c.}}, \\ 
\\ 
\widetilde{h}_{\mu \nu }^{\mathbf{a.c.}}=\breve{k}\dint \dfrac{T_{\mu \nu }^{%
\mathbf{EM}}\left( ct-\left\vert \mathbf{x-x}^{\prime }\right\vert ,\mathbf{x%
}^{\prime }\right) }{\left\vert \mathbf{x-x}^{\prime }\right\vert }d\mathbf{x%
}^{\prime }, \\ 
\end{array}
& \text{ \ }(4.2.3.5)%
\end{array}%
$

\bigskip 

plus a general solution of the homogeneous wave equation that we comlete
ignore in this consideration.In the linear approximation, all terms of $%
O(c^{-4})$ are neglected in the metric tensor. Thus from Eqs.(4.2.3.5) for
the sources under consideration here one obtain

\bigskip 

\bigskip $%
\begin{array}{cc}
\begin{array}{c}
\\ 
\widetilde{h}_{00}^{\mathbf{a.c.}}=\dfrac{4\Phi ^{\mathbf{a.c.}}\left( t,%
\mathbf{x}\right) }{c^{2}}, \\ 
\\ 
\widetilde{h}_{0i}^{\mathbf{a.c.}}=-\dfrac{2A_{i}^{\mathbf{a.c.}}\left( t,%
\mathbf{x}\right) }{c^{2}}, \\ 
\\ 
\widetilde{h}_{ij}=O(c^{-4});i,j\neq 0. \\ 
\end{array}
& \text{ \ }(4.2.3.6)%
\end{array}%
$

\bigskip 

\bigskip Then the spacetime metric in the linear approximation is\bigskip

$%
\begin{array}{cc}
\begin{array}{c}
\\ 
ds^{2}=-c^{2}\left( 1-2\dfrac{\Phi ^{\mathbf{a.c.}}\left( t,\mathbf{x}%
\right) }{c^{2}}\right) -\dfrac{4}{c}\left( \mathbf{A}^{\mathbf{a.c.}}\left(
t,\mathbf{x}\right) \cdot d\mathbf{x}\right) dt+ \\ 
\\ 
+\left( 1+2\dfrac{\Phi ^{\mathbf{a.c.}}\left( t,\mathbf{x}\right) }{c^{2}}%
\right) \delta _{ij}dx^{i}dx^{j}, \\ 
\end{array}
& \text{ \ }(4.2.3.7)%
\end{array}%
$

\bigskip 

The geodesic equation for the motion of a free test particle is

\bigskip 

\bigskip $%
\begin{array}{cc}
\begin{array}{c}
\\ 
\dfrac{dU^{\mu }}{d\tau }=\breve{\Gamma}_{\rho \sigma }^{\mu }U^{\rho
}U^{\sigma }, \\ 
\end{array}
& \text{ \ }(4.2.3.8)%
\end{array}%
$

where $\tau /c$ is the proper time and $U^{\mu }=dx^{\mu }/d\tau $ is the
unit four-velocity vector of the test particle. The Christoffel symbols are
given by

\bigskip 

\bigskip $%
\begin{array}{cc}
\begin{array}{c}
\\ 
c^{2}\breve{\Gamma}_{0\mu }^{0}=\Phi _{,\mu }^{\mathbf{a.c.}};c^{2}\breve{%
\Gamma}_{ij}^{0}=2A_{\left( i,j\right) }^{\mathbf{a.c.}}+\delta _{ij}\Phi
_{,0}^{\mathbf{a.c.}}, \\ 
\\ 
c^{2}\breve{\Gamma}_{00}^{i}=-\Phi _{,i}^{\mathbf{a.c.}}-2A_{i,0}^{\mathbf{%
a.c}};c^{2}\breve{\Gamma}_{0j}^{i}=\delta _{ij}\Phi _{,0}^{\mathbf{a.c.}%
}+\epsilon _{ijk}B^{k} \\ 
\\ 
c^{2}\breve{\Gamma}_{jk}^{i}=\delta _{ij}\Phi _{,k}^{\mathbf{a.c.}}+\delta
_{ik}\Phi _{,j}^{\mathbf{a.c.}}-\delta _{jk}\Phi _{,i}^{\mathbf{a.c.}} \\ 
\end{array}
& \text{ \ }(4.2.3.9)%
\end{array}%
$

\bigskip

The geodesic equation can be reduced via $U^{\mu }=\gamma (1,\mathbf{\beta }%
) $ with $\mathbf{\beta }=\mathbf{V}/c$ to

\bigskip

\bigskip $%
\begin{array}{cc}
\begin{array}{c}
\\ 
\dfrac{c}{\gamma }\dfrac{d\gamma }{d\tau }=\left( 1-\beta ^{2}\right) \Phi
_{,0}^{\mathbf{a.c.}}+2\beta ^{i}\left[ \Phi _{,i}^{\mathbf{a.c.}}-A_{\left(
i,j\right) }^{\mathbf{a.c.}}\beta ^{j}\right] , \\ 
\\ 
\dfrac{dV^{i}}{dt}=\left( 1+\beta ^{2}\right) \Phi _{,i}^{\mathbf{a.c.}%
}-2\left( \mathbf{\beta \times B}\right) _{i}+A_{i,0}^{\mathbf{a.c}}-\beta
^{i}\left( 3-\beta ^{2}\right) \Phi _{,0}^{\mathbf{a.c.}}+ \\ 
\\ 
+2\beta ^{i}\beta ^{j}\left[ A_{\left( j,k\right) }^{\mathbf{a.c.}}\beta
^{k}-2\Phi _{,j}^{\mathbf{a.c.}}\right] . \\ 
\end{array}
& \text{ \ }(4.2.3.10)%
\end{array}%
$

\bigskip Moreover, $U_{\mu }^{\mu }U=-1$ implies that

$%
\begin{array}{cc}
\begin{array}{c}
\\ 
\dfrac{1}{\gamma ^{2}}=1-\beta ^{2}-\dfrac{2}{c^{2}}\left( 1+\beta
^{2}\right) \Phi +\dfrac{4}{c^{2}}\mathbf{\beta }\cdot \mathbf{A}^{a.c.} \\ 
\end{array}
& \text{ \ }(4.2.3.11)%
\end{array}%
$

For a stationary source ($\partial _{t}\Phi \simeq 0$ and $\partial _{t}%
\mathbf{A}^{a.c.}\simeq 0$), equations of motion (4.2.3.10) reduces to

\bigskip

$%
\begin{array}{cc}
\begin{array}{c}
\\ 
m\dfrac{d\mathbf{V}}{dt}=m\breve{k}\mathbf{E}-2m\breve{k}\dfrac{\mathbf{V}}{c%
}\mathbf{\times B,} \\ 
\\ 
\breve{k}=\dfrac{e}{m}. \\ 
\end{array}
& \text{ \ }(4.2.3.12)%
\end{array}%
$

\bigskip 

\bigskip where velocity-dependent terms of order higher that $\beta =V/c$
are neglected.

Thus

\bigskip $%
\begin{array}{cc}
\begin{array}{c}
\\ 
m\dfrac{d\mathbf{V}}{dt}=e\mathbf{E}-2e\dfrac{\mathbf{V}}{c}\mathbf{\times B.%
} \\ 
\end{array}
& \text{ \ }(4.2.3.13)%
\end{array}%
$

In the case of a general nonstationary source, however, the equations of
motion (4.2.3.10) does not reduces to the Lorentz force law.

\bigskip

\subsection{IV.3.General Variational Action Principle in Bimetric Theory of
Gravitational-Inertial Field.}

\bigskip

Let's consider bimetric space-time $\Re _{2}\left( g_{\mu \nu },\breve{g}%
_{\mu \nu }\right) $ with $R\left( g_{\mu \nu }\right) \neq 0,$ $\breve{R}%
\left( \breve{g}_{\mu \nu }\right) \neq 0.$

Action integral takes the form

\bigskip

\bigskip $%
\begin{array}{cc}
\begin{array}{c}
\\ 
\mathbf{S=}\dfrac{1}{c}\dint \tciLaplace \sqrt{-\breve{g}}d^{4}x, \\ 
\\ 
\tciLaplace \left( g_{\mu \nu },\breve{g}_{\mu \nu }\right) =\tciLaplace
_{m}\left( g_{\mu \nu },\breve{g}_{\mu \nu }\right) +\tciLaplace
_{f_{1}}\left( g_{\mu \nu },\breve{g}_{\mu \nu }\right) +\tciLaplace
_{f_{2}}\left( \breve{g}_{\mu \nu }\right) , \\ 
\end{array}
& \text{ \ \ \ \ }(4.3.1)%
\end{array}%
$

\bigskip

where

\bigskip 

\bigskip\ $%
\begin{array}{cc}
\begin{array}{c}
\\ 
\tciLaplace _{m}=\left[ \left( c^{2}+\Xi \right) \rho -p\right] _{g}\cdot
K,K=\sqrt{\dfrac{-g}{\breve{g}}}, \\ 
\\ 
\tciLaplace _{f_{1}}=\dfrac{c^{4}}{16\pi G}f_{1}\left( \chi ^{\alpha \beta
},g_{\alpha \beta ||\lambda };\breve{h}^{\alpha \beta }\right) , \\ 
\\ 
\tciLaplace _{f_{2}}=\dfrac{c^{4}}{16\pi G}f_{2}\left( \breve{g}_{\alpha
\beta },\breve{g}_{\alpha \beta ;\lambda }\right) , \\ 
\\ 
g_{\mu \lambda }\chi ^{\lambda \nu }=\delta _{\mu }^{\nu };\text{ }\breve{g}%
_{\mu \lambda }\breve{h}^{\lambda \nu }=\delta _{\mu }^{\nu }, \\ 
\\ 
g^{\alpha \beta }=g_{\mu \lambda }\breve{h}^{\mu \alpha }\breve{h}^{\nu
\beta }\neq \breve{\chi}^{\alpha \beta }, \\ 
\\ 
\chi _{\alpha \beta }=\chi ^{\mu \nu }\breve{g}_{\mu \alpha }\breve{g}_{\nu
\beta }\neq g_{\alpha \beta }. \\ 
\\ 
\Xi =\int_{0}^{p\left( \rho \right) }\dfrac{dp}{p\left( \rho \right) }, \\ 
\\ 
p=p\left( \rho \right) ,\rho =\rho \left( p\right) , \\ 
\\ 
g=\det \left\Vert g_{\mu \lambda }\right\Vert <0,\breve{g}=\det \left\Vert 
\breve{g}_{\mu \lambda }\right\Vert <0. \\ 
\end{array}
& \text{ \ \ }(4.3.2)%
\end{array}%
$

\bigskip

Here, a subscripts $g$ and $\breve{g}$ stands for specifying that the
labelled quantity is defined by curved space-time metric $%
ds_{1}^{2}=g_{\alpha \beta }dx^{\alpha }dx^{\beta }$ and $ds_{2}^{2}=\breve{g%
}_{\alpha \beta }dx^{\alpha }dx^{\beta }$ accordingly.

The $\breve{g}$-covariant derivatives are denoted by a bislash ($||$)
followed by a certain (Greek) subscript, or (equivalently) by a derivative
symbol ($\breve{D}_{\alpha }$) followed by the same subscript. For example

\bigskip

\bigskip\ $\ 
\begin{array}{cc}
\begin{array}{c}
\\ 
g_{\alpha \beta ||\lambda }=\breve{D}_{\lambda }g_{\alpha \beta }=g_{\alpha
\beta ,\lambda }-\breve{G}_{\alpha \lambda }^{\text{ }\sigma }g_{\alpha
\beta }-\breve{G}_{\beta \lambda }^{\text{ }\sigma }g_{\sigma \alpha }, \\ 
\\ 
\breve{G}_{\mu \nu }^{\lambda }=\dfrac{1}{2}\breve{h}^{\lambda \sigma
}\left( \breve{g}_{\mu \sigma ,\eta }+\breve{g}_{\nu \sigma ,\rho }-\breve{g}%
_{\mu \nu ,\sigma }\right) . \\ 
\end{array}
& \text{ \ \ }(4.3.3)%
\end{array}%
$

For obtaining the field equations and the energy canonical tensor one obtain
to the following identity

\bigskip 

\bigskip $%
\begin{array}{cc}
\begin{array}{c}
\\ 
\dfrac{1}{\sqrt{-g}}\delta \left( \sqrt{-g}L\right) =D_{\lambda }q^{\lambda
}- \\ 
\\ 
\dfrac{1}{2}K\left( T_{\alpha \beta }+\dfrac{c^{4}}{8\pi G}E_{\alpha \beta
}\right) \delta \chi ^{\alpha \beta }-\dfrac{1}{2}\Im _{\alpha \beta }\delta
h^{\alpha \beta }, \\ 
\end{array}
& \text{ \ }(4.3.4)%
\end{array}%
$

where

\bigskip 

\bigskip\ $\ \ \ \ \ 
\begin{array}{cc}
\begin{array}{c}
\\ 
q^{\lambda }=\mathbf{P}^{\alpha \beta |||\lambda }\delta g_{\alpha \beta
}+2g_{\sigma \nu }\mathbf{P}^{\mu \nu ||\eta }\Omega _{\nu \mu ||\alpha
\beta }\delta \breve{h}^{\alpha \beta }, \\ 
\\ 
\mathbf{P}^{\alpha \beta |||\lambda }=\dfrac{\partial L_{f_{1}}}{\partial
g_{\alpha \beta ||\lambda }}, \\ 
\\ 
\breve{\Omega}_{\nu \mu ||\alpha \beta }^{\lambda \sigma }=\dfrac{1}{4}%
\left( \delta _{\beta }^{\lambda }\delta _{\nu }^{\sigma }\breve{g}_{\mu
\alpha }+\delta _{\lambda }^{\alpha }\delta _{\nu }^{\sigma }\breve{g}_{\mu
\beta }+\delta _{\beta }^{\lambda }\delta _{\mu }^{\sigma }\breve{g}_{\nu
\alpha }+\right. \\ 
\\ 
+\delta _{\alpha }^{\lambda }\delta _{\mu }^{\sigma }\breve{g}_{\nu \beta }-%
\breve{h}^{\lambda \sigma }\breve{g}_{\mu \alpha }\breve{g}_{\nu \beta }-%
\breve{h}^{\lambda \sigma }\breve{g}_{\mu \beta }\breve{g}_{\nu \alpha }),
\\ 
\\ 
\left[ T_{\alpha \beta }\right] _{g}=\left\{ \left( c^{2}+\Xi \right) \rho
U_{\alpha }U_{\beta }-p\gamma _{\alpha \beta }\right\} _{g}, \\ 
\\ 
\widetilde{E}_{\alpha \beta }=\widetilde{R}_{\alpha \beta }-\dfrac{1}{2}%
g_{\alpha \beta }\left( \chi ^{\mu \nu }\widetilde{R}_{\mu \nu }\right) , \\ 
\\ 
\widetilde{R}_{\alpha \beta }=-\dfrac{1}{K}\left\{ \left( \dfrac{\partial
f_{1}}{\partial \chi _{\alpha \beta }}-\dfrac{1}{2}g_{\alpha \beta }\chi
^{\mu \nu }\dfrac{\partial f_{1}}{\partial \chi ^{\mu \nu }}\right) \right. +
\\ 
\\ 
\left. \left( g_{\mu \alpha }g_{\nu \beta }-\dfrac{1}{2}g_{\alpha \beta
}g_{\mu \nu }\right) \breve{D}_{\lambda }p^{\mu \nu ||\lambda }\right\} , \\ 
\\ 
p^{\alpha \beta |||\lambda }=\dfrac{16\pi G}{c^{4}}\mathbf{P}^{\alpha \beta
|||\lambda }, \\ 
\\ 
\Im _{\alpha \beta }=D_{\sigma }Q_{\alpha \beta }^{\sigma }-2\left( \dfrac{%
\partial L_{f_{1}}}{\partial \breve{h}^{\alpha \beta }}-\dfrac{1}{2}\breve{g}%
_{\alpha \beta }L_{f_{1}}\right) , \\ 
\\ 
\breve{Q}_{\alpha \beta }^{\sigma }=4g_{\lambda \zeta }\mathbf{P}^{\mu \zeta
||\nu }\breve{\Omega}_{\mu \nu ||\alpha \beta }^{\lambda \sigma }. \\ 
\end{array}
& \text{ \ }(4.3.5)%
\end{array}%
$

\bigskip

\bigskip

\bigskip \bigskip

\subsection{V.\textbf{4.Linear Post-Newtonian Approximation to} \textbf{%
Bimetric Theory of Gravitational-Inertial Field. Gravito-Inertial
Ectromagnetism (GIEM).}}

\bigskip \bigskip \bigskip \bigskip

\subsection{V.\textbf{4.1.Linear Post-Newtonian Approximation to} \textbf{%
Bimetric Theory of Gravitational-Inertial Field. Gravito-Inertial
Ectromagnetism in a purely inertial approximation.}}

\bigskip

In Maxwell's electromagnetism, the combined dynamics of charged particles
and electromagnetic field are consistently described by Maxwell's field
equations and the Lorentz force law. Well-known that general relativity does
indeed contain induction effects. These effects turn out to be, despite the
differences, on the whole closely analogous to electromagnetic induction
effects.

Let's consider the curved bimetric accelerated spacetime $\tciLaplace
(V_{4},g,\breve{g})$ ($K_{ij}\left( g\right) \simeq 0$ $\left\Vert
K_{ij}\left( g\right) \right\Vert \ll \left\Vert \breve{R}_{ij}\left( \breve{%
g}\right) \right\Vert $) generated by an external pure nongravitational
"nonrelativistic" Lorentz force [53] and sufficiently small gravitational
force [54]. Suppose that the gravitational-Inertial field $g_{ij}(t,x)$ is
governed by either:

(1) massive gravitational source with mass density $\rho (t,x),$

(2) ectromagnetic field $\left\{ \mathbf{A}(t,x),\Phi (t,x)\right\} $ and

(3) charged massive particles with mass density $\mu (t,x)$ and charge
density $\rho _{ch}(t,x).$

Here $\Phi (t,x)$ is the electric potential and $\mathbf{A}(t,x)$ is the
magnetic vector potential. In the linear approximation, pure inertial field
tensor $\breve{g}$ can be written as

$\ 
\begin{array}{cc}
\begin{array}{c}
\\ 
\breve{g}_{\mu \nu }=\eta _{\mu \nu }+\breve{h}_{\mu \nu }, \\ 
\end{array}
& \text{ \ }(4.4.1.1)%
\end{array}%
$

where $\eta _{\mu \nu }$ is the Minkowski metric tensor with signature $+2$
and $\breve{h}_{\mu \nu }$ is a first-order perturbation.Under
transformation of the background coordinates $x^{\mu }$\bigskip $=(ct,%
\mathbf{\vec{x}}),$ $x^{\mu }\rightarrow $ $x^{\mu }-\varepsilon ^{\mu },$%
the pure inertial field potentials $\breve{h}_{\mu \nu }$ transform as

$%
\begin{array}{cc}
\begin{array}{c}
\\ 
\breve{h}_{\mu \nu }\rightarrow \breve{h}_{\mu \nu }+\varepsilon _{\mu ,\nu
}+\varepsilon _{\nu ,\mu }. \\ 
\end{array}
& \text{ \ }(4.4.1.2)%
\end{array}%
$

Henceforth, the potentials are considered to be gauge dependent, while the
background global inertial coordinate system is in effect fixed. The
accelerated spacetime curvature $\breve{R}\left( \breve{g}\right) $ is,
however, gauge invariant. It is useful to introduce the

trace-reversed pure inertial potentials

\bigskip $%
\begin{array}{cc}
\begin{array}{c}
\\ 
\widetilde{h}_{\mu \nu }\triangleq \breve{h}_{\mu \nu }-\dfrac{1}{2}h\eta
_{\mu \nu }, \\ 
\\ 
h=tr\left( h_{\mu \nu }\right) . \\ 
\end{array}
& \text{ \ }(4.4.1.3)%
\end{array}%
$

By imposing the transverse gauge condition $\widetilde{h}_{,\nu }^{\mu \nu
}=0,$the gravitational-inertial field equations take the form\bigskip

\bigskip\ $%
\begin{array}{cc}
\begin{array}{c}
\\ 
\square \widetilde{h}_{\mu \nu }=-\breve{k}T_{\mu \nu }^{\mathbf{a.c.}}-%
\dfrac{4G}{c^{4}}T_{\mu \nu }. \\ 
\end{array}
& \text{ }(4.4.1.4)%
\end{array}%
$

Where $T_{\mu \nu }^{\mathbf{a.c.}}\triangleq T_{\mu \nu }^{\mathbf{EM}}$ is
the corresponding electromagnetic stress-energy tensor.

\bigskip The general solution of (4.4.1.4) is given by the special retarded
solution

$\ \ 
\begin{array}{cc}
\begin{array}{c}
\\ 
\widetilde{h}_{\mu \nu }=\widetilde{h}_{\mu \nu }^{\mathbf{a.c.}}+\widetilde{%
h}_{\mu \nu }^{\mathbf{Gr.}}\simeq \widetilde{h}_{\mu \nu }^{\mathbf{a.c.}},
\\ 
\\ 
\widetilde{h}_{\mu \nu }^{\mathbf{a.c.}}=\breve{k}\dint \dfrac{T_{\mu \nu }^{%
\mathbf{EM}}\left( ct-\left\vert \mathbf{x-x}^{\prime }\right\vert ,\mathbf{x%
}^{\prime }\right) }{\left\vert \mathbf{x-x}^{\prime }\right\vert }d\mathbf{x%
}^{\prime }, \\ 
\\ 
\widetilde{h}_{\mu \nu }^{\mathbf{Gr.}}=\dfrac{4G}{c^{4}}\dint \dfrac{T_{\mu
\nu }^{\mathbf{Gr.}}\left( ct-\left\vert \mathbf{x-x}^{\prime }\right\vert ,%
\mathbf{x}^{\prime }\right) }{\left\vert \mathbf{x-x}^{\prime }\right\vert }d%
\mathbf{x}^{\prime }, \\ 
\\ 
\dfrac{4G}{c^{4}}\ll \breve{k}, \\ 
\end{array}
& \text{ \ }(4.4.1.5)%
\end{array}%
$

plus a general solution of the homogeneous wave equation that we ignore in
this consideration.In the linear GIEM approximation, all terms of $O(c^{-4})$
are neglected in the metric tensor. Thus from Eqs.(4.4.1.4) for the sources
under consideration here one obtain

\bigskip

\bigskip $\ 
\begin{array}{cc}
\begin{array}{c}
\\ 
\widetilde{h}_{00}^{\mathbf{a.c.}}=\dfrac{4\Phi ^{\mathbf{a.c.}}\left( t,%
\mathbf{x}\right) }{c^{2}};\widetilde{h}_{00}^{\mathbf{Gr.}}=\dfrac{4\Phi ^{%
\mathbf{Gr.}}\left( t,\mathbf{x}\right) }{c^{2}}, \\ 
\\ 
\widetilde{h}_{0i}^{\mathbf{a.c.}}=-\dfrac{2\mathbf{A}^{\mathbf{a.c.}}\left(
t,\mathbf{x}\right) }{c^{2}};\widetilde{h}_{0i}^{\mathbf{Gr.}}=-\dfrac{2%
\mathbf{A}^{\mathbf{Gr.}}\left( t,\mathbf{x}\right) }{c^{2}}, \\ 
\\ 
\widetilde{h}_{ij}=O(c^{-4});i,j\neq 0. \\ 
\end{array}
& \text{ \ }(4.4.1.6)%
\end{array}%
$

\bigskip

Where:

\begin{itemize}
\item $\Phi ^{\mathbf{a.c.}}\left( t,\mathbf{x}\right) $ is the
inertiaelectric potential,\ 

\item $\mathbf{A}^{\mathbf{a.c.}}\left( t,\mathbf{x}\right) $ is the
inertiamagnetic vector potential,

\item $\Phi ^{\mathbf{Gr.}}\left( t,\mathbf{x}\right) $ is the
gravitoelectric potential,

\item $\mathbf{A}^{\mathbf{Gr.}}\left( t,\mathbf{x}\right) $ is the
gravitomagnetic vector potential.
\end{itemize}

Where far from the gravitational source potentials $\Phi ^{\mathbf{Gr.}%
}\left( t,\mathbf{x}\right) $ and $\mathbf{A}^{\mathbf{Gr.}}\left( t,\mathbf{%
x}\right) $ can be expressed as [54]:

\bigskip

$\ 
\begin{array}{cc}
\begin{array}{c}
\\ 
\Phi ^{\mathbf{Gr.}}\left( t,\mathbf{x}\right) =\dfrac{GM}{r}, \\ 
\\ 
\mathbf{A}^{\mathbf{Gr.}}\left( t,\mathbf{x}\right) =\dfrac{G}{c}\dfrac{%
\mathbf{J\times x}}{r^{3}}, \\ 
\end{array}
& \text{ \ }(4.4.1.7)%
\end{array}%
$

\bigskip

Here $M$ and $\mathbf{J}$ are the inertial mass and angular momentum of the
source, $r=|\mathbf{x}|,$ $r\gg GM/c^{2}$and $r\gg J/(Mc).$

The spacetime metric in the linear GIEM approximation is\bigskip

\bigskip $%
\begin{array}{cc}
\begin{array}{c}
\\ 
ds^{2}=-c^{2}\left( 1-2\dfrac{\Phi ^{\mathbf{Gr.I}}\left( t,\mathbf{x}%
\right) }{c^{2}}\right) -\dfrac{4}{c}\left( \mathbf{A}^{\mathbf{Gr.I}}\left(
t,\mathbf{x}\right) \cdot d\mathbf{x}\right) dt+ \\ 
\\ 
+\left( 1+2\dfrac{\Phi ^{\mathbf{Gr.I}}\left( t,\mathbf{x}\right) }{c^{2}}%
\right) \delta _{ij}dx^{i}dx^{j}, \\ 
\\ 
\Phi ^{\mathbf{Gr.I}}\left( t,\mathbf{x}\right) =\Phi ^{\mathbf{a.c.}}\left(
t,\mathbf{x}\right) +\Phi ^{\mathbf{Gr.}}\left( t,\mathbf{x}\right) , \\ 
\\ 
\mathbf{A}^{\mathbf{Gr.I}}\left( t,\mathbf{x}\right) =\mathbf{A}^{\mathbf{%
a.c.}}\left( t,\mathbf{x}\right) +\mathbf{A}^{\mathbf{Gr.}}\left( t,\mathbf{x%
}\right) . \\ 
\end{array}
& \text{ \ }(4.4.1.8)%
\end{array}%
$

\bigskip

\bigskip Let us note that the gauge condition implies that

\bigskip $%
\begin{array}{cc}
\begin{array}{c}
\\ 
\dfrac{1}{c}\partial _{t}\Phi ^{\mathbf{Gr.}}+\nabla \cdot \left( \dfrac{1}{2%
}\mathbf{A}^{\mathbf{Gr.}}\right) =0, \\ 
\\ 
\dfrac{1}{c}\partial _{t}\Phi ^{\mathbf{a.c.}}+\nabla \cdot \left( \dfrac{1}{%
2}\mathbf{A}^{\mathbf{a.c.}}\right) =0, \\ 
\\ 
\dfrac{1}{c}\partial _{t}\Phi ^{\mathbf{Gr.I}}+\nabla \cdot \left( \dfrac{1}{%
2}\mathbf{A}^{\mathbf{Gr.I}}\right) =0. \\ 
\end{array}
& \text{ \ }(4.4.1.9)%
\end{array}%
$

This is related to the conservation of mass-energy of the
gravitational-inertional sources via Eq.(4.4.1.4).That is,let

\bigskip

$\ 
\begin{array}{cc}
\begin{array}{c}
\\ 
T^{(\mathbf{Gr.)}00}=\left( \rho +\mu \right) c^{2} \\ 
\\ 
T^{(\mathbf{Gr.)}0i}=cj^{\left( \mathbf{Gr.}\right) i}, \\ 
\end{array}
& \text{ \ }(4.4.1.10)%
\end{array}%
$

\bigskip

where $j^{\left( \mathbf{Gr.}\right) \nu }=\left( c\rho ,\mathbf{j}\right) $
is the mass-energy current of the gravitational source.

Hence equations (4.4.1.9) is equivalent to

\bigskip

\bigskip\ $%
\begin{array}{cc}
\begin{array}{c}
\\ 
j_{;\mu }^{\left( \mathbf{Gr.I}\right) \mu }=0, \\ 
\\ 
j^{\left( \mathbf{Gr.I}\right) \mu }=j^{\left( \mathbf{Gr.}\right) \mu
}+j^{\left( \mathbf{EM}\right) \mu }. \\ 
\end{array}
& \text{ \ }(4.4.1.11)%
\end{array}%
$

\bigskip

Thus one can to define:

\begin{itemize}
\item the gravitoelectric field $\mathbf{E}^{\mathbf{Gr.}},$

\item the gravitomagnetic field $\mathbf{B}^{\mathbf{Gr.}}$

\item the inertialelectric field $\mathbf{E}^{\mathbf{a.c.}},$

\item the inertialmagnetic field $\mathbf{B}^{\mathbf{a.c.}}$

\item the gravitoinertialelectric field $\mathbf{E}^{\mathbf{Gr.I}}$

\item the gravitoinertialmagnetic field $\mathbf{B}^{\mathbf{Gr.I}}$
\end{itemize}

in complete analogy with electrodynamics

\bigskip

$%
\begin{array}{cc}
\begin{array}{c}
\mathbf{E}^{\mathbf{Gr.}}=-\nabla \Phi ^{\mathbf{Gr.}}+\dfrac{1}{c}\partial
_{r}\left( \dfrac{1}{2}\mathbf{A}^{\mathbf{Gr.}}\right) , \\ 
\\ 
\mathbf{B}^{\mathbf{Gr.}}=\nabla \times \mathbf{A}^{\mathbf{Gr.}}\mathbf{,}
\\ 
\\ 
\mathbf{E}^{\mathbf{a.c.}}\mathbf{=}-\nabla \Phi ^{\mathbf{a.c.}}+\dfrac{1}{c%
}\partial _{r}\left( \dfrac{1}{2}\mathbf{A}^{\mathbf{a.c.}}\right) , \\ 
\\ 
\mathbf{B}^{\mathbf{a.c.}}=\nabla \times \mathbf{A}^{\mathbf{a.c.}}. \\ 
\\ 
\mathbf{E}^{\mathbf{Gr.I}}=\mathbf{E}^{\mathbf{Gr.}}+\mathbf{E}^{\mathbf{a.c.%
}}. \\ 
\end{array}
& \text{ \ }(4.4.1.11)%
\end{array}%
$

\bigskip

\bigskip From Eq.(4.4.1.11) one obtain

\bigskip\ $%
\begin{array}{cc}
\begin{array}{c}
\\ 
\nabla \times \mathbf{E}^{\mathbf{GI}}=-\dfrac{1}{c}\partial _{r}\left( 
\dfrac{1}{2}\mathbf{B}^{\mathbf{GI}}\right) , \\ 
\\ 
\nabla \cdot \left( \dfrac{1}{2}\mathbf{B}^{\mathbf{GI}}\right) =0. \\ 
\end{array}
& \text{ \ }(4.4.12)%
\end{array}%
$

\bigskip Eq.(4.4.12) and the gravitational-inertial field equations
(4.4.1.4) imply

\bigskip

$%
\begin{array}{cc}
\begin{array}{c}
\\ 
\nabla \cdot \mathbf{E}^{\mathbf{Gr.I}}=4\pi G\left( \rho +\mu \right) +4\pi 
\breve{k}\rho _{ch.}, \\ 
\\ 
\nabla \times \left( \dfrac{1}{2}\mathbf{B}^{\mathbf{Gr.I}}\right) =\dfrac{1%
}{c}\partial _{t}\mathbf{E}^{\mathbf{Gr.I}}+\dfrac{4\pi G}{c}\cdot \mathbf{j}%
^{\mathbf{Gr.}}+\dfrac{4\pi \breve{k}}{c}\cdot \mathbf{j}^{\mathbf{EM}}. \\ 
\end{array}
& \text{ \ }(4.4.13)%
\end{array}%
$

\bigskip

\bigskip \bigskip

\subsection{\textbf{IV.5. Bimetric theory of gravitational-inertial field in
a purely inertial field approximation.}}

\bigskip \bigskip \bigskip

\subsection{IV.\textbf{5.1.Bimetric theory of gravitational-inertial field
in a purely inertial field approximation. Rosen type approximation.}}

\bigskip \bigskip \bigskip

\subsection{\textbf{IV.5.2.Bimetric theory of gravitational-inertial field
in a purely inertial field approximation. Einstein type approximation.}}

\bigskip

\bigskip

\bigskip The Gravitational-Inertial field equations in Einstein approximation%
\textbf{\ }is

\bigskip $\ 
\begin{array}{cc}
\begin{array}{c}
\\ 
\breve{R}_{\mu \nu }-\dfrac{1}{2}g_{\mu \nu }\breve{R}\simeq \breve{\kappa}%
\breve{T}_{\mu \nu }, \\ 
\end{array}
& \text{ \ }(4.5.2.1)%
\end{array}%
$

where $\breve{R}_{\mu \nu }$ is the Ricci tensor and where $\breve{T}_{\mu
\nu }$ is the energy-momentum tensor which

in our farther consideration is the one for electromagnetism

\bigskip $\ $

$%
\begin{array}{cc}
\begin{array}{c}
\\ 
\breve{T}_{\mu \nu }=F_{\mu \rho }F_{\nu }^{\rho }-\dfrac{1}{4}\breve{g}%
_{\mu \nu }F_{\rho \sigma }F^{\rho \sigma }, \\ 
\end{array}
& \text{ \ }(4.5.2.2)%
\end{array}%
$

\bigskip

where $F_{\mu \nu }$ is the electromagnetic field strength tensor.Note that $%
\breve{T}_{\mu \nu }$ has zero trace,

$\breve{T}=\breve{g}^{\mu \nu }F_{\mu \nu }=0.$Eq. (4.5.2.2) allows us to
rewrite the Eq. (4.5.2.1) in the following form

\bigskip

\bigskip $%
\begin{array}{cc}
\begin{array}{c}
\\ 
\breve{R}_{\mu \nu }\simeq \breve{\kappa}\breve{T}_{\mu \nu }. \\ 
\end{array}
& (4.5.2.3)%
\end{array}%
$

\bigskip Finally, the Maxwell's equations are

\bigskip

\bigskip $%
\begin{array}{cc}
\begin{array}{c}
\\ 
\breve{g}^{\mu \nu }\nabla _{\mu }F_{\nu \sigma }=0, \\ 
\\ 
\nabla _{\lbrack \mu }F_{\nu \rho ]}=0. \\ 
\end{array}
& (4.5.2.4)%
\end{array}%
$

\bigskip 

\bigskip \bigskip

\subsection{\textbf{IV.5.3. Gravitational-inertial black hole in a purely
inertial field approximation. Einstein type approximation.}}

\bigskip 

In General Relativity one of famous static solutions to the Einstein's field
equations is the Reissner-Nordstrom metric describing the geometry of the
spacetime surrounding a non-rotating charged spherical black hole. In this
section we obtain   

completely \textit{purely inertial analog} of the Reissner-Nordstrom black
hole in Einstein approximation.

Canonical form for the metric in 4D spherical coordinates $\left( t,r,\theta
,\phi \right) $ is

\bigskip

\bigskip $%
\begin{array}{cc}
\begin{array}{c}
\\ 
d\breve{s}^{2}=-e^{2\alpha \left( r,t\right) }dt^{2}+e^{2\beta \left(
r,t\right) }dr^{2}+r^{2}d\Omega ^{2}, \\ 
\\ 
d\Omega ^{2}=d\theta ^{2}+\sin ^{2}\theta d\phi . \\ 
\end{array}
& \text{ \ }(4.5.3.1)%
\end{array}%
$

\bigskip

Since there is spherical symmetry, the only non-zero components of the
magnetic and electric fields are the radial components which should be
independent of $\theta $ and $\phi .$Therefore the radial component of the
electric field has a form of

\bigskip

\bigskip $%
\begin{array}{cc}
\begin{array}{c}
\\ 
E_{r}=F_{tr}=-F_{rt}=f_{1}\left( r,t\right)  \\ 
\end{array}
& \text{ \ }(4.5.3.2)%
\end{array}%
$

\bigskip

The radial component of the magnetic field has a form of

\bigskip

\bigskip $%
\begin{array}{cc}
\begin{array}{c}
\\ 
B_{r}=\dfrac{2\breve{g}_{rr}}{\sqrt{\left\vert g\right\vert }}F_{\theta \phi
}, \\ 
\\ 
F_{\theta \phi }=-F_{\phi \theta }=f_{2}\left( r,t\right) r^{2}\sin
^{2}\theta . \\ 
\end{array}
& \text{ \ }(4.5.3.3)%
\end{array}%
$

\bigskip

All the remaining components of the electromagnetic field strength tensor
are either zero or related to these two through symmetries. Therefore for
the electromagnetic

field strength tensor one obtain

\bigskip

$%
\begin{array}{cc}
F_{\mu \nu }=%
\begin{bmatrix}
0 & f_{1}\left( r,t\right)  & 0 & 0 \\ 
-f_{1}\left( r,t\right)  & 0 & 0 & 0 \\ 
0 & 0 & 0 & f_{2}\left( r,t\right) r^{2}\sin ^{2}\theta  \\ 
0 & -f_{2}\left( r,t\right) r^{2}\sin ^{2}\theta  & 0 & 0%
\end{bmatrix}
& \text{ \ }(4.5.3.4)%
\end{array}%
$

\bigskip

For the $\theta \theta $-component of the the Riemann tensor $\breve{R}_{\mu
\nu }$ and of the electromagnetic stress tensor $\breve{T}_{\mu \nu }$ one
obtain

\bigskip

$%
\begin{array}{cc}
\begin{array}{c}
\\ 
R_{\theta \theta }=e^{-2\beta \left( r,t\right) }\left[ r\left( \partial
_{r}\beta \left( r,t\right) -\partial _{r}\alpha \left( r,t\right) \right) -1%
\right] +1 \\ 
\\ 
\breve{T}_{\theta \theta }=\dfrac{1}{2}r^{2}f_{2}\left( r,t\right)
+f_{1}\left( r,t\right) e^{-2\left( \alpha \left( r,t\right) +\beta \left(
r,t\right) \right) }, \\ 
\\ 
\alpha \left( r,t\right) =\alpha \left( r\right) =-\beta \left( r\right) .
\\ 
\end{array}
& \text{ \ }(4.5.3.5)%
\end{array}%
$

\bigskip

Now lets solve the Maxwell equations for the form of the electromagnetic
field strength tensor given in Eq.(4.5.3.4).Finally, for the electromagnetic
field strength tensor one obtain

\bigskip

\bigskip $%
\begin{array}{cc}
F_{\mu \nu }=\dfrac{1}{\sqrt{4\pi }}%
\begin{bmatrix}
0 & Qr^{-2} & 0 & 0 \\ 
-Qr^{-2} & 0 & 0 & 0 \\ 
0 & 0 & 0 & P\sin \theta  \\ 
0 & 0 & -P\sin \theta  & 0%
\end{bmatrix}
& \text{ \ }(4.5.3.6)%
\end{array}%
$

\bigskip Lets consider the $\theta \theta $ \ component of the\ Eq.(4.5.2.3).

$%
\begin{array}{cc}
\begin{array}{c}
\\ 
\breve{R}_{\theta \theta }\simeq 8\pi \breve{\kappa}\breve{T}_{\theta \theta
}. \\ 
\end{array}
& (4.5.3.7)%
\end{array}%
$

\bigskip Substituting Eq.(4.5.3.5) into Eq.(4.5.3.7) we obtain

$%
\begin{array}{cc}
\begin{array}{c}
\\ 
\partial _{r}\left( re^{2\alpha }\right) =1-\dfrac{\breve{\kappa}}{r^{2}}%
\left( Q^{2}+P^{2}\right) . \\ 
\end{array}
& \text{ \ }(4.5.3.6)%
\end{array}%
$

\bigskip 

By integration we obtain

\bigskip 

\bigskip $%
\begin{array}{cc}
\begin{array}{c}
\\ 
e^{2\alpha \left( r\right) }=1+\dfrac{const}{r}+\dfrac{\breve{\kappa}}{r^{2}}%
\left( Q^{2}+P^{2}\right) . \\ 
\end{array}
& \text{ \ }(4.5.3.7)%
\end{array}%
$

\bigskip Take into account Eq.(4.2.2.7.a) we obtain 

\bigskip $%
\begin{array}{cc}
\begin{array}{c}
\\ 
e^{2\alpha \left( r\right) }=1+\dfrac{2\sqrt{\breve{\kappa}}Q}{r}+\dfrac{%
\breve{\kappa}}{r^{2}}\left( Q^{2}+P^{2}\right) , \\ 
\\ 
\breve{\kappa}=\left( \dfrac{e}{mc^{2}}\right) ^{2}, \\ 
\end{array}
& \text{ \ }(4.5.3.7)%
\end{array}%
$

or

\bigskip 

\bigskip $%
\begin{array}{cc}
\begin{array}{c}
\\ 
e^{2\alpha \left( r\right) }=1+\dfrac{2\mu Q}{r}+\dfrac{\mu ^{2}}{r^{2}}%
\left( Q^{2}+P^{2}\right) , \\ 
\\ 
\mu =\dfrac{e}{mc^{2}}. \\ 
\end{array}
& \text{ \ }(4.5.3.7)%
\end{array}%
$

Finally, upon substitution of Eq.(4.5.3.7) into Eq.(4.5.3.1) the metric of
the purely inertial Reissner-Nordstrom black hole  is readily found:

\bigskip $%
\begin{array}{cc}
\begin{array}{c}
\\ 
d\breve{s}^{2}=-\Delta dt^{2}+\Delta ^{-1}dr^{2}+r^{2}d\Omega ^{2}, \\ 
\\ 
\Delta =1+\dfrac{2\mu Q}{r}+\dfrac{\mu ^{2}}{r^{2}}\left( Q^{2}+P^{2}\right)
, \\ 
\\ 
\mu =\dfrac{e}{mc^{2}}. \\ 
\end{array}
& \text{ \ }(4.5.3.8)%
\end{array}%
$

\bigskip 

Note that in the absence of charges, this should reduce to the flat metric
and hence 

purely inertial analog of the Schwarzschild black hole is absent.

\bigskip

\subsection{V.\textbf{Noninertial Pure Accelerated Curved Reference Frame in
Bimetric theory of gravitational-inertial field.}}

\bigskip

Recall the basic concept and definitions of the accelerated reference frame
in canonical GTR [43],[44],[45].Let us considered flat (curved) basic
Lorentzian space-time $\tciLaplace _{4}=\tciLaplace \left( V_{4},g\right) $
and any timelike congruence $\mathcal{C}$, in a certain domain $\mathcal{A}%
_{4}$ $\mathcal{\subseteqq }$ $V_{4},$ defined in a coordinate system $%
\left\{ x^{\alpha }\right\} $ by

\bigskip\ $\ 
\begin{array}{cc}
\begin{array}{c}
\\ 
\mathcal{C}\text{ }:x^{\alpha }=x^{\alpha }\left( t,\lambda ^{i}\right) , \\ 
\\ 
\alpha ,\beta =0,1,2,3;i=1,2,3 \\ 
\end{array}
& \left( 5.1\right) 
\end{array}%
$

where $\left\{ \lambda ^{i}\right\} $ are three parametr marking the
specific curve in $\mathcal{C}$ an $t$ is any parametr along these curves in 
$\mathcal{C}.$

Some canonical intristict element defined by the congruence $\mathcal{C}$
are: \ \ \ \ \ \ \ \ \ \ \ \ \ \ \ \ \ \ \ \ \ \ \ \ \ \ \ \ \ \ \ \ \ \ \ \ 
\textbf{(i)} The quotient space $V_{3}\left( \mathcal{C}\right) $ associated
to $\mathcal{C}$ or the internal space of $\mathcal{C}$ given by the

equivalence relation: $V_{3}\left( \mathcal{C}\right) \triangleq V_{3}/%
\mathcal{C}.$ \ \ \ \ \ \ \ \ \ \ \ \ \ \ \ \ \ \ \ \ \ \ \ \ \ \ \ \ \ \ \
\ \ \ \ \ \ \ \ \ \ \ \ \ \ \ \ \ \ \ \ \ \ \ \ \ \ \ \ \ \ \ \ \ \ \ \ \ \
\ \ \ \ \ \ \ \ \ \ \ \ \ \ \ \ \ \ \ \ \ \textbf{(ii)} The natural
projection $j:\mathcal{A}_{4}\rightarrow V_{3}/\mathcal{C}$

\bigskip

\bigskip $\ 
\begin{array}{cc}
\begin{array}{c}
\\ 
x^{\alpha }\rightarrow \xi ^{i}=\lambda ^{i}\left( x^{\alpha }\right) \\ 
\end{array}
& \left( 5.2\right)%
\end{array}%
$ \ \ \ 

\ \ \ 

where $\lambda ^{i}\left( x^{\alpha }\right) $ are the inverted functions of
(2.1.1) and $\left\{ \xi ^{i}\right\} $ are a coordinate sistem of $%
V_{3}\left( \mathcal{C}\right) .$The pull-back and push-forward of $j$ allow
us to define certain objects on $\mathcal{A}_{4}$ and $V_{3}\left( \mathcal{C%
}\right) $ respectively. For instance, three one-form fields $\left\{ d\xi
^{i}\right\} $ of the naturalco- basis in $V_{3}\left( \mathcal{C}\right) $
can be pulleed back to the tree one-form fields $\left\{ \omega ^{i}\right\} 
$ defined by

\bigskip

\bigskip $%
\begin{array}{cc}
\begin{array}{c}
\\ 
\omega ^{i}=\dfrac{\partial \lambda ^{i}\left( x^{\alpha }\right) }{\partial
x^{\alpha }}dx^{\alpha }. \\ 
\end{array}
& \left( 5.3\right) 
\end{array}%
$

Thus 3-dimensional subspase $\Delta _{3}$ of $T_{V_{4}}^{\ast }$ spanned by $%
\left\{ \omega ^{i}\right\} $, i.e.$\Delta _{3}\triangleq \mathbf{span}%
\left\{ \omega ^{1},\omega ^{2},\omega ^{3}\right\} $ is invariantly
characterized by $\mathcal{C}$. Concerning the push-forward projection,
defined for example as follows

\bigskip

$%
\begin{array}{cc}
\begin{array}{c}
\\ 
j^{\prime }:T_{x}\rightarrow T_{j\left( x\right) }, \\ 
\\ 
t^{\alpha }\left( x\right) \rightarrow t^{i}\left[ \lambda \left( x\right) %
\right] =t^{\alpha }\left( x\right) \omega _{\alpha }^{i} \\ 
\end{array}
& \left( 5.4\right)%
\end{array}%
$

\bigskip

\textbf{(iii)} The proper time $\tau $ and 4-velocity vector $%
\overrightarrow{u}$ defined respectively by formulae

\bigskip

\bigskip $%
\begin{array}{cc}
\begin{array}{c}
\\ 
\tau =\int \sqrt{-g_{\alpha \beta }\dfrac{\partial x^{\alpha }}{\partial t}%
\dfrac{\partial x^{\beta }}{\partial t}}+C\left( \lambda ^{\alpha }\right) ,
\\ 
\end{array}
& \left( 5.5\right)%
\end{array}%
$

\bigskip and

$%
\begin{array}{cc}
\begin{array}{c}
\\ 
\overrightarrow{\mathbf{u}}=\left( \dfrac{1}{\sqrt{-g_{\alpha \beta }\dfrac{%
\partial x^{\alpha }}{\partial t}\dfrac{\partial x^{\beta }}{\partial t}}}%
\dfrac{\partial x^{\sigma }}{\partial t}\right) \dfrac{\partial }{\partial
x^{\sigma }}. \\ 
\end{array}
& \left( 5.6\right)%
\end{array}%
$

\bigskip

Where $C\left( \lambda ^{\alpha }\right) $ being\ an arbitrary function of $%
\ \arg $uments $\left\{ \lambda ^{\alpha }\right\} $ [43].Notice that $%
\overrightarrow{\mathbf{u}}$ is orthogonal to the three $\omega ^{i},i=1,2,3$
which means that the functions $\lambda ^{i}\left( x^{\alpha }\right) $ are
three independent first integrals of $\overrightarrow{\mathbf{u}}.$Of
course, given any timelike unit vector $\overrightarrow{\mathbf{u}}$ one can
build its associated congruences locally by constructing the curves tangent
to $\overrightarrow{\mathbf{u}}$ by means of the canonical integration.
Therefore one can identifies a timelike congruence and its tangent
for-velocity vector field.The canonical kinematical quantities associated
with $\overrightarrow{\mathbf{u}}$ \ are intistinc\ to\ congruence $\mathcal{%
C.}$ For example, the projection tensor orthogonal to $\overrightarrow{%
\mathbf{u}}:$ $P_{\alpha \beta }=g_{\alpha \beta }+u_{\alpha }u_{\beta },$
the acceleration $\overrightarrow{\mathbf{a}},$ the deformation tensor $%
\Sigma _{\alpha \beta }$ and the rotation tensor $\omega _{\alpha \beta }.$

\bigskip \bigskip

Let us considered flat basic Lorentzian space-time $\tciLaplace
_{4}=\tciLaplace \left( V_{4},\eta _{\alpha \beta }\right) $ where $\eta
_{\alpha \beta }=diag\left( 1,-1,-1,-1\right) $.Any timelike congruence $%
\mathcal{C}$, in a certain domain $\mathcal{A}_{4}$ $\mathcal{\subseteqq }$ $%
V_{4},$ formed by Eq.(5.1) by using a regular parametrized timelike curves $%
\Gamma \left( s,\lambda ^{i}\right) .$It is also convenient to restrict
ourselves to timelike curves $x^{\alpha }=x^{\alpha }\left( s\right) $ i.e.
those for which

\bigskip $\eta _{\alpha \beta }\dfrac{dx^{\alpha }}{ds}\dfrac{dx^{\beta }}{ds%
}=1,$where now $s$ denotes the arc length parameter in the sense of
Minkowski metric $\eta _{\alpha \beta }.$Accordingly,if we denote the tetrad
vectors by $u_{\left( A\right) }^{\alpha }$ ($A=0,1,2,3$)\bigskip , then the
orthonormality conditions read \ $u_{\left( A\right) }^{\alpha }u_{\left(
A\right) \beta }=\eta _{\alpha \beta }u_{\left( A\right) }^{\alpha
}u_{\left( B\right) }^{\beta }=\eta _{AB}.$

If we chose $u_{\left( 0\right) }^{\alpha }=\dfrac{dx^{\alpha }}{ds}$ then
we can easily construct an orthonormal basis of vectors

$\left\{ u_{\left( A\right) }^{\alpha }\right\} $,defined along the curve,
which obey the following four-dimensional Serret-Frenet equations,given in
matrix representation by

\bigskip\ $%
\begin{array}{rr}
\begin{array}{r}
\\ 
\left[ 
\begin{array}{r}
\dfrac{du_{0}(s)}{ds} \\ 
\dfrac{du_{1}(s)}{ds} \\ 
\dfrac{du_{2}(s)}{ds} \\ 
\dfrac{du_{3}(s)}{ds}%
\end{array}%
\right] =\left[ 
\begin{array}{rrrr}
0 & \kappa (s) & 0 & 0 \\ 
\kappa (s) & 0 & \tau _{1}(s) & 0 \\ 
0 & -\tau _{1}(s) & 0 & \tau _{2}(s) \\ 
0 & 0 & -\tau _{2}(s) & 0%
\end{array}%
\right] \left[ 
\begin{array}{r}
u_{0}(s) \\ 
u_{1}(s) \\ 
u_{2}(s) \\ 
u_{3}(s)%
\end{array}%
\right] \\ 
\end{array}
& \left( 5.7\right)%
\end{array}%
$

\textbf{Theorem.5.1.} [46] Given differentiable functions $\kappa (s)>0,\tau
_{1}(s)$ and $\tau _{2}(s),$there exists a regular parametrized timelike
curve $\Gamma $ such that $\kappa (s)$ is the curvature, $\tau _{1}(s)$ and $%
\tau _{2}(s)$ are, respectively, the first and second torsion of $\Gamma .$%
Any other curve $\widetilde{\Gamma }$ satisfying the same conditions,
differs from $\Gamma $ by a Poincar\'{e} transformation, i.e. by a
transformation of the type $x_{\mu }^{\prime }=\Lambda _{\mu }^{\nu }x_{\nu
}+a_{\mu },$ where $\mu $ represents a proper Lorentz matrix and $a_{\mu }$
is a constant four-vector.\bigskip

$\ \ $

\bigskip

\bigskip

As we have seen above, the canonical introduction of reference (comovin)
frame formulation the crucial role in their mathematical discription has to
by played in a study of congruences by the world lines of particles forming
bodies of reference, i.e. the physical time lines for the chosen reference
frame.The congruence concept is essential because for the sake of regularity
of the mathematical description of the frame,these lines have not to
mutually intersect, and they must cover completely the space-time region
under consideration so that at every world point one has to find one and
only one line passing throus it.Exactly the same approach is used for
description of a regular continuous media,i.e. in relativistic gydrodynamics
of a perfect (magnetic) fluid [44].

\bigskip

\bigskip

\bigskip

\begin{remark}
5.1.\textbf{\ }However it is important to note that in canonical GTR in
contrast with GIFT, curvature of the basic Lorentzian space-time $%
\tciLaplace \left( V_{4},g\right) $ does not depend from accelerations of
the particles forming fluid or bodies of reference.
\end{remark}

\bigskip

\bigskip

\bigskip

\subsection{V.1. Non Inertial Anholonomic Accelerated Frame of Refferences
in Finsler-Lagrange Approximation. Hollands type \textbf{comovin frame.}}

\bigskip

Holland was studied a unified formalism which uses a anholonomic frame
(nonintegrable $1$-form) on space-time, \textit{a sort of plastic deformation%
}, arising from consideration of a charged particle moving in an external
electromagnetic field in the background space-time viewed as a strained
medium [10]-[11]. In fact, Ingarden [12] was first to point out that the
Lorentz force law, in this case, can be written as a geodesic equation on a
Finsler space called Randers space [13] i.e., the physical space with a
metric:

\bigskip

\bigskip\ $\ 
\begin{array}{cc}
\begin{array}{c}
\\ 
ds=\sqrt{g_{ij}dx^{i}dx^{j}}+a_{k}\left( x\right) dx^{k}, \\ 
\\ 
\det \left\Vert g_{ij}\right\Vert \neq 0. \\ 
\end{array}
& \text{ }(5.1.1)%
\end{array}%
$

\bigskip

The metric given by Eq.(5.1.1) is defined by the pair $\left(
g_{ij},a_{k}\right) $ of the tensor field $g_{ij}$ and vector field $a_{k},$
where $g_{ij}$ influences the local inhomogeneity of the space, $a_{k}$
changes the local anisotropy. \bigskip

\begin{remark}
5.1.1.Note that the additional term in the geodesic equation acts as
repulsive force against the gravity [21].
\end{remark}

\bigskip

For complete references on these Finsler spaces see [14]-[17].This results
in geometrical entities which depend on the electromagnetic field (vector
potential), particle (velocity) and background space-time parameters. The
Finsler structure implies the existence of a global anholonomic (Holland)
frame which in turn yields a connection with torsion and vanishing Finsler
curvatures.His differential geometric method is based on fundamental work of
S. Amari on a Finsler approach to crystal dislocation theory [19]. \bigskip
Amari and Holland's idea\ conduct one to considered Holland type frames \ as
non inertial anholonomic accelerated frame of refferences in
Finsler-Lagrange approximation.

\bigskip

\bigskip \bigskip

\subsection{V.2. Non Inertial Anholonomic Accelerated Frame of Refferences
in Riemannian Approximation.\textbf{Bravais Type Relativistic Comovin Frame
with Curvature and Torsion.}}

\bigskip\ 

\bigskip\ \ \ \ \ \ \ \ \ \ \ \ \ \ \ \ \ \ \ \ \ \ \ \ \ \ \ \ \ \ \ \ \ \
\ \ \ \ \ \ \ \ \ \ \ \ \ \ \ \ \ \ \ \ \ \ \ \ \ \ \ \ \ \ \ \ \ \ \ \ \ \
\ \ \ \ \ \ \ \ \ \ \ \ \ \ \ \ \ \ \ \ \ \ \ \ \ \ \ \ \ \ \ \ \ \ \ \ \ \
\ \ \ \ \ \ \ \ \ \ \ \ \ \ \ 

\bigskip\ \ \ \ \ \ \ \ \ \ \ \ \ \ \ \ \ \ \ \ \ \ \ \ \ \ \ \ \ \ \ \ \ \
\ \ \ \ \ \ \ \ \ \ \ \ \ \ \ \ \ \ \ \ \ \ \ \ \ \ \ \ \ \ \ \ \ \ \ \ \ \
\ \ \ \ \ \ \ \ \ \ \ \ \ \ \ \ \ \ \ \ \ \ \ \ \ \ \ \ \ \ \ \ \ \ \ \ \ \
\ \ \ \ \ \ \ \ \ \ \ \ \ \ \ \ \ \ \ \ \ \ \ \ 

\ \ \ \ \ \ \ \ \ \ \ \ \ \ \ \ \ \ \ \ \ \ \ \ \ \ \ \ \ \ \ \ \ \ \ \ \ \
\ \ \ \ \ \ \ \ \ \ \ \ \ \ \ \ \ \ \ \ \ \ \ \ \ \ \ \ \ \ \ \ \ \ \ \ \ \
\ \ \ \ \ \ \ \ \ \ \ \ \ \ \ \ \ \ \ \ \ \ \ \ \ \ \ \ \ \ \ \ \ \ \ \ \ \
\ \ \ \ \ \ \ \ \ \ \ \ \ \ \ \ \ \ \ \ \ \ \ \ \ Let $M$ be a
differentiable manifold of dimension $n.$ At a point $p\in M,$ let $\left\{
e^{i}\right\} \bigskip ,i=$ $1,...,n$ constitute the basis of the cotangent
space $T_{p}^{\ast }\left( M\right) $ \bigskip and let $\left\{
e_{i}\right\} $ be the base vectors of the tangent space $T_{p}\left(
M\right) .$ The local coordinate form of the bases of $T_{x}^{\ast }\left(
M\right) $ and $T_{x}\left( M\right) $ at $p=x$ are $\left\{ dx^{i}\right\} $
and $\left\{ \partial _{i}=\dfrac{\partial }{\partial x^{i}}\right\} $
respectively. Let $\omega _{k}^{i}$ be the connection 1-form of $M.$ Then
the description of $M$ is given by Cartan's structure equations:

\bigskip\ $%
\begin{array}{cc}
\begin{array}{c}
\\ 
T^{i}=De^{i}=de^{i}+\omega _{k}^{i}\wedge e^{k}=\frac{1}{2}%
T_{kl}^{i}e^{k}\wedge e^{l}, \\ 
\\ 
R_{k}^{i}=D\omega _{k}^{i}=d\omega _{k}^{i}+\omega _{l}^{i}\wedge \omega
_{k}^{l}=\tfrac{1}{2}R_{klm}^{i}e^{l}\wedge e^{m}. \\ 
\end{array}
& (5.2.1)%
\end{array}%
$

The integrability conditions of the above equations are given by

\bigskip

\bigskip\ $%
\begin{array}{cc}
\begin{array}{c}
\\ 
DT^{i}=R_{k}^{i}\wedge e^{k}, \\ 
\\ 
DR_{k}^{i}=0. \\ 
\end{array}
& \text{ }(5.2.2)%
\end{array}%
$

These are known as Bianchi identities. The Cartan equations and Bianchi
identities are present in both Yang-Mills and gravity-type gauge theories.
However, the latter type has the following additional structural features. A
symmetric metric tensor $g=g_{ik}e^{i}\otimes e^{k},$ $g_{ik}=g_{ki}$ $%
=e_{i}\cdot e_{k}$ is introduced on $M.$ In local coordinates, the metric is
used to describe the distance element: $ds^{2}=g_{ik}dx^{i}dx^{k}.$The
inverse metric $g^{kl}$ is such that $g^{kl}g_{li}=\delta _{i}^{k}.$The
metric and the connection are so far two independent fields, defined at each
point of $M.$ A manifold in which the covariant derivative of the metric
tensor vanishes is singled out by the property that the angle between two
vectors and their lengths remain unchanged by the operation of parallel
displacement of vectors on $M$. It is this property which guarantees locally
Euclidean structure of the manifold. A connection is called metric
compatible if

\bigskip

$%
\begin{array}{cc}
\begin{array}{c}
\\ 
Dg_{ik}=dg_{ik}-g_{il}\omega _{k}^{l}-g_{kl}\omega _{i}^{l}=0. \\ 
\end{array}
& \text{ }(5.2.3)%
\end{array}%
$

\bigskip

In general, the connection $\omega _{k}^{i}$ can have a torsion-free part $%
\tilde{\omega}_{k}^{i}$ and an additional part $\tau _{k}^{i}$ which
represents the non-Riemannian part, called the contorsion 1-form. The local
coordinate representations of these geometrical objects are:

$\ 
\begin{array}{cc}
\begin{array}{c}
\\ 
\omega _{l}^{k}=\Gamma _{ml}^{k}dx^{k},\tilde{\omega}_{l}^{k}=\left\{ \QATOP{%
k}{ml}\right\} dx^{m}, \\ 
\\ 
\tau _{l}^{k}=S_{ml}^{k}dx^{m},T^{k}=\frac{1}{2}T_{ml}^{k}dx^{m}\wedge
dx^{l}. \\ 
\end{array}
& \text{ }(5.2.4)%
\end{array}%
$

\bigskip

Here $\left\{ \QATOP{k}{ml}\right\} =g^{ks}(\partial _{m}g_{sl}-\partial
_{s}g_{lm}+\partial _{l}g_{ms}),$ and $%
S_{ml}^{k}=g^{ks}(S_{msl}-S_{slm}+S_{lms})$ are respectively the Christoffel
symbol of the second kind and the contorsion tensor. We next relate the
above structure to that of a non inertial accelerated frame (comoving to
accelerated body).

\begin{definition}
5.2.1.Non inertial accelerated frame (comoving to accelerated body) of
refferences is identified with a four dimensional differentiable manifold $M$
embedded in the real four-dimensional linear space $%
%TCIMACRO{\U{211d} }%
%BeginExpansion
\mathbb{R}
%EndExpansion
^{4}.$The current coordinates of the manifold of the accelerated frame
(accelerated deformed body) $M^{\prime }$ are $x^{i}$ ( $%
i,j,k,l,m,n,...=1,2,3,4$) and the cartesian coordinates of the \textit{%
anholonomy-free configuration} (reference manifold $M$) are $x^{a}$ ($%
a,b,c,d,...=1,2,3,4$).
\end{definition}

\bigskip

\begin{definition}
\bigskip\ 5.2.2.The current configuration of the accelerated frame (comoving
to accelerated deformed body) $M^{\prime }$ is anholonomy-free iff functions 
$x^{i}=x^{i}(x^{a})$ and $x^{a}=x^{a}(x^{i})$ are well behaved,
single-valued and differentiable functions of their respective arguments.
The matrix $\beta _{a}^{i}=\partial _{a}x^{i}=\dfrac{\partial x^{i}}{%
\partial x^{a}}$ is the holonomy deformation (distortion) matrix.Its inverse
matrix is $\beta _{i}^{a}=\partial _{i}x^{a}=\dfrac{\partial x^{a}}{\partial
x^{i}}.$
\end{definition}

\bigskip

Anholonomy-free manifold $M$ (defect-free body) is characterized by a global
coordinate basis of the reference manifold (reference body). The metric $%
e_{a}\cdot e_{b}=\delta _{ab}$ is Euclidean and the connection $\omega
_{b}^{a}=\Gamma _{cb}^{a}dx^{c}$ vanishes identically.The metric and the
connection of the current configuration are

\bigskip

\bigskip $%
\begin{array}{cc}
\begin{array}{c}
\\ 
g_{ik}=\beta _{i}^{a}\beta _{k}^{b}\delta _{ab}, \\ 
\\ 
\omega _{k}^{i}=\beta _{a}^{i}d\beta _{k}^{a}. \\ 
\end{array}
& \text{ \ }(5.2.5)%
\end{array}%
$

Holonomic (defect-free body) manifold $M$ is characterized by a global
coordinate basis $e^{i}=dx^{i},$ and a (metric compatible) flat connection $%
\omega _{k}^{i}=\beta _{a}^{i}d\beta _{k}^{a}.$These equations may be
regarded as a set of differential equations for $e^{i}$ and $\omega
_{k}^{i}. $ In this case the torsion and curvature tensors are zero and the
integrability equations are (5.2.1) with the right sides set equal to zero.
Torsion and curvature in general case represent anholonomic deformation of
the accelerated frame (or defects). Anholonomic deformation (defects) are
obstructions to diffeomorphisms from $M$ to $M^{\prime }.$

In order to relate the mathematical structure to the description of the
accelerated frame (accelerated deformed body), consider the infinitesimal
transformation

\bigskip

$%
\begin{array}{cc}
\begin{array}{c}
\\ 
x^{a}\rightarrow x^{m}=\left( x^{a}+u^{a}\left( x^{b}\right) \right) \delta
_{a}^{m} \\ 
\end{array}
& \text{ }(5.2.6)%
\end{array}%
$

\bigskip

where the total displacement $u^{a}$ consists of an holonomy (elastic) part
and an anholonomy (plastic) part. The elastic part is integrable and the
plastic part is not. The total deformation (distortion) tensors are

\bigskip $\ 
\begin{array}{cc}
\begin{array}{c}
\\ 
\beta _{a}^{i}=\delta _{a}^{i}+\partial _{a}u_{i}, \\ 
\\ 
\beta _{i}^{a}=\delta _{i}^{a}-\partial _{i}u_{a}. \\ 
\end{array}
& \text{ \ }(5.2.7)%
\end{array}%
$

\bigskip

The metric gets related to the total "strain" tensor

\bigskip $\ \ \ \ \ \ \ \ \ \ \ \ \ \ \ \ \ \ \ \ \ \ \ \ \ \ \ \ \ \ \ \ \
\ \ \ \ \ \ \ \ \ $

$\ \ \ \ \ \ \ \ \ \ \ $

$%
\begin{array}{cc}
\begin{array}{c}
\\ 
g_{ik}=\beta _{ai}\beta _{k}^{a}= \\ 
\\ 
\delta _{ik}-\partial _{i}u_{k}-\partial _{k}u_{i}= \\ 
\\ 
\delta _{ik}-2e_{ik}. \\ 
\end{array}
& \text{ }(5.2.8)%
\end{array}%
$

\bigskip

\textbf{1.} The present internal geometry of the accelerated frame (comoving
to accelerated deformed body) is Riemannian if :

$\bigskip $

\ $%
\begin{array}{cc}
\begin{array}{c}
\\ 
Dg=0, \\ 
\\ 
R_{k}^{i}=D\omega _{k}^{i}\neq 0, \\ 
\\ 
DR_{k}^{i}=0.\  \\ 
\end{array}
& \text{ }(5.2.9)%
\end{array}%
$

\bigskip

\textbf{2.} The present internal geometry of the accelerated frame (comoving
to accelerated deformed body) is non Riemannian (tele-parallel) if :

$\bigskip $

$\ 
\begin{array}{cc}
\begin{array}{c}
\\ 
Dg=0, \\ 
\\ 
R_{k}^{i}=0, \\ 
\\ 
DR_{k}^{i}=0, \\ 
\\ 
DT^{i}=0, \\ 
\\ 
T^{i}=De^{i}\neq 0. \\ 
\end{array}
& \text{ }(5.2.10)%
\end{array}%
$

\bigskip

\textbf{3.} In general case the present internal geometry of the accelerated
frame (comoving to accelerated deformed body) is characterized by
Eq.(5.2.1), the Bianchi identities Eq.(5.2.3)\ is non Riemannian. It is also
called Riemann-Cartan geometry.

\bigskip

\begin{definition}
5.2.3.Cartan's structure equations are nothing but the very definition of
"dislocation" and "disclination" density tensors of manifold:
\end{definition}

\bigskip

\bigskip \bigskip $%
\begin{array}{cc}
\begin{array}{c}
\\ 
\alpha ^{ij}=\varepsilon ^{ikm}T_{km}^{\text{ }j}, \\ 
\\ 
\theta ^{ij}=\dfrac{1}{2}\varepsilon ^{imn}\varepsilon ^{jkl}R_{klmn}. \\ 
\end{array}
& \text{\ }(5.2.11)%
\end{array}%
$

\bigskip

In general case (in particular the presence of "defects"), the coordinate
system of $M^{\text{ }\prime }$ is nonholonomic. Denoting the anholonomic
coordinates by $e^{a}$ instead of $e^{i},$we may write $e^{a}=\beta
_{a}^{\alpha }dx^{a}$ but $\beta _{a}^{\alpha }$ is no longer a gradient
field.$\bigskip $

\bigskip

\bigskip

\subsection{\textbf{VI.Gauge theories} \textbf{of accelerated comovin frame.}%
}

\textbf{\bigskip }

\bigskip

\subsection{\textbf{VI.1.General mathematical structure of gauge theories.}}

\bigskip

Gauge theories are divided into two different classes:

\begin{itemize}
\item \textbf{1.}Yang-Mills type gauge theories and
\end{itemize}

\begin{itemize}
\item \textbf{2.}Gravity type gauge theories.

\item \textbf{3.}Mixed type gauge theories.
\end{itemize}

Let us first consider their general mathematical structure. We shall refrain
from a terse mathematical presentation of the principal fibre-bundle
structure since this can be done away with.Let $u^{i}(x),i=1,2,...,n$ be a
system of initial fields, called matter fields. Here $x$ is a space-time
point on a base Lorentzian manifold $M.$ To each point $x$ of $M$ is
attached a fibre space $V$ whose elements are values of $u^{i}.$ This may be
regarded as an \textit{internal} \textit{space.} The functions $u_{i}(x)$
are cross sections on the fibre-bundle $M\times V.$ Further we assume that a
space-time group $\mathbf{P}_{0}$ and an internal group $\mathbf{G}_{0}$ act
on $M$ and $V$ respectively.

\begin{itemize}
\item The group $\mathbf{P}_{0}$ could be, for example, the Poincar\'{e}
group or one of its sub-groups (translation, rotation,etc.).

\item The group $\mathbf{G}_{0}$ could be another Lie group such as a
rotation or a unitary group.

\item The group actions are $\mathbf{P}_{0}$ $:M\rightarrow M$ and $\mathbf{G%
}_{0}$ $:V\rightarrow V.$ Thus both groups are continuous transformation
groups. Both these groups are said to act globally, i.e., their actions do
not depend on $x.$
\end{itemize}

\bigskip

Let a matter field model be given by a Lagrangian $\tciLaplace
_{0}=\tciLaplace _{0}(\partial u,u)$ which is invariant with respect to $%
\mathbf{P}_{0}$ and $\mathbf{G}_{0}$. This global symmetry is a necessary
prerequisite of any gauge theory. The basic idea of gauging is to extend the
global invariance group G$_{0}$ or P$_{0}$ to a local gauge group $\mathbf{G}
$ (or $\mathbf{P}$) by allowing the transformations $\mathbf{G}\times
V\rightarrow V$ and $\mathbf{P}\times M\rightarrow M$ to be $x$
dependent.The gauge theory based on $\mathbf{G}_{0}$ $\mathbf{\rightarrow G}$
is of Yang-Mills type and that based on $\mathbf{P}_{0}$ $\mathbf{%
\rightarrow P}$ is of gravity type. A mixed type could be based on gauging
of both $\mathbf{G}_{0}$ $\mathbf{\rightarrow G}$ and $\mathbf{P}_{0}$ $%
\mathbf{\rightarrow P}$. In order to ensure local invariance, the Lagrangian
must contain, in addition to fields $u^{i},$ a set of connection fields or
gauge potentials $A_{\mu }(x).$ these are a set of compensating fields
coupled (minimally) to the matter fields $u^{i}.$The values of $A_{\mu }$
belong to the Lie algebra $\tciLaplace \left( \mathbf{G}_{0}\right) $ of $%
\mathbf{G}_{0}$ (or $\mathbf{P}_{0}$).

These fields are called connections on the corresponding principal fibre
bundles.

To obtain a closed system of equations for ui and $A_{\mu },$ the gauge
approach prescribes two recipes. Firstly, the derivatives $\partial _{\mu }$
are to be replaced by covariant derivatives $D_{\mu }=\partial _{\mu
}+A_{\mu }(x).$

Secondly, the new Lagrangian $\widetilde{\tciLaplace }$ is supposed be given
by $\widetilde{\tciLaplace }=\tciLaplace _{0}(Du,u)+\tciLaplace _{1}(F)$
(minimal coupling) where $F_{\lambda \mu }=D_{\lambda }A_{\mu }$ is the
Yang-Mills field (curvature field associated with the connection field). The
piece $\tciLaplace _{1}$ is usually chosen as $\mathbf{Tr}(FF^{\dag }).$

\bigskip

\bigskip

\subsection{\textbf{VI.2. Charged Particles }as Defects In Bimetric
Lorentzian Manifold}

\bigskip

Kleinert [34],[48] demonstrates that a space with torsion and curvature can
be generated from a Minkowski space via singular coordinate transformations $%
x^{i}=x^{i}\left( x^{\mu }\right) $ and is completely equivalent to a
crystal which has undergone plastic deformation being filled with
dislocations and disclinations. Typycally canonical transformation
associated with dislocation can be described multivalued function [34]:

$%
\begin{array}{cc}
\begin{array}{c}
\\ 
x^{\bar{1}}=x^{1}, \\ 
\\ 
x^{\bar{2}}=x^{2}-\dfrac{b}{2\pi }\tan ^{-1}\left( \dfrac{x^{2}}{x^{1}}%
\right) , \\ 
\end{array}
& (6.2.1)%
\end{array}%
$

\bigskip where the function $\tan ^{-1}\left( \dfrac{x^{2}}{x^{1}}\right) $
defined to be equal $\pm \pi $ for $x^{1}<0,x^{2}=\pm \epsilon .$

\bigskip $%
\begin{array}{cc}
\begin{array}{c}
\\ 
dx^{\bar{1}}=dx^{1}, \\ 
\\ 
\left( dx_{\epsilon }^{\bar{2}}\right) _{\epsilon }=dx^{2}- \\ 
\\ 
\dfrac{b}{2\pi }\left( x^{2}dx^{1}-x^{1}dx^{2}\right) \left( \dfrac{1}{%
\left( x^{1}\right) ^{2}+\left( x^{2}\right) ^{2}+\epsilon }\right)
_{\epsilon } \\ 
\end{array}
& (6.2.2)%
\end{array}%
$

\bigskip with the components of the Colombeau vielbein $\left( e_{\mu
,\epsilon }^{i}\right) _{\epsilon }=\left( \dfrac{\partial x_{\epsilon }^{i}%
}{\partial x_{\mu }}\right) _{\epsilon }$

$%
\begin{array}{cc}
\begin{array}{c}
\\ 
\left( e_{\mu ,\epsilon }^{i}\right) _{\epsilon }= \\ 
\\ 
\begin{bmatrix}
1 & 0 \\ 
\dfrac{b}{2\pi }\left( \dfrac{x^{2}}{\left( x^{1}\right) ^{2}+\left(
x^{2}\right) ^{2}+\epsilon }\right) _{\epsilon } & 1-\dfrac{b}{2\pi }\left( 
\dfrac{x^{1}}{\left( x^{1}\right) ^{2}+\left( x^{2}\right) ^{2}+\epsilon }%
\right)%
\end{bmatrix}
\\ 
\end{array}
& (6.2.3)%
\end{array}%
$

\bigskip

\bigskip $%
\begin{array}{cc}
\begin{array}{c}
\\ 
S_{12}^{\text{ \ \ \ }\bar{2}}=\partial _{1}e_{\text{ \ }2}^{\bar{2}%
}-\partial _{2}e_{\text{ \ }1}^{\bar{2}}= \\ 
\\ 
\\ 
\end{array}
& (6.2.4)%
\end{array}%
$

The associated Cartan curvature tensor $R_{\mu \nu \lambda }^{\text{ \ \ \ }%
\kappa }:$

\bigskip

\bigskip\ $%
\begin{array}{cc}
\begin{array}{c}
\\ 
R_{\mu \nu \lambda }^{\text{ \ \ \ }\kappa }=e_{i}^{\text{ }\kappa }\left(
\partial _{\mu }\partial _{\nu }-\partial _{\nu }\partial _{\mu }\right)
e_{\lambda }^{i} \\ 
\end{array}
& (6.2.5)%
\end{array}%
$

vanishes, making the connection affine-flat. The antisymmetric part of the
connection

\bigskip $%
\begin{array}{cc}
\begin{array}{c}
\\ 
S_{\alpha \beta }^{\text{ \ \ }\gamma }=\dfrac{1}{2}\left( \Gamma _{\alpha
\beta }^{\text{ \ \ }\gamma }-\Gamma _{\beta \alpha }^{\text{ \ \ \ }\gamma
}\right) , \\ 
\end{array}
& \text{ }(6.2.6)%
\end{array}%
$

which is a tensor, is nonzero giving rise to a nonvanishing Riemann
curvature tensor $\widetilde{R}_{\mu \nu \lambda }^{\text{ \ \ \ }\kappa
}\neq 0.$The latter is formed by canonical manner from the Levi-Cevita
connection $\widetilde{\Gamma }_{\mu \nu \lambda },$ also called Christoffel
symbol:

\bigskip

$\ 
\begin{array}{cc}
\begin{array}{c}
\\ 
\widetilde{R}_{\mu \nu \lambda }^{\text{ \ \ \ }\kappa }=\partial _{\mu }%
\widetilde{\Gamma }_{\nu \lambda }^{\text{ \ }\kappa }-\partial _{\nu }%
\widetilde{\Gamma }_{\mu \lambda }^{\text{ \ }\kappa }-\left[ \widetilde{%
\Gamma }_{\mu },\widetilde{\Gamma }_{\nu }\right] _{\lambda }^{\text{ \ }%
\kappa }, \\ 
\\ 
\widetilde{\Gamma }_{\mu \nu \lambda }=\dfrac{1}{2}\left( \partial _{\mu
}g_{\nu \lambda }+\partial _{\nu }g_{\mu \lambda }-\partial _{\lambda
}g_{\mu \nu }\right) , \\ 
\\ 
\left[ \widetilde{\Gamma }_{\mu },\widetilde{\Gamma }_{\nu }\right]
_{\lambda }^{\text{ \ }\kappa }=\widetilde{\Gamma }_{\mu \lambda }^{\text{ \ 
}\sigma }\widetilde{\Gamma }_{\nu \sigma }^{\text{ \ }\kappa }-\widetilde{%
\Gamma }_{\nu \lambda }^{\text{ \ }\sigma }\widetilde{\Gamma }_{\mu \sigma
}^{\text{ \ }\kappa }, \\ 
\end{array}
& \text{ \ }(6.2.7)%
\end{array}%
$

\bigskip

where

\bigskip $%
\begin{array}{cc}
\begin{array}{c}
\\ 
g_{\mu \nu }=e_{\mu }^{i}e_{\nu }^{i} \\ 
\end{array}
& \text{ \ }(6.2.8)%
\end{array}%
$

is the Riemann metric in the space of anholonomic coordinates. The relation
between the two curvature tensors is

\bigskip

\bigskip $%
\begin{array}{cc}
\begin{array}{c}
\\ 
R_{\mu \nu \lambda }^{\text{ \ \ \ \ }\kappa }-\widetilde{R}_{\mu \nu
\lambda }^{\text{ \ \ \ \ }\kappa }=D_{\mu }K_{\nu \lambda }^{\text{ \ }%
\kappa }-D_{\nu }K_{\mu \lambda }^{\text{ \ }\kappa }-\left[ K_{\mu },K_{\nu
}\right] _{\lambda }^{\text{ \ }\kappa }, \\ 
\\ 
\left[ K_{\mu },K_{\nu }\right] _{\lambda }^{\text{ \ }\kappa }=K_{\mu
\lambda }^{\text{ \ }\sigma }K_{\nu \sigma }^{\text{ \ }\kappa }-K_{\nu
\lambda }^{\text{ \ }\sigma }K_{\mu \sigma }^{\text{ \ }\kappa }, \\ 
\\ 
K_{\mu \nu }^{\lambda }=\Gamma _{\mu \nu }^{\lambda }-\widetilde{\Gamma }%
_{\mu \nu }^{\lambda }. \\ 
\end{array}
& \text{ \ }(6.2.9)%
\end{array}%
$

\bigskip

Where the symbols $D_{\mu }$ denote the covariant derivatives formed with
the Christoffel symbol. From either of the two curvature tensors,i.e. Cartan
curvature tensor $R_{\mu \nu \lambda }^{\text{ \ \ \ \ }\kappa }$ and
Riemann curvature tensor $\widetilde{R}_{\mu \nu \lambda }^{\text{ \ \ \ \ }%
\kappa }$ one can form the oncecontracted tensors of rank 2, the Ricci
tensors and the curvature scalars:\bigskip

\bigskip $%
\begin{array}{cc}
\begin{array}{c}
\\ 
R_{\mu \nu }=R_{\mu \nu \lambda }^{\text{ \ \ \ \ }\mu };R=g^{\nu \lambda
}R_{\nu \lambda }, \\ 
\\ 
\widetilde{R}_{\mu \nu }=\widetilde{R}_{\mu \nu \lambda }^{\text{ \ \ \ \ }%
\mu };\widetilde{R}=g^{\nu \lambda }\widetilde{R}_{\nu \lambda }. \\ 
\end{array}
& \text{ \ }(6.2.10)%
\end{array}%
$

It is possible to map a flat $x$-space locally into a curved $y$-space with $%
\widetilde{R}_{\mu \nu \lambda }^{\text{ \ \ \ }\kappa }\neq 0$ via an
infinitesimal anholonomic transformation

\bigskip $%
\begin{array}{cc}
\begin{array}{c}
\\ 
dx^{i}=e_{\mu }^{i}\left( y\right) dy^{\mu } \\ 
\end{array}
& \text{ \ }(6.2.11)%
\end{array}%
$

\bigskip with coefficient functions $e_{\mu }^{i}\left( y\right) $ which are
not integrable,i.e.

\bigskip $%
\begin{array}{cc}
\begin{array}{c}
\\ 
\partial _{\mu }e_{\nu }^{i}\left( y\right) -\partial _{\nu }e_{\mu
}^{i}\left( y\right) \neq 0. \\ 
\end{array}
& \text{ \ }(6.2.12)%
\end{array}%
$

\bigskip

\bigskip \bigskip

\subsection{\textbf{VI.3.}Gravity type gauge theories \textbf{of accelerated
comovin frame formed by elastic media }with defects.}

\bigskip

\bigskip

Let's consider infinite three dimensional elastic media with defects.A
theory of relativistic elastic media with defects based on gravity type
gauge theories in three dimensional case (two space plus one time) is
considered by Katanaev and Volovich [41].They introduce a metric affine
space with a metric constructed from distortion $e_{\mu }^{i}$ and a $%
\mathbf{SO}(3)$ - connection $\omega _{\mu }^{ij}$. Here the index $\mu $ is
a general curvilinear coordinate label of the material manifold and $i$
labels the coordinate $X^{i}$ of the current configuration manifold. Using
simple and physically reasonable assumptions they define a two-parameter
static Lagrangian which is the sum of the Hilbert-Einstein Lagrangian for
the distorsion and the square of the antisymmetric part of the Ricci tensor
[42]:

\bigskip $%
\begin{array}{cc}
\begin{array}{c}
\\ 
\dfrac{1}{e}\tciLaplace =-\kappa \widetilde{R}+2\gamma R_{ij}^{A}R^{Aij}, \\ 
\\ 
e=\det \left( e_{\mu }^{i}\right) ,%
\end{array}
& \text{ }(6.3.1)%
\end{array}%
$

which is the sum of the Hilbert--Einstein Lagrangian for the vielbein and
the square of the antisymmetric part of the Ricci tensor. The vielbein $%
e_{\mu }^{i}$ and $\mathbf{SO}(3)$ connection

$\omega _{\mu }^{ij}$ are basic and independent variables in the geometric
approach.

\textbf{Remark.6.3.1.}Note that $\widetilde{R}=\widetilde{R}\left( e\right) $
and $R=R\left( e,\omega \right) $ are constructed from differentmetrical
connections and \bigskip the identity (6.3.2) is valid in the
Riemann--Cartan geometry in an arbitrary number of dimensions:

\bigskip

$%
\begin{array}{cc}
\begin{array}{c}
\\ 
\widetilde{R}\left( e\right) =R\left( e,\omega \right) +\dfrac{1}{4}%
T_{ijk}T^{ijk}- \\ 
\\ 
-\dfrac{1}{2}T_{ijk}T^{kij}-T_{i}T^{i}-\dfrac{2}{e}\partial _{\mu }\left(
eT^{\mu }\right) , \\ 
\\ 
e=\det \left( e_{\mu }^{i}\right) . \\ 
\end{array}
& \text{ }(6.3.2)%
\end{array}%
$

\bigskip

One assume that equations of equilibrium must be covariant under general
coordinate transformations and local rotations, be at most of the second
order, and follow from a variational principle. The expression for the free
energy leading to the equilibrium equations must then be equal to a volume
integral of the scalar function (the Lagrangian) that is quadratic in
torsion and curvature tensors. There are three independent invariants
quadratic in the torsion tensor and three independent invariants quadratic
in the curvature tensor in three dimensions [41]. It is possible to add the
scalar curvature and a \textquotedblleft cosmological\textquotedblright\
constant $\Lambda $. One thus obtain

a general eight-parameter Lagrangian [41]:

\bigskip

\bigskip $%
\begin{array}{cc}
\begin{array}{c}
\\ 
\dfrac{1}{\left( e_{\epsilon }\right) _{\epsilon }}\left( \tciLaplace
_{\epsilon }\right) _{\epsilon }=-\kappa \left( R_{\epsilon }\right)
_{\epsilon }+ \\ 
\\ 
\dfrac{1}{4}\left( T_{\epsilon ,ijk}\right) _{\epsilon }\cdot \left( \beta
_{1}\left( T_{\epsilon }^{ijk}\right) _{\epsilon }+\beta _{2}\left(
T_{\epsilon }^{kij}\right) _{\epsilon }+\beta _{3}\left( T_{\epsilon
}^{j}\right) _{\epsilon }\cdot \delta ^{ik}\right) + \\ 
\\ 
+\dfrac{1}{4}\left( R_{\epsilon ,ijkl}\right) _{\epsilon }\left( \gamma
_{1}\left( R_{\epsilon }^{ijkl}\right) _{\epsilon }+\gamma _{2}\left(
R_{\epsilon }^{klij}\right) _{\epsilon }+\gamma _{3}\left( R_{\epsilon
}^{ik}\right) _{\epsilon }\cdot \delta ^{jl}\right) -\Lambda , \\ 
\\ 
\left( e_{\epsilon }\right) _{\epsilon }=\det \left( e_{\epsilon ,\mu
}^{i}\right) _{\epsilon },%
\end{array}
& \text{ \ }(6.3.3)%
\end{array}%
$\ \ \ \ \ \ \ \ \ \ \ \ \ \ \ \ \ \ \ \ \ \ \ \ \ \ \ \ \ \ \ \ \ \ \ \ \ \
\ \ \ \ 

\bigskip

where $\kappa ,\beta _{1,2,3}$ and $\gamma _{1,2,3}$ are some constants, and
we have introduced the trace of the generalized torsion tensor $\left(
T_{\epsilon ,j}\right) _{\epsilon }=\left( T_{\epsilon ,ij}^{i}\right)
_{\epsilon }$ and the generalized Ricci tensor $\left( R_{\epsilon
,ik}\right) _{\epsilon }=\left( R_{\epsilon ,ijk}^{\text{ \ \ }j}\right)
_{\epsilon }.$

The particular feature of three dimensions is that the full curvature tensor
is in a one to one correspondence with Ricci tensor $\left( R_{\epsilon
,ijkl}\right) _{\epsilon }$ and has three irreducible components.

Therefore, the Lagrangian contains only three independent invariants
quadratic in curvature tensor. We do not need to add the Hilbert--Einstein
Lagrangian $\widetilde{R}$, also yielding second-order equations, to the
free energy given by Eq.(6.3.3)

The Lagrangian (6.3.3) gives equations of the accelerated relativistic media
[42]:

\bigskip

\bigskip $%
\begin{array}{cc}
\begin{array}{c}
\\ 
\dfrac{1}{\left( e_{\epsilon }\right) _{\epsilon }}\dfrac{\delta \left(
\tciLaplace _{\epsilon }\right) _{\epsilon }}{\delta \left( e_{\epsilon ,\mu
}^{i}\right) _{\epsilon }}=-\kappa \left[ \left( R_{\epsilon }\right)
_{\epsilon }\cdot \left( e_{\epsilon ,i}^{\mu }\right) _{\epsilon }-2\left(
R_{\epsilon ,i}^{\mu }\right) _{\epsilon }\right] + \\ 
\\ 
\beta _{1}\left[ \nabla _{\nu }\left( T_{\epsilon ,i}^{\nu \mu }\right)
_{\epsilon }-\dfrac{1}{4}\left( T_{\epsilon ,jkl}\right) _{\epsilon }\cdot
\left( T_{\epsilon }^{jkl}\right) _{\epsilon }\cdot \left( e_{\epsilon
,i}^{\mu }\right) _{\epsilon }+\left( T_{\epsilon }^{\mu jk}\right)
_{\epsilon }\cdot \left( T_{\epsilon ,ijk}\right) _{\epsilon }\right] + \\ 
\\ 
\beta _{2}\left[ -\dfrac{1}{2}\nabla _{\nu }\left( \left( T_{\epsilon
,i}^{\mu \nu }\right) _{\epsilon }-\left( T_{\epsilon ,i}^{\nu \mu }\right)
_{\epsilon }\right) -\dfrac{1}{4}\left( T_{\epsilon ,jkl}\right) _{\epsilon
}\cdot \left( T_{\epsilon }^{ljk}\right) _{\epsilon }\cdot \left(
e_{\epsilon ,i}^{\mu }\right) _{\epsilon }\right. \\ 
\\ 
\left. -\dfrac{1}{2}\left( T_{\epsilon }^{j\mu k}\right) _{\epsilon }\cdot
\left( T_{\epsilon ,kij}\right) _{\epsilon }+\dfrac{1}{2}\left( T_{\epsilon
}^{jk\mu }\right) _{\epsilon }\cdot \left( T_{\epsilon ,kij}\right)
_{\epsilon }\right] + \\ 
\\ 
\beta _{3}\left[ -\dfrac{1}{2}\nabla _{\nu }\left( \left( T_{\epsilon }^{\nu
}\right) _{\epsilon }\cdot \left( e_{\epsilon ,i}^{\mu }\right) _{\epsilon
}-\left( T_{\epsilon }^{\mu }\right) _{\epsilon }\cdot \left( e_{\epsilon
,i}^{\nu }\right) _{\epsilon }\right) \right. \\ 
\\ 
\left. -\dfrac{1}{4}\left( T_{\epsilon ,j}\right) _{\epsilon }\cdot \left(
T_{\epsilon }^{j}\right) _{\epsilon }\cdot \left( e_{\epsilon ,i}^{\mu
}\right) _{\epsilon }+\dfrac{1}{2}\left( T_{\epsilon }^{\mu }\right)
_{\epsilon }\cdot \left( T_{\epsilon ,i}\right) _{\epsilon }+\dfrac{1}{2}%
\left( T_{\epsilon }^{j}\right) _{\epsilon }\cdot \left( T_{\epsilon
,ij}^{\mu }\right) _{\epsilon }\right] + \\ 
\\ 
\gamma _{1}\left[ -\dfrac{1}{4}\left( R_{\epsilon ,jklm}\right) _{\epsilon
}\cdot \left( R_{\epsilon }^{jklm}\right) _{\epsilon }\cdot \left(
e_{\epsilon ,i}^{\mu }\right) _{\epsilon }+\left( R_{\epsilon }^{\mu
jkl}\right) _{\epsilon }\cdot \left( R_{\epsilon ,ijkl}\right) _{\epsilon }%
\right] + \\ 
\\ 
\gamma _{2}\left[ -\dfrac{1}{4}\left( R_{\epsilon ,jklm}\right) _{\epsilon
}\cdot \left( R_{\epsilon }^{lmjk}\right) _{\epsilon }\cdot \left(
e_{\epsilon ,i}^{\mu }\right) _{\epsilon }+\left( R_{\epsilon
}^{klmj}\right) _{\epsilon }\cdot \left( R_{\epsilon ,ijkl}\right)
_{\epsilon }\right] + \\ 
\\ 
\gamma _{3}\left[ -\dfrac{1}{4}\left( R_{\epsilon ,jk}\right) _{\epsilon
}\cdot \left( R_{\epsilon }^{jk}\right) _{\epsilon }\cdot \left( e_{\epsilon
,i}^{\mu }\right) _{\epsilon }+\dfrac{1}{2}\left( R_{\epsilon }^{\mu
j}\right) _{\epsilon }\cdot \left( R_{\epsilon ,ij}\right) _{\epsilon
}\right. + \\ 
\\ 
\left. +\dfrac{1}{2}\left( R_{\epsilon }^{jk}\right) _{\epsilon }\cdot
\left( R_{\epsilon ,jik}^{\mu }\right) _{\epsilon }\right] + \\ 
\\ 
+\Lambda \left( e_{\epsilon ,i}^{\mu }\right) _{\epsilon }=\left( \mathbf{T}%
_{\epsilon ,ij}\right) _{\epsilon }, \\ 
\end{array}
& \text{ \ }(6.3.4)%
\end{array}%
$

\bigskip where $\mathbf{T}_{ij}$

\bigskip\ \ \ \ \ \ \ \ \ \ \ \ \ \ \ \ \ \ $\ \ 
\begin{array}{cc}
\begin{array}{c}
\\ 
\dfrac{1}{\left( e_{\epsilon }\right) _{\epsilon }}\dfrac{\delta \left(
\tciLaplace _{\epsilon }\right) _{\epsilon }}{\delta \left( \omega
_{\epsilon ,\mu }^{ij}\right) _{\epsilon }}= \\ 
\\ 
\kappa \left[ \dfrac{1}{2}\left( T_{\epsilon ,ij}^{\text{ \ }\mu }\right)
_{\epsilon }+\left( T_{\epsilon ,i}\right) _{\epsilon }\cdot \left(
e_{_{\epsilon ,}\text{ \ \ }j}^{\mu }\right) _{\epsilon }\right] +\beta _{1}%
\dfrac{1}{2}\left( T_{\text{ \ }\epsilon ,ji}^{\mu }\right) _{\epsilon } \\ 
\\ 
+\beta _{2}\dfrac{1}{4}\left[ \left( T_{\epsilon ,i\text{ \ }j}^{\mu
}\right) _{\epsilon }-\left( T_{\epsilon ,ij}^{\text{ \ }\mu }\right)
_{\epsilon }\right] + \\ 
\\ 
\beta _{3}\dfrac{1}{4}\left( T_{\epsilon ,j}\right) _{\epsilon }\cdot \left(
e_{_{\epsilon ,}\text{ \ \ }i}^{\mu }\right) _{\epsilon }+\gamma _{1}\dfrac{1%
}{2}\nabla _{\nu }\left( R_{_{\epsilon },\text{ \ \ \ }ij}^{\nu \mu }\right)
_{\epsilon }+\gamma _{2}\dfrac{1}{2}\nabla _{\nu }\left( R_{\epsilon ,ij}^{%
\text{ \ \ }\nu \mu }\right) _{\epsilon }+ \\ 
\\ 
\gamma _{3}\dfrac{1}{4}\nabla _{\nu }\left[ \left( R_{_{\epsilon ,}\text{ \
\ }i}^{\nu }\right) _{\epsilon }\cdot \left( e_{_{\epsilon ,}\text{ \ \ }%
j}^{\mu }\right) _{\epsilon }-\left( R_{_{\epsilon ,}\text{ \ \ }i}^{\mu
}\right) _{\epsilon }\cdot \left( e_{_{\epsilon ,}\text{ \ \ }j}^{\nu
}\right) _{\epsilon }\right] -\left( i\leftrightarrow j\right) =0, \\ 
\end{array}
& \text{ \ \ }(6.3.5)%
\end{array}%
$

\bigskip

where the covariant derivative acts with the $\mathbf{SO}(3)$ connection on
the Latin indices and with the Christoffel symbols on the Greek ones.

\bigskip

\textbf{Remark.6.3.2.}The most general Lagrangian () depends on eight
constants: $\kappa ,\beta _{1},$ $\beta _{2},\beta _{3},\gamma _{1},\gamma
_{2},\gamma _{3},\Lambda $ and leads to very complicated equations of
accelerated relativistic media. Note that \bigskip these constants is not
apriori fixed and we do not know precisely what values of the constants
describe this or that accelerated relativistic media and corresponding
comovin frame.

Therefore, we make physically reasonable assumptions to simplify matters at
least for the case $R_{\mu \nu }^{\text{ \ \ \ }ij}=0.$

\textbf{Remark.6.3.3.}Note that curvature $R_{\mu \nu }^{\text{ \ \ \ }ij}$
of the comovin frame of the any accelerated relativistic media in contrast
with [42] satisfies the inequality $R_{\mu \nu }^{\text{ \ \ \ }ij}\neq 0.$

\bigskip

Thus, one obtain that equations of the accelerated relativistic media must

admit the following three types of solutions:

\bigskip

\begin{itemize}
\item \textbf{1.}There are solutions describing the relativistic accelerated
media with only "dislocations":\bigskip\ 

$\ \ \ \ \ \ \ \ \ \ \ \ \ \ \ \ \ \ \ $
\end{itemize}

\bigskip\ $%
\begin{array}{cc}
\begin{array}{c}
\\ 
\ \left( R_{\epsilon ,\mu \nu }^{\text{ \ \ \ }ij}\right) _{\epsilon }=0, \\ 
\\ 
\left( T_{\epsilon ,\mu \nu }^{\text{ \ \ \ }i}\right) _{\epsilon }\neq 0,
\\ 
\\ 
\left( \widetilde{R}_{\epsilon ,\mu \nu }^{\text{ }}\right) _{\epsilon }\neq
0. \\ 
\end{array}
& \text{ \ }(6.3.6)%
\end{array}%
$

\begin{itemize}
\item \bigskip \textbf{2.}There are solutions describing the relativistic
accelerated media with only "disclinations":
\end{itemize}

\bigskip\ $%
\begin{array}{cc}
\begin{array}{c}
\\ 
\left( \ R_{\mu \nu }^{\text{ \ \ \ }ij}\right) _{\epsilon }\neq 0, \\ 
\\ 
\left( T_{\mu \nu }^{\text{ \ \ \ }i}\right) _{\epsilon }=0, \\ 
\\ 
\left( \widetilde{R}_{\mu \nu }^{\text{ }}\right) _{\epsilon }\neq 0. \\ 
\end{array}
& \text{ \ }(6.3.7)%
\end{array}%
$

\begin{itemize}
\item 3.There are solutions describing the relativistic accelerated media
almost without \ "dislocations" and "disclinations":
\end{itemize}

\bigskip $%
\begin{array}{cc}
\begin{array}{c}
\\ 
\left\Vert R_{\mu \nu }^{\text{ \ \ \ }ij}\right\Vert \simeq 0, \\ 
\\ 
\left\Vert T_{\mu \nu }^{\text{ \ \ \ }i}\right\Vert \simeq 0, \\ 
\\ 
\widetilde{R}_{\mu \nu }^{\text{ }}\neq 0. \\ 
\end{array}
& \text{ \ }(6.3.8)%
\end{array}%
$

\bigskip

\bigskip Substitution of the condition $R_{\mu \nu }^{\text{ \ \ \ }ij}=0$
into Eq.(6.3.4 ) for the $\mathbf{SO}(3)$ connection gives

$%
\begin{array}{cc}
\begin{array}{c}
\\ 
(12\kappa +2\beta _{1}-\beta _{2}-2\beta _{3})T_{i}=0, \\ 
\\ 
(\kappa -\beta _{1}-\beta _{2})T^{\ast }=0, \\ 
\\ 
(4\kappa +2\beta _{1}-\beta _{2})W_{ijk}=0. \\ 
\end{array}
& \text{ \ }(6.3.9)%
\end{array}%
$

Here $T_{i},T^{\ast },$ and $W_{ijk}$ are the irreducible components of the
torsion tensor.In a general case of dislocations all irreducible components
of torsion tensor differ from

zero ($T_{i},T^{\ast },W_{ijk}\neq 0$) and Eqs. (6.3.9) have a unique
solution

\bigskip

\bigskip $%
\begin{array}{cc}
\begin{array}{c}
\\ 
\beta _{1}=\kappa ,\beta _{2}=2\kappa ,\beta _{3}=4\kappa . \\ 
\end{array}
& \text{ \ }(6.3.10)%
\end{array}%
$

For these coupling constants, the first four terms in Lagrangian (6.3.3) are
equal to the Hilbert--Einstein Lagrangian $\kappa \widetilde{R}\left(
e\right) $ up to a total divergence due to identity(6.3.2). \ Equation
(6.3.2) then reduces to the Einstein equations with a cosmological constant

\bigskip $%
\begin{array}{cc}
\begin{array}{c}
\\ 
\widetilde{R}_{\mu \nu }-\dfrac{1}{2}g_{\mu \nu }\widetilde{R}-\dfrac{%
\Lambda }{2\kappa }g_{\mu \nu }=0 \\ 
\end{array}
& \text{ \ }(6.3.11)%
\end{array}%
$

\bigskip

According to the second condition,the equations of accelerated equilibrium
must allow solutions with zero torsion $T_{\mu \nu }^{\text{ \ \ \ }i}=0.$In
this case, the curvature tensor has additional symmetry $R_{ijkl}=R_{klij},$
and Eq.(6.3.5 ) becomes

\bigskip

\bigskip $%
\begin{array}{cc}
\begin{array}{c}
\\ 
\left( \gamma _{1}+\gamma _{2}+\dfrac{\gamma _{3}}{4}\right) \nabla _{\nu
}\left( R_{\text{ \ \ \ \ }i}^{\mathbf{S}\text{ }\nu }e_{\text{ \ \ }j}^{\mu
}-R_{\text{ \ \ \ \ }i}^{\mathbf{S}\text{ }\mu }e_{\text{ \ \ }j}^{\nu }+R_{%
\text{ \ \ \ \ }j}^{\mathbf{S}\text{ }\mu }e_{\text{ \ \ }i}^{\nu }\right) +
\\ 
\\ 
+\dfrac{1}{6}\left( \gamma _{1}+\gamma _{2}+4\gamma _{3}\right) \left( e_{%
\text{ \ \ }i}^{\nu }e_{\text{ \ \ }j}^{\mu }-e_{\text{ \ \ }i}^{\mu }e_{%
\text{ \ \ }j}^{\nu }\right) \nabla _{\nu }R\simeq 0, \\ 
\end{array}
& \text{ \ }(6.3.12)%
\end{array}%
$

\bigskip where

$%
\begin{array}{cc}
\begin{array}{c}
\\ 
R_{ij}=R_{ij}^{\mathbf{S}}+R_{ij}^{\mathbf{A}}+\dfrac{1}{3}R\delta _{ij}, \\ 
\\ 
R_{ij}^{\mathbf{S}}=R_{ji}^{\mathbf{S}},R_{\text{ \ \ }i}^{\mathbf{S}%
}=0,R_{ij}^{\mathbf{A}}=-R_{ji}^{\mathbf{A}}. \\ 
\end{array}
& \text{ \ }(6.3.13)%
\end{array}%
$

\bigskip

Note that for almost zero torsion, the Ricci tensor is symmetrical,i.e.$%
R_{ij}^{\mathbf{A}}=0.$

Contraction of Eq.(6.3.12) with $e_{\mu }^{j}$ gives

\bigskip

\bigskip $%
\begin{array}{cc}
\begin{array}{c}
\\ 
\left( \gamma _{1}+\gamma _{2}+\dfrac{\gamma _{3}}{4}\right) \nabla _{\nu
}R_{\text{ \ \ \ \ }\mu }^{\mathbf{S}\text{ }\nu }+ \\ 
\\ 
\dfrac{1}{3}\left( \gamma _{1}+\gamma _{2}+4\gamma _{3}\right) \nabla _{\mu
}R=0. \\ 
\end{array}
& \text{ \ }(6.3.14)%
\end{array}%
$

Note that in the general case of nonvanishing curvature, the covariant
derivatives $\nabla _{\nu }R_{\text{ \ \ \ \ }\mu }^{\mathbf{S}\text{ }\nu }$
and $\nabla _{\mu }R$ differ from zero and are independent. Therefore, one
obtain two equations for the coupling constants,

\bigskip $%
\begin{array}{cc}
\begin{array}{c}
\\ 
\gamma _{1}+\gamma _{2}+\dfrac{\gamma _{3}}{4}=0, \\ 
\\ 
\gamma _{1}+\gamma _{2}+4\gamma _{3}=0, \\ 
\end{array}
& \text{ \ }(6.3.15)%
\end{array}%
$

\bigskip which have a unique solution

$%
\begin{array}{cc}
\begin{array}{c}
\\ 
\gamma _{1}=-\gamma _{2}=\gamma ,\gamma _{3}=0. \\ 
\end{array}
& \text{ \ }(6.3.16)%
\end{array}%
$

The last requirement for the noexistence of solutions with zero curvature
and torsion is satisfied only for the non zero cosmological constant

\bigskip

\bigskip $%
\begin{array}{cc}
\begin{array}{c}
\\ 
\dfrac{1}{e}\tciLaplace =-\kappa \widetilde{R}+2\gamma
R_{ij}^{A}R^{Aij}+\Lambda , \\ 
\\ 
\Lambda \neq 0.%
\end{array}
& \text{ \ \ \ \ }(6.3.17)%
\end{array}%
$

\bigskip $\ \ \ \ \ \ \ \ \ \ \ \ \ \ \ \ \ \ \ \ \ \ \ \ \ \ \ \ \ \ $

\bigskip $\ \ \ \ \ \ \ \ \ \ \ \ \ \ \ \ $

$\ \ \ \ \ \ \ \ \ \ \ \ \ \ \ \ \ \ \ \ \ \ \ \ \ \ \ \ \ \ \ \ \ \ $

\bigskip $\ \ \ \ \ \ \ \ \ \ \ \ \ \ \ \ \ \ \ \ \ \ \ \ \ \ \ \ \ \ \ \ \
\ \ \ $

\bigskip

\subsection{VII.Bimetrical interpretation some exact solutions of the
Einstein field equations.}

\bigskip

\bigskip

\bigskip

\bigskip

\subsection{VII.1. General consideration.}

\bigskip

Let us consider that in the Minkovsky space with the signature the continuum
medium moves in some force field, the motion law of this continuum in the
Lagrange variables has the form [70]:

\bigskip 

$%
\begin{array}{cc}
\begin{array}{c}
\\ 
x^{\mu }=x^{\mu }(y^{k},\xi ^{0}), \\ 
\end{array}
& \text{ \ }(7.1.1)%
\end{array}%
$

\bigskip

where $x^{\mu }$ are the Euler coordinates, $y^{k}$ are the Lagrange
coordinates constant along each fixed world line of the medium particle, is
the some time parameter. The greek indexes are changed from zero to three,
the latin indexes are changed from unit to three. We consider that the
medium particles do not interact with each other and they interact only with
the external field.

\qquad Similarly to electrodynamics the actions for the probe particle in
the force field we specify in the form

\bigskip 

\bigskip\ $\ 
\begin{array}{cc}
\begin{array}{c}
\\ 
S=-mc\int_{a}^{b}(ds+\alpha {A_{\mu }}dx^{\mu }), \\ 
\\ 
\qquad \alpha =\dfrac{e}{mc^{2}}, \\ 
\end{array}
& \text{ \ }(7.1.2)%
\end{array}%
$

\bigskip

where for each medium particle the ds interval along the world lines is $%
ds=V_{\mu }dx{^{\mu }},$ $V^{\mu }$ is the four dimensional velocity.From
the action variation the motion equation follows

\bigskip

\bigskip $%
\begin{array}{cc}
\begin{array}{c}
\\ 
\dfrac{DV_{\mu }}{ds}=\alpha {F_{\mu \nu }}V^{\nu }, \\ 
\end{array}
& \text{ \ }(7.1.3)%
\end{array}%
$

\bigskip where the field tensor ${F_{\mu \nu }}$ is determined as

\bigskip $%
\begin{array}{cc}
\begin{array}{c}
\\ 
F_{\mu \nu }=\nabla _{\mu }A{_{\nu }}-\nabla _{\nu }A{_{\mu }}=\dfrac{%
\partial A{_{\nu }}}{\partial x_{\mu }}-\dfrac{\partial A{_{\mu }}}{\partial
x_{\nu }} \\ 
\end{array}
& \text{ \ }(7.1.4)%
\end{array}%
$

On the other hand one can to introduce the effective interval$\ d\widetilde{s%
}=ds+\alpha A{_{\mu }}dx^{\mu }\ \ \ \ \ \ \ \ \ \ \ \ \ \ \ \ \ \ \ \ \ \ \
\ \ \ \ \ \ \ \ \ \ \ \ \ \ \ $

so (7.1.2) is represented in the form

$%
\begin{array}{cc}
\begin{array}{c}
\\ 
S=-mc\int d\text{ }\widetilde{s}, \\ 
\end{array}
& \text{ \ }(7.1.5)%
\end{array}%
$

variation of this action results in the motion of the probe particle on the
geodesic line in some pseudo Riemannian space $\Re \ \ \ \ \ \ \ $

$\bigskip $

$%
\begin{array}{cc}
\begin{array}{c}
\\ 
\dfrac{dU_{\mu }}{d\widetilde{s}}+\tilde{\Gamma}_{\mu ,\nu \epsilon }U^{\nu
}U^{\epsilon }=0. \\ 
\end{array}
& \text{ \ }(7.1.6)%
\end{array}%
$

\bigskip

\textbf{Claim.7.1.1.} Suppose that the dynamic equations (7.1.3) and (7.1.6)
is equivalent,i.e. $\mathbf{(7.1.3)}\iff \mathbf{(7.1.6).}$

It follows from the expression for the effective interval $d\widetilde{s}$
along the geodesic line that

\bigskip\ $\ \ \ \ \ \ \ \ \ \ \ \ \ \ \ \ \ \ \ \ \ \ \ \ \ \ \ \ \ \ \ \ \
\ \ \ \ \ \ \ \ \ \ \ \ \ \ \ \ \ \ $

$%
\begin{array}{cc}
\begin{array}{c}
\\ 
d\widetilde{s}=(V_{\mu }+\alpha A{_{\mu }})dx^{\mu }\equiv U{_{\mu }}dx^{\mu
},\qquad U_{\mu }\equiv V{_{\mu }}+\alpha A{_{\mu }}, \\ 
\\ 
U^{\mu }=\dfrac{dx^{\mu }}{d\widetilde{s}}=\dfrac{dx^{\mu }}{ds}\dfrac{ds}{d%
\widetilde{s}}=PV{^{\mu }},\qquad P\equiv (1+{\alpha }A_{\epsilon }V{%
^{\epsilon }})^{-1}. \\ 
\end{array}
& \text{ \ }(7.1.7)%
\end{array}%
$

\bigskip

Besides the connection between covariant $U_{\nu }$ and the contravariant $%
U^{\mu }$ vectors of the 4-velocity in the pseudo Riemannian space has the
form

\bigskip

$\bigskip 
\begin{array}{cc}
\begin{array}{c}
\\ 
U_{\nu }=g_{\nu \mu }U^{\mu }=V_{\nu }+\alpha A{_{\nu }}. \\ 
\end{array}
& \text{ \ }(7.1.8)%
\end{array}%
$

\bigskip

Conditions (7.1.3), (7.1.4), (7.1.6), (7.1.7) and (7.1.8) will be is self
consistent if and only if the metric tensor $g_{\nu \mu }$ of the pseudo
Riemannian space $\Re $ will have the form:

\bigskip

$\bigskip 
\begin{array}{cc}
\begin{array}{c}
\\ 
g_{\nu \mu }={\gamma }_{\mu \nu }+{\alpha ^{2}}A_{\mu }A{_{\nu }}+\alpha
A_{\mu }V{_{\nu }}+\alpha A_{\nu }V{_{\mu }}, \\ 
\end{array}
& \text{ \ }(7.1.9)%
\end{array}%
$

\bigskip

Thus, one can consider the motion of the probe particle as motion in Rosen
bimetric space $\Re _{2}=\Re \left( g_{\nu \mu },\gamma _{\mu \nu }\right) $
i.e. from two points of view:

\begin{itemize}
\item The motion on the world line in the Minkovsky space in the force field
(7.1.3) with the metrics $\gamma _{\mu \nu }$.

\item The motion in the Riemannian space on the geodesic line with the
metrics $g_{\nu \mu }$ determined in accordance with (7.1.9).
\end{itemize}

The correlations between the 4-velocities in the different spaces are
determined with the formulas (7.1.7) and (7.1.8). Herewith in the two spaces
the general coordination has been selected. Unlike electrodynamics the
tensor field $F_{\mu \nu }$structure in (7.1.4) has not been concreted, that
is for the tensor field $F_{\mu \nu }$ equations are not specified.

Let the probe particles move in the Einstein gravitational field. Then the
\textquotedblleft charge\textquotedblright\ $e=m$, and the metrics (7.1.9)
has to satisfy to the Einstein equations with the dusty stress energy tensor.

\bigskip

\bigskip $%
\begin{array}{cc}
\begin{array}{c}
\\ 
R_{\mu \nu }-\dfrac{1}{2}g_{\mu \nu }R=\dfrac{8\pi {\kappa }}{c^{4}}%
\varepsilon U_{\mu }U{_{\nu }}. \\ 
\end{array}
& \text{ \ }(7.1.10)%
\end{array}%
$

\bigskip

If as a result of the solution of the equations (7.1.10) obtained $g_{\mu
\nu }$ and $U_{\nu }$will provide the equalities (7.1.8) and (7.1.9), then
we can to find the field of 4-velocity $V_{\mu }$, the potentials $A_{\mu }$
and the field tensor $F_{\mu \nu }$in the Minkovsky space, that is the
mapping of the curvature field of the Riemannian space on the force field of
the plane space- time will be constructed.

Let us ascertain the connection between the congruencies of the world lines
in the Minkovsky space and the congruencies of the geodesic lines in the
Riemannian space which in the general coordination are determined with the
Eq.(7.1.1). Because of the Eq.(7.1.9) in the space-time two metric tensors $%
g_{\mu \nu }$ and $\gamma _{\mu \nu }$ have been introduced, and,
consequently, two connections $\tilde{\Gamma}^{\epsilon }{}_{\mu \nu }$ and $%
\Gamma ^{\epsilon }{}_{\mu \nu }$ exist, the first connection relates to the
pseudo Riemannian space $\Re $, and the second one relates to the Minkovsky
space $M_{4}$. In the Minkovsky space the curvature coordinates can be
introduced. Thus, in the general coordination two different covariant
derivatives $\tilde{\nabla}_{\nu }$ and $\nabla _{\nu }$ arise.\bigskip From
the Eq.(7.1.8) we have

\bigskip $%
\begin{array}{cc}
\begin{array}{c}
\\ 
\tilde{\nabla}_{\nu }U_{\mu }=-S^{\epsilon }{}_{\nu \mu }U_{\epsilon
}+\nabla _{\nu }V{_{\mu }}+\alpha {\nabla _{\nu }}A_{\mu }, \\ 
\\ 
S^{\epsilon }{}_{\nu \mu }=\tilde{\Gamma}^{\epsilon }{}_{\nu \mu }-\Gamma
^{\epsilon }{}_{\nu \mu }, \\ 
\end{array}
& \text{ \ }(7.1.11)%
\end{array}%
$

where $S^{\epsilon }{}_{\nu \mu }$ is the tensor of the affine connectivity
deformation.From (7.1.11) we find

\bigskip

$%
\begin{array}{cc}
\begin{array}{c}
\\ 
2\tilde{\nabla}_{[\nu }U_{\mu ]}=2\nabla _{\lbrack \nu }V{_{\mu ]}}-\alpha F{%
_{\mu \nu }}, \\ 
\end{array}
& \text{ \ }(7.1.12)%
\end{array}%
$

\bigskip

\bigskip For geodesic congruensies without rotations the equalities take
place

\bigskip

$%
\begin{array}{cc}
\begin{array}{c}
\\ 
\tilde{\nabla}_{[\nu }U_{\mu ]}=0;\qquad 2\nabla _{\lbrack \nu }V{_{\mu ]}}%
=\alpha F{_{\mu \nu }}. \\ 
\end{array}
& \text{ \ }(7.1.13)%
\end{array}%
$

\bigskip

Convoluting (7.1.13) with $V^{\nu }$ we once again obtain the equation
(7.1.13). From the equalities (7.1.13) and (7.1.7) we have

\bigskip

\bigskip $%
\begin{array}{cc}
\begin{array}{c}
\\ 
U_{\mu }=\dfrac{\partial \Phi }{\partial x^{\mu }}=V_{\mu }+\alpha A_{\mu }
\\ 
\end{array}
& \text{ \ }(7.1.14)%
\end{array}%
$

that permits the representation of the (7.1.9) metrics in the form

\bigskip

\bigskip $%
\begin{array}{cc}
\begin{array}{c}
\\ 
g_{\mu \nu }=\gamma _{\mu \nu }+\dfrac{\partial \Phi }{\partial x^{\mu }}%
\dfrac{\partial \Phi }{\partial x^{\nu }}-V_{\mu }V_{\nu } \\ 
\end{array}
& \text{ \ }(7.1.15)%
\end{array}%
$

\bigskip For the contravariant components we have

$%
\begin{array}{cc}
\begin{array}{c}
\\ 
g^{\mu \nu }=\gamma ^{\mu \nu }+{P^{2}}V{^{\mu }}V{^{\nu }}\left( 1+\gamma
^{\alpha \beta }\dfrac{\partial \Phi }{\partial x^{\alpha }}\dfrac{\partial
\Phi }{\partial x^{\beta }}\right) - \\ 
\\ 
\\ 
\end{array}
& \text{ \ }(7.1.16)%
\end{array}%
$

\bigskip

where in accordance with (7.1.7)

\bigskip

\bigskip $%
\begin{array}{cc}
\begin{array}{c}
\\ 
P=(1+{\alpha }{\gamma _{\epsilon \sigma }}A{^{\epsilon }}V{^{\sigma }})^{-1}=
\\ 
\end{array}
& \text{ \ }(7.1.17)%
\end{array}%
$

$\bigskip $It follows from the equalities (7.1.9), (7.1.14) and (7.1.15)

$%
\begin{array}{cc}
\begin{array}{c}
\\ 
g_{\mu \nu }-U_{\mu }U{_{\nu }}=\gamma _{\mu \nu }-V_{\mu }V_{\nu }, \\ 
\end{array}
& \text{ \ }(7.1.18)%
\end{array}%
$

\bigskip

i.e. is the projection operators determining the space geometry of the
hypersurfaces orthogonal to the world lines in the Minkovsky space and the
hupersurfaces orthogonal to the geodesic lines in the Riemannian space are
the invariants of the correspondence.$\ \ \ \ \ \ \ \ \ \ \ \ \ \ \ \ \ \ \
\ \ \ \ \ \ \ \ \ \ \ $

$\ \ \ \ \ \ \ \ \ \ \ \ \ \ \ \ $

$\ \ \ \ \ \ \ \ \ \ \ \ \ \ \ \ \ \ \ \ \ \ \ \ \ \ \ \ \ \ \ \ \ \ \ \ \ \
\ \ \ \ \ \ \ \ \ \ \ \ \ \ \ \ \ \ \ \ \ \ \ \ \ \ \ \ \ \ \ \ $

\bigskip

\subsection{VII.2. Bimetrical interpretation of the Shvartzshild solution.}

\bigskip

Let us consider some particular cases of the general mapping considered
above.Let in the Minkovsky space the dust continuum moves on the radius to
the centre. We consider the case of the stationary motion that means time
independence of the velocity field in the Euler variables and the potentials 
$A_{\mu }$. In the GIFT language this corresponds to the constant
gravitational-inertional field.

In order to the metric tensor (7.1.15) does not obviously depend from the
time and pass at the infinity to the Galilean form it is necessary that the
velocity at the infinity becomes zero. Thus the next equalities have to be
satisfied:

\bigskip

\bigskip $%
\begin{array}{cc}
\begin{array}{c}
\\ 
\Phi =x^{0}+\Psi (x^{k}), \\ 
\\ 
V_{a}=-V\left( r\right) n_{a}=-V\left( r\right) \dfrac{x_{a}}{r} \\ 
\end{array}
& \text{ \ }(7.1.19)%
\end{array}%
$

\bigskip

Using formulas (7.1.15) and (7.1.19) we find the expressions for
three-dimensional metric tensor: $\tilde{\gamma}%
_{kl}=-g_{kl}+g_{0k}g_{0l}/g_{00};$three-dimensional vector: $%
g_{l}=g_{0l}/g_{00}=g_{0l}/h;$

\bigskip three-dimensional antisymmetric tensor: $f_{kl}=\partial
g_{l}/\partial x^{k}-\partial g^{k}/\partial x_{l}.$As a result we have

\bigskip $%
\begin{array}{cc}
\begin{array}{c}
\\ 
g_{00}=1-V^{2},\text{ }g_{l}=n_{l}\dfrac{\frac{\partial \Phi }{\partial r}%
+V_{0}V}{h},\text{ }f_{kl}=0,\text{ }V_{0}^{2}-V^{2}=1, \\ 
\\ 
\tilde{\gamma}_{kl}=\delta _{kl}+D\left( r\right) n_{k}n_{l},\text{ } \\ 
\\ 
D=\dfrac{2V^{2}+2V_{0}V\frac{\partial \Phi }{\partial r}+V^{2}\left( \frac{%
\partial \Phi }{\partial r}\right) ^{2}}{1-V^{2}} \\ 
\\ 
\tilde{\gamma}^{kl}=-g^{kl}=\delta ^{kl}+Tn^{k}n^{l},\text{ }n^{k}=n_{k},%
\text{ }V^{0}=V_{0},\text{ }\tilde{\gamma}_{kl}\tilde{\gamma}^{ln}=\delta
_{n}^{k}, \\ 
\\ 
T=\dfrac{2V^{2}+2V_{0}V\frac{\partial \Phi }{\partial r}+V^{2}\left( \frac{%
\partial \Phi }{\partial r}\right) ^{2}}{\left( V^{0}+V\frac{\partial \Phi }{%
\partial r}\right) ^{2}}. \\ 
\end{array}
& \text{ \ }(7.1.20)%
\end{array}%
$

\bigskip

Einstein equations for the case of the constant gravitational field in the
vacuum (we consider that the dusty medium is strongly discharged and itself
does not create the field) will result to two independent expressions:

\bigskip

\bigskip $%
\begin{array}{cc}
\begin{array}{c}
\\ 
\dfrac{\partial }{\partial r}\left( \dfrac{r^{2}\frac{\partial F}{\partial r}%
}{\sqrt{1+D}}\right) =0,\text{ }F=\sqrt{h}=\sqrt{1-V^{2}}, \\ 
\\ 
D+\dfrac{r}{2}\dfrac{\partial D}{\partial r}\dfrac{1}{1+D}=\dfrac{r}{F}%
\dfrac{\partial F}{\partial r} \\ 
\end{array}
& \text{ \ }(7.1.21)%
\end{array}%
$

\bigskip the solution of which has the form

\bigskip $%
\begin{array}{cc}
\begin{array}{c}
\\ 
D=\dfrac{{r_{g}}/{r}}{1-{r_{g}}/{r}},F=\sqrt{g_{00}}=\sqrt{1-\dfrac{r_{g}}{r}%
},r_{g}\equiv \dfrac{2kM}{c^{2}}. \\ 
\end{array}
& \text{ \ }(7.1.22)%
\end{array}%
$

From correlations (7.1.20) and (7.1.22) we find zero and radial field
components of the 4-velocity in the Minkovsky space in the Euler variables
and also function $\Phi :$

\bigskip $%
\begin{array}{cc}
\begin{array}{c}
\\ 
V_{0}=V^{0}=\left( 1+\dfrac{r_{g}}{r}\right) ^{1/2},V^{1}=V=-\sqrt{\dfrac{%
r_{g}}{r}}, \\ 
\\ 
V_{0}+V\dfrac{\partial \Phi }{\partial r}=V^{\epsilon }\dfrac{\partial \Phi 
}{\partial x^{\epsilon }}=\dfrac{\partial \Phi }{\partial s}=1. \\ 
\end{array}
& \text{ \ }(7.1.23)%
\end{array}%
$

Thus,$\Phi /c=\tau =s/c$ coincides with the own time of the basis particles
in the Minkovsky and Riemann space.It follows from (7.1.23), (7.1.7) and
(7.1.14) that $(1+\alpha {A_{\mu }}V^{\mu })=P^{-1}=1,$this results in the
equality of the contravariant components of 4-velocities $U^{\mu }=V^{\mu }$
of the basis particles in the plane and curved space-time. Covariant
components $U_{\mu }$ and $V_{\mu }$ are connected with the equation
(7.1.14). Integrating equation (7.1.23) for $\Phi $ taking into account
(7.1.19) we find

\bigskip

\bigskip $%
\begin{array}{cc}
\begin{array}{c}
\\ 
\Phi =c\tau =s=x^{0}+\dfrac{2}{3}r_{g}\left( \dfrac{r}{r_{g}}+1\right)
^{3/2}-\dfrac{2}{3}\dfrac{r^{3/2}}{r_{g}{}^{1/2}}. \\ 
\end{array}
& \text{ \ }(7.1.24)%
\end{array}%
$

\bigskip

Using (7.1.20), (7.1.23) and (7.1.24) we obtain the expression for the
interval element of the \textquotedblleft original\textquotedblright\ in the
spherical Euler coordinates and time $T$ of the Minkovsky space

\bigskip

\bigskip $%
\begin{array}{cc}
\begin{array}{c}
\\ 
d\tilde{s}^{2}=c^{2}dT^{2}\left( 1-\dfrac{r_{g}}{r}\right) - \\ 
\\ 
dr^{2}\left\{ 2\left[ \dfrac{r}{r_{g}}\left( \dfrac{r}{r_{g}}+1\right)
^{1/2}-2\dfrac{r}{r_{g}}+\dfrac{r_{g}}{r}\right] \right\} + \\ 
\\ 
2cdTdr\left[ \left( 1+\dfrac{r}{r_{g}}\right) ^{1/2}-\left( \dfrac{r}{r_{g}}%
\right) ^{1/2}-\dfrac{r_{g}}{r}\left( 1+\dfrac{r}{r_{g}}\right) ^{1/2}\right]
- \\ 
\\ 
r^{2}(\sin ^{2}\Theta {d\varphi ^{2}}+d\Theta ^{2}). \\ 
\end{array}
& \text{ \ }(7.1.25)%
\end{array}%
$

\bigskip

Known the field of the 4-velocity in the Euler variables we find the motion
law of the continuum in the Lagrange variables (7.1.1) selecting as a time
parameter $\xi ^{0}$ the own time $\tau =\Phi /c=s/c$. From (7.1.23) we have 
$dr/ds=V=-(r_{g}/r)^{1/2}.$Integrating we obtain $%
R-s=2/3(r^{3/2}/r_{g}{}^{1/2})$,where $R$ is the constant of integration.
Taking into account (7.1.24) as a result we find

\bigskip

$%
\begin{array}{cc}
\begin{array}{c}
\\ 
r=\left[ \dfrac{3}{2}(R-c\tau )\right] ^{2/3}r_{g}{}^{1/3}, \\ 
\\ 
x^{0}=cT=R-\dfrac{2}{3}r_{g}\left\{ \left[ \dfrac{3}{2r_{g}}(R-c\tau )\right]
^{2/3}+1\right\} ^{3/2}, \\ 
\end{array}
& \text{ \ }(7.1.26)%
\end{array}%
$

\bigskip

that determines the sought motion law in the Lagrange variables,
substitution of this law to the expression (7.1.25) results in the Lemetr
interval element.Formulas (7.1.23), (7.1.26) determine the kinematics of the
dust medium moving with the acceleration on the radius to the centre in the
Minkovsky space in the gravitational field of the central body. For the
field of the three-dimensional velocity $v,$ 4-acceleration $g,$
three-dimensional acceleration $a$ and the three-dimensional force $N$ we
have:

\bigskip

$%
\begin{array}{cc}
\begin{array}{c}
\\ 
\dfrac{dr}{dT}=v=-c\left( 1+\dfrac{r}{r_{g}}\right) ^{-1/2},\dfrac{1}{c^{2}}%
\dfrac{d^{2}{r}}{d{\tau }^{2}}=g=-\dfrac{r_{g}}{2{r}^{2}}, \\ 
\\ 
a=\dfrac{d^{2}{r}}{dT^{2}}=-\dfrac{{c^{2}}r_{g}}{2r^{2}}\left( 1+\dfrac{r_{g}%
}{r}\right) ^{-2}, \\ 
\\ 
N=\dfrac{d}{dT}\left[ \dfrac{mv}{\left( 1-\dfrac{v^{2}}{c^{2}}\right) ^{1/2}}%
\right] =-\dfrac{m{r_{g}}c^{2}}{2r^{2}(1+r_{g}/r)^{1/2}}. \\ 
\end{array}
& \text{ \ }(7.1.27)%
\end{array}%
$

\bigskip

Movement of the Lemetr basis in the Minkovsky space is described with the
functions continuous in the region $0<r<\infty $ not having the
particularities at the gravitational radius. Three-dimensional velocity v
and three-dimensional acceleration a are restricted at the origin of the
coordinates,$v(0)=-c,a(0)=$ $-c^{2}/(2r_{g}).$ . The value of the
three-dimensional force $N$ (7.1.27) influencing on the probe mass from the
side of the central body is smaller then in the Newton gravitation theory

\bigskip

\bigskip $%
\begin{array}{cc}
\begin{array}{c}
\\ 
N=-\dfrac{kmM}{r^{2}\left( 1+\dfrac{2kM}{c^{2}{r}}\right) ^{1/2}}. \\ 
\end{array}
& \text{ \ }(7.1.28)%
\end{array}%
$

It is evident that the space components of the 4-velocity $cV^{1}$ (7.1.23)
and 4-acceleration $gc^{2}$ (7.1.27) exactly coincide with the usual
velocity and acceleration in the non-relativistic Newton mechanics, when the
radial fall of the dust having zero velocity at the infinity on the force
centre is considered. From the formulas (7.1.23) and (7.1.27) we find the
time of the basis particles fall from the distance $r_{1}>r$ up to $r\geq 0$
in accordance with the clock of the falling particle and in accordance with
the Minkovsky space clock $T$

\bigskip

\bigskip $%
\begin{array}{cc}
\begin{array}{c}
\\ 
\delta \tau =\dfrac{2}{3}\left[ \dfrac{r_{1}}{c}\left( \dfrac{r_{1}}{r_{g}}%
\right) ^{1/2}-\dfrac{r}{c}\left( \dfrac{r}{r_{g}}\right) ^{1/2}\right] , \\ 
\end{array}
& \text{ \ }(7.1.29)%
\end{array}%
$

\bigskip

\bigskip $%
\begin{array}{cc}
\begin{array}{c}
\\ 
\delta {T}=\dfrac{2}{3}\left[ \left( 1+\dfrac{r_{1}}{r_{g}}\right)
^{3/2}-\left( 1+\dfrac{r}{r_{g}}\right) ^{3/2}\right] \dfrac{r_{g}}{c}. \\ 
\end{array}
& \text{ \ }(7.1.30)%
\end{array}%
$

Eqn. (7.1.29) coincides with the result of the Newton theory.It follows from
the formulas (7.1.29), (7.1.30) that the time of the particle fall is finite
for any $r$ from the range $0\leq {r}\leq {r_{1}}$,both in accordance with
the clock of the fallen particle and in accordance with the clock of the
Minkovsky space.

\qquad Usually in GR the time coordinate $t$ including in the Shvartzshild
solution is introduced as a time of the external observer. The connection
between the $t$ coordinate and $T$ time of the Minkovsky space is determined
with the formula

\bigskip

\bigskip $%
\begin{array}{cc}
\begin{array}{c}
\\ 
T=t-\dfrac{1}{c}\dint \left[ \left( 1+\dfrac{r}{r_{g}}\right) ^{1/2}\left( 1-%
\dfrac{r_{g}}{r}\right) -\left( \dfrac{r}{r_{g}}\right) ^{1/2}\right] \left(
1-\dfrac{r_{g}}{r}\right) ^{-1}dr= \\ 
\\ 
t-\dfrac{r_{g}}{c}\left[ \dfrac{2}{3}\left( 1+\dfrac{r}{r_{g}}\right)
^{3/2}-2\left( \dfrac{r}{r_{g}}\right) ^{1/2}\left( 1+\dfrac{1}{3}\dfrac{r}{%
r_{g}}\right) -\ln \left\vert \dfrac{1-\left( \frac{r}{r_{g}}\right) ^{1/2}}{%
1+\left( \frac{r}{r_{g}}\right) ^{1/2}}\right\vert \right] \\ 
\end{array}
& \text{ \ }(7.1.31)%
\end{array}%
$

\bigskip

Substitution of the formula to the interval (7.1.25) forms the Schwarzschild
interval.

The velocity field $dr/dt$ of the Lemetr basis in the Schwarzschild metrics
is connected with the velocity field $dr/dT=v$ (7.1.7) in the Minkovsky
space with the

\bigskip equation

\bigskip $%
\begin{array}{cc}
\begin{array}{c}
\\ 
\dfrac{dr}{dt}=\dfrac{dr}{dT}\dfrac{dT}{dt}=\dfrac{dr}{dT}\left( \dfrac{dT}{%
dt}+\dfrac{dT}{dr}\dfrac{dr}{dt}\right) \\ 
\end{array}
& \text{ \ }(7.1.32)%
\end{array}%
$

\bigskip

Whence using (4.21.31) we find

$%
\begin{array}{cc}
\begin{array}{c}
\\ 
\dfrac{dr}{dt}=\dfrac{\dfrac{dr}{dT}\dfrac{dT}{dt}}{1-\dfrac{dr}{dT}\dfrac{dT%
}{dr}} \\ 
\end{array}
& \text{ \ }(7.1.33)%
\end{array}%
$

\bigskip

that coincides with the \textquotedblleft coordinate\textquotedblright\
parabolic velocity of the free fall in the Shvartzshild field obtained from
the equations for the geodesic. If the \textquotedblleft
coordinate\textquotedblright\ velocity in the Shvartsshild field goes to
zero when approximation to the gravitational radius then the velocities of
the particles in the Minkovsky space in the force field (7.1.28) are always
smaller than the light velocity in the vacuum and their tend to the light
velocity when $r\rightarrow 0,$ and at the gravitational radius $|v|=c/\sqrt{%
2}$.It follows from (4.21.33) that if the external observer uses the time
Shvartsshild coordinate as a time of the removed observer then the
approximation to the gravitational radius demands the infinite value $t$ [7,
60]. The later becomes clear from the form of the formula (4.21.31) when at $%
r\rightarrow r{_{g}},t\rightarrow \infty $ at any finite $T.$ From our view
point $T$ should be taken as the time of the removed observer, $T$ in
accordance with the image construction is the time in the Minkovsky space
and interval (4.21.25) is written in the \textquotedblleft
primary\textquotedblright\ coordinate system where the radial $r,$angle $%
\Theta ,\varphi $ and time $T$ coordinates have evident metric sense and
they determine the interval in the Minkovsky space in the form

\bigskip

\bigskip $%
\begin{array}{cc}
\begin{array}{c}
\\ 
ds^{2}=c^{2}{dT^{2}}-dr^{2}-r^{2}(\sin ^{2}{\Theta }d{\varphi }^{2}+d{\Theta 
}^{2}). \\ 
\end{array}
& \text{ \ }(7.1.34)%
\end{array}%
$

At ${r_{g}}/r\ll {1}$ the interval element (7.1.25) passes to the interval
of the plane space-time (7.1.34). Naturally besides interval (7.1.25) one
can to consider any other coordinate systems but from our view point the
coordinates entering to (7.1.25) coincide with the STR Galilean coordinates
and so they are stood out with their clarity from all other coordinate
systems.

\qquad As is well known when moving the particle in the constant field its
energy is kept $W_{0}$, is the time component of the covariant 4-vector of
the pulse. From (7.1.14), (7.1.24) we have for the basis particles

\bigskip

\bigskip $%
\begin{array}{cc}
\begin{array}{c}
\\ 
W_{0}=m{_{0}}c^{2}U{_{0}}=m_{0}{c^{2}}=m_{0}c{^{2}}(V_{0}+\alpha A{_{0}}).
\\ 
\end{array}
& \text{ \ }(7.1.35)%
\end{array}%
$

\bigskip

whence using (7.1.23), (7.1.24), (7.1.35), (7.1.19) we find

\bigskip

$%
\begin{array}{cc}
\begin{array}{c}
\\ 
\alpha {A_{0}}=1-\left( 1+\dfrac{r_{g}}{r}\right) ^{1/2}, \\ 
\\ 
\alpha {A_{k}}=\dfrac{\partial \Phi }{\partial x^{k}}-V_{k}=\left[ \left( 1+%
\dfrac{r_{g}}{r}\right) ^{1/2}-\left( \dfrac{r}{r_{g}}\right) ^{1/2}-\left( 
\dfrac{r_{g}}{r}\right) ^{1/2}\right] n_{k} \\ 
\end{array}
& \text{ \ }(7.1.36)%
\end{array}%
$

\bigskip

It follows from (7.1.36) that $\alpha {A_{\mu }}V^{\mu }=0,$ that is in
agreement with (7.1.23).

Thus, the solution of the Einstein equations determined the metric $g_{\mu
\nu }$ (7.1.25) in the coordinates of the Minkovsky space, field velocity $%
V_{\mu }$ and vector-potential $A_{\mu }$.

From (7.1.36) we find the tensor of the constant gravitational field $F_{\mu
\nu }$ in the Minkovsky space

\bigskip

\bigskip $%
\begin{array}{cc}
\begin{array}{c}
\\ 
F_{\mu \nu }=\left( \dfrac{{\partial }A_{\nu }}{{\partial }x^{\mu }}-\dfrac{{%
\partial }A_{\mu }}{{\partial }x^{\nu }}\right) ,\qquad F_{kl}=0, \\ 
\\ 
F_{0k}=-\dfrac{{\partial }A_{0}}{{\partial }x^{k}}=-\dfrac{{r_{g}}n_{k}}{%
2\alpha {r^{2}}\sqrt{1+(r_{g}/r)}}. \\ 
\end{array}
& \text{ \ }(7.1.37)%
\end{array}%
$

Similarly to the electrodynamics one can see that the tensor $F_{\mu \nu }$
for the case of the spherical symmetry does not contain the analogue of the
\textquotedblleft magnetic\textquotedblright\ field $\vec{H}$. The intensity
of the gravitational field $E_{k}$ taking into account (7.1.28) has the form

\bigskip $%
\begin{array}{cc}
\begin{array}{c}
\\ 
E_{k}=F_{0k}=\dfrac{N}{m_{0}}n_{k}=-\dfrac{kM{n_{k}}}{r^{2}\left( 1+\dfrac{%
2kM}{c^{2}{r}}\right) ^{1/2}}. \\ 
\end{array}
& \text{ \ }(7.1.38)%
\end{array}%
$

\bigskip Let us introduce the \textquotedblleft induction\textquotedblright\
vector $D_{k}=\varepsilon {E_{k},}$ where

\bigskip $%
\begin{array}{cc}
\begin{array}{c}
\\ 
\varepsilon \equiv -\left( 1+\dfrac{2kM}{c^{2}{r}}\right) ^{1/2}\dfrac{1}{k}%
,\qquad D_{k}=\dfrac{M}{r^{2}}n_{k}. \\ 
\end{array}
& \text{ \ }(7.1.39)%
\end{array}%
$

Thus, for the case of the spherically-symmetrical gravistatic field outside
of the creating mass the expressions are valid

\bigskip

\bigskip $%
\begin{array}{cc}
\begin{array}{c}
\\ 
{\vec{\nabla}}\times {\vec{E}}=0,\qquad {\vec{\nabla}}\cdot {\vec{D}}%
=0,\qquad {\vec{H}}=0. \\ 
\end{array}
& \text{ \ }(7.1.40)%
\end{array}%
$

Whence the energy density of the gravistatic field $\rho $ in analogy with
the electrostatics is calculated in accordance with the formula

\bigskip

$%
\begin{array}{cc}
\begin{array}{c}
\\ 
\rho =\dfrac{ED}{8\pi }=-\dfrac{{k}M^{2}}{8\pi {r^{4}}\left( 1+\dfrac{2kM}{%
c^{2}r}\right) ^{1/2}}. \\ 
\end{array}
& \text{ \ }(7.1.41)%
\end{array}%
$

Note that energy density has no a particularity at the gravitational
radius.Field energy $W$ outside of the sphere with the radius $r_{0}$ is
determined with the equation

\bigskip\ 

$%
\begin{array}{cc}
\begin{array}{c}
\\ 
W=\int\limits_{r_{0}}^{\infty }{\rho }4\pi {r^{2}}dr=-\dfrac{Mc^{2}}{2}\left[
\left( 1+\dfrac{r_{g}}{r_{0}}\right) ^{1/2}-1\right] , \\ 
\end{array}
& \text{ \ }(7.1.42)%
\end{array}%
$

which passes to the Newton expression $W=-(kM^{2})/(2r_{0})$ at $r{_{g}}%
/r\ll 1$.

Calculation of the known GR effects in accordance with the metrics (7.1.25)
connected with the path form results in the same result as in the
Shvartsshild field. The difference reveals in the expressions depending on
the time and on the time derivatives. For the light beams from (7.1.25)
spreading on the radius at $d{\tilde{s}}^{2}=0$ we have:

$%
\begin{array}{cc}
\begin{array}{c}
\\ 
\left( \dfrac{dr}{dT}\right) _{1}=c_{1}(r)=c\left[ 1-\left( \dfrac{r}{r_{g}}%
\right) ^{1/2}\right] \\ 
\\ 
\left[ \left( 1+\dfrac{r}{r_{g}}\right) ^{1/2}\left( \left( \dfrac{r}{r_{g}}%
\right) ^{1/2}-1\right) -\dfrac{r}{r_{g}}\right] ^{-1}, \\ 
\end{array}
& \text{ \ }(7.1.43)%
\end{array}%
$

\bigskip

\bigskip $%
\begin{array}{cc}
\begin{array}{c}
\\ 
\left( \dfrac{dr}{dT}\right) _{2}=c_{2}(r)=c\left[ 1+\left( \dfrac{r}{r_{g}}%
\right) ^{1/2}\right] \\ 
\\ 
\left[ \dfrac{r}{r_{g}}-\left( 1+\dfrac{r}{r_{g}}\right) ^{1/2}\left( \left( 
\dfrac{r}{r_{g}}\right) ^{1/2}+1\right) \right] ^{-1}, \\ 
\end{array}
& \text{ \ }(7.1.44)%
\end{array}%
$

where (7.1.43) corresponds to the velocity of the spreading beams, and
(7.1.44) corresponds to the velocity of the converging ones. At $r<r_{g}$
the expressions (7.1.43), (7.1.44) are negative that is the beams spread
only in one direction inside.

Note that ${c_{1}}(r_{g})=0.$So the time of the light signal spreading from $%
r=r_{g}$ to $r_{0}>r_{g}$ tends to infinity.

\bigskip

${c_{1}}{}|_{r>r_{g}}>0;\quad |{c_{1}}|\leq {c}$ equal sign takes place at $%
r\rightarrow {0};\quad r\rightarrow \infty $.

${c_{1}}{}|_{r>r_{g}}<0;\quad |{c_{1}}|\geq {c}$ equal sign takes place at $%
r\rightarrow {0};\quad r\rightarrow \infty $.

$|{c_{2}}|$ has a maximum at the $r=3r_{g},{c_{2}}\left( 3r_{g}\right)
=c\left( 7+\sqrt{3}\right) /11.$

For converging beams the time of the signal spreading between any $0\leq
r_{1}<\infty $ and $0\leq r_{2}<\infty $ is finite. If $r_{g}/r\ll 1,$ then

\bigskip 

\bigskip $%
\begin{array}{cc}
\begin{array}{c}
\\ 
c_{1}\simeq \left( 1-0.5\left( r_{g}/r\right) ^{1/2}-r_{g}/r\right) c, \\ 
\\ 
c_{2}\simeq -\left( 1-0.5\left( r_{g}/r\right) ^{1/2}-r_{g}/r\right) c. \\ 
\end{array}
& \text{ \ }%
\end{array}%
$

\bigskip

\section{References}

\bigskip

\begin{itemize}
\item \lbrack 1] \ Davtyan O.K.,Theory of Gravitational-Inertial Field of \ 

\ \ \ \ \ \ Universe.I.Gravitational-Inertial Field Equations.

\ \ \ \ \ \ Annalen der Physik.Volume 490 Issue 4, Pages

\ \ \ \ \ \ 247 - 267.

\item \lbrack 2] \ Davtyan O.K.,Theory of Gravitational-Inertial Field of

\ \ \ \ \ \ \ Universe. II.On Critical Systems in Universe. Annalen der

\ \ \ \ \ \ \ Physik.Volume 490 Issue 4, Pages 268 - 280.

\item \lbrack 3] \ Davtyan O.K.,Theory of Gravitational-Inertial Field of

\ \ \ \ \ \ \ Universe. III.The Structure of \ Critical Systems.Annalen

\ \ \ \ \ \ \ der Physik. Volume 491 Issue 3, Pages 217 - 226.

\item \lbrack 4] \ Fock V. A., (1964). "The Theory of Space, Time and

\ \ \ \ \ \ \ Gravitation".Macmillan.

\item \lbrack 5] \ Rosen N., Gen. Relativ. Gravit. 4:435, (1973).

\item \lbrack 6] \ Rosen N., Ann. Phys. (New York) 84, 455,(1974).

\item \bigskip \lbrack 7] \ Rosen N., Gen. Relativity and Gravitation 9,339
(1978).

\item \lbrack 8] \ Katore S.D., Rane R. S., Magnetized cosmological models

\ \ \ \ \ \ \ in bimetric theory of gravitation, Prama. Volume 67, Number 2,

\ \ \ \ \ \ \ August, 2006.

\item \lbrack 9] \ Sahoo P.K., On Kantowski-Sachs Cosmic Strings Coupled with

\ \ \ \ \ \ \ Maxwell Fields in Bimetric Relativity, IJTP. Volume 49, Number
1,

\ \ \ \ \ \ \ January, 2010.

\item \bigskip \lbrack 10] Holland P.R., Electromagnetism, particles and
anholonomy,

\ \ \ \ \ \ Phys. Lett. 91,\ \ (1982),\ 275-278.

\item \lbrack 11] Holland P.R. and Philippidis C., Anholonomic deformations

\ \ \ \ \ \ in the ether: a significance for the electrodynamic potentials,

\ \ \ \ \ \ Quantum Implications, (B.J. Hiley and F.D. Peat, eds.),

\ \ \ \ \ \ Routledge and Kegan Paul,London and New \ York, 1987,

\ \ \ \ \ \ pp.295--311.

\item \bigskip \lbrack 12] Ingarden R.S., Differential geometry and physics,

\ \ \ \ \ \ Tensor, N.S.30, (1976), 201--209.

\item \lbrack 13] Randers G., On an asymmetric metric in the four-spaces of

\ \ \ \ \ \ \ general relativity, Phys. Rev.59, (1941), 195--199.

\item \lbrack 14] Antonelli P.L., Ingarden R.S. and Matsumoto M., The Theory

\ \ \ \ \ \ \ of Sprays and Finsler Spaces with Applications in Physics and

\ \ \ \ \ \ \ Biology,Kluwer Academic Press,\ Dordrecht 1993, pp.350.

\item \lbrack 15] Matsumoto M., Theory of Finsler Spaces with $(\alpha
,\beta )$-metric,

\ \ \ \ \ \ \ \ Rep. Math. Phys.31,\ (1992), 43--83.

\item \lbrack 16] Miron R., General Randers spaces, in Lagrange and Finsler

\ \ \ \ \ \ \ Geometry.\ Application to Physics and Biology, (P. Antonelli
and

\ \ \ \ \ \ \ R. Miron, eds.),Kluwer Academic Press,\ Dordrecht (1996),

\ \ \ \ \ \ 123--140.

\item \lbrack 18] Antonelli P.L., Bucataru I.,On Holland's frame for Randers
space
\end{itemize}

\ \ \ \ \ \ \ \ \ \ \ \ and its applications in physics, Proceedings of the
Colloquium on

\ \ \ \ \ \ \ \ \ \ \ \ Differential \ Geometry, 25--30 July, 2000,
Debrecen, Hungary.

\begin{itemize}
\item \lbrack 19] Amari S., A theory of deformations and stresses of
Ferromagnetic

\ \ \ \ \ \ \ substances by Finsler geometry, RAAG Memoirs of the Unifying

\ \ \ \ \ \ \ Study of Basic Problems in Engineering and Physical Sciences
by \ \ 

\ \ \ \ \ \ \ Means of Geometry (K. Kondo, ed.), vol 3,\ Gakujutsu Bunken, \ 

\ \ \ \ \ \ \ Fukyu-Kai, Tokyo,1962, pp. 193--214.

\item \lbrack 20] Vacaru S.I.,Einstein Gravity, Lagrange--Finsler Geometry,
and

\ \ \ \ \ \ \ \ Nonsymmetric Metrics. SIGMA 4 (2008), 071, 29 pages.

\ \ \ \ \ \ \ \ http://arxiv.org/PS\_cache/arxiv/pdf/0806/0806.3810v2.pdf

\item \lbrack 21] Vacaru S.,Dehnen H., Locally Anisotropic Structures and
Nonlinear

\ \ \ \ \ \ \ \ \ Connections in Einstein and Gauge Gravity, Gen. Rel. Grav.
35

\ \ \ \ \ \ \ \ \ (2003) 209-250.

\item \lbrack 22] Vacaru S. and Goncharenko Yu., Yang-Mills Fields and Gauge
Gravity

\ \ \ \ \ \ \ \ on Generalized Lagrange and Finsler Spaces, Int. J. Theor.
Phys.

\ \ \ \ \ \ \ \ 34 (1995) 1955--1980.\ 

\ \ \ \ \ \ 

\item \lbrack 23]\ Stavrinos P., Gravitational and cosmological
considerations based

\ \ \ \ \ \ \ on the Finsler and Lagrange metric structures.Nonlinear
Analysis

\ \ \ \ \ \ \ 71 (2009) pp.1380-1392.

\item \lbrack 24] \ Changa Z.,Li X.,Modified Friedmann model in
Randers--Finsler

\ \ \ \ \ \ \ space of approximate Berwald type as a possible alternative to

\ \ \ \ \ \ \ dark energy hypothesis. Physics Letters B, Volume 676, Issues

\ \ \ \ \ \ \ 4-5, 8 June 2009, Pages 173-176.

\item \lbrack 25] Broda B., Przanovski M., Electromagnetic field as a
non-linear

\ \ \ \ \ \ \ \ connection.\ Acta Physika Polonica. Vol. B17 (1986).

\item \lbrack 26] Vacaru S.I.,Finsler-Lagrange Geometries and Standard
Theories in \ 

\ \ \ \ \ \ \ Physics:New Methods in Einstein and String Gravity.

\ \ \ \ \ \ \ http://arxiv.org/abs/0707.1524

\item \lbrack 27] Munteanu G., Complex spaces in Finsler, Lagrange, and
Hamilton \ 

\ \ \ \ \ \ geometries.Fundamental Theories of Physics, Vol.141.2004,XI,

\ \ \ \ \ \ 228 p.,

\item \lbrack 28] \ Anastasiei M., (Editor), Antonelli P.L., (Editor)
Finsler and Lagrange

\ \ \ \ \ \ \ \ Geometries.\ Springer.(July 31, 2003) 322p.,

\item \lbrack 29] Janner A., Ascher E., Bravais classes of two-dimensional
relativistic

\ \ \ \ \ \ \ lattices.Physica Volume 45, Issue 1,17 November 1969,

\ \ \ \ \ \ \ Pages 33-66.

\item \lbrack 30] Trz\c{e}sowski A., Effective dislocation lines in
continuously dislocated

\ \ \ \ \ \ \ \ crystals.I. Material anholonomity, arXiv.org e-Print
Archive, 2007:\ 

\ \ \ \ \ \ \ \ http://arxiv.org/abs/0709.1793.

\item \lbrack 31] Trz\c{e}sowski A., Effective dislocation lines in
continuously dislocated

\ \ \ \ \ \ \ crystals.II.Congruences of effective dislocations, arXiv.org
e-Print

\ \ \ \ \ \ \ Archive, 2010:\ http://arxiv.org/abs/0709.1798.

\item \lbrack 32] Trz\c{e}sowski A.,Effective dislocation lines in
continuously dislocated

\ \ \ \ \ \ \ crystals.Part. III. Kinematics, Journal of Technical Physics,
49,

\ \ \ \ \ \ \ 79-99, 2008.

\item \lbrack 33] \ Kleinert H.,Gauge fields in condensed matter, Vol.
II,Part III.

\ \ \ \ \ \ \ Stresses and defects,World Scientific, Singapore, 1989.

\item \lbrack 34] \ Kleinert H., Gauge fields in condensed matter, Vol.
II,Part

\ \ \ \ \ \ \ IV.Differential geometry of defects and gravity with torsion,

\ \ \ \ \ \ \ World Scientific, Singapore,1989.

\item \lbrack 35] \ M\"{o}ller C., The Theory of Relativity, Clarendon
Press, 1955,

\ \ \ \ \ \ \ Science.\ 386 pages.

\ \ \ \ \ \ \ \ \ http://www.archive.org/details/theoryofrelativi029229mbp

\item \lbrack 36] \ Will C.M.,1993, Theory and Experiment in Gravitational
Physics.

\ \ \ \ \ \ \ \ \ Cambridge University Press, Cambridge, 375 pages.

\item \lbrack 37] \ Will C. M., Gravitational radiation from binary systems
in

\ \ \ \ \ \ \ \ alternative metric theories of gravity - Dipole radiation
and the

\ \ \ \ \ \ \ \ binary pulsarAstrophysical Journal, Part 1, vol. 214, June
15,

\ \ \ \ \ \ \ 1977, p. 826-839.

\item \lbrack 38] \ Dobrescu I., Ionescu-Pallas N.,Variational and
Conservative

\ \ \ \ \ \ \ \ Principles in Rosen's Theory of Gravitation, Balkan Journal
of \ 

\ \ \ \ \ \ \ \ Geometry and Its Applications, Vol. 4, No. 1, 1999

\item \lbrack 39] \ Ashby N.,Bertotti B.,1984, Phys. Rev. Lett., 52, 485.

\item \lbrack 40] \ Bertotti B., Grishchuk L.P., 1990, Class. Quantum Grav.,
7, 1733.

\item \lbrack 41] \ Katanaev M.O., Volovich I.V., Theory of defects in
solids and \ 

\ \ \ \ \ \ \ \ three-dimensional gravity, Ann.Phys:216(1992) 1.

\item \lbrack 42] \ Katanaev M.O., Gometric theory of defects,

\ \ \ \ \ \ \ \ \ cond-mat/0407469 (2004).

\item \lbrack 43] Hildebrant S.R, Senovilla M.M, About congruences of
reference.

\ \ \ \ \ \ "Some topics on general relativity and gravitational radiation".

\ \ \ \ \ \ Proceedings of the "Spanish Relativity Meeting'96" Valencia

\ \ \ \ \ \ (Spain), Sep.10-13,1996, 312 pages. Editor S\'{a}ez D.,
Publisher:

\ \ \ \ \ \ Atlantica S\'{e}guier Fronti\'{e}res,1997.ISBN 2863322184.

\item \lbrack 44] Mit\'{s}kevich N.V., Relativistic physics in arbitrary
reference frames.

\ \ \ \ \ \ \ Nova Science Publishers (February 7,2005),166 pages.

\item \lbrack 45] Liosa J.,Soler D., Connections associated with inertial
forces.

\ \ \ \ \ \ "Some topics on general relativity and gravitational radiation".

\ \ \ \ \ \ \ Proceedings of the "Spanish Relativity Meeting'96" Valencia

\ \ \ \ \ \ \ (Spain), Sep.10-13,1996, 312 pages.Editor S\'{a}ez D.,
Publisher:

\ \ \ \ \ \ \ \ Atlantica S\'{e}guier Fronti\'{e}res,1997.ISBN 2863322184.

\item \lbrack 46] \ Barreda M.,Olivert J., L-Rigidity In The Post-Newtonial

\ \ \ \ \ \ \ \ Approximation. "Some topics on general relativity and

\ \ \ \ \ \ \ \ gravitational radiation".Proceedings of the "Spanish
Relativity

\ \ \ \ \ \ \ \ Meeting'96" Valencia (Spain), Sep.10-13,1996, 312 pages.

\ \ \ \ \ \ \ \ Editor S\'{a}ez D., Publisher: Atlantica S\'{e}guier Fronti%
\'{e}res,

\ \ \ \ \ \ \ \ 1997.ISBN 2863322184.

\item \lbrack 47] \ Formiga J. B.,Romero C.,On the differential geometry of
curves

\ \ \ \ \ \ \ \ \ in Minkowski space.American Journal of Physics -- November

\ \ \ \ \ \ \ \ \ 2006 -- Volume 74, Issue 11, pp.1012-1016.

\ \ \ \ \ \ \ \ \ http://arxiv.org/abs/gr-qc/0601002v1

\item \lbrack 48] Kleinert H.,Nonabelian Bosonization as a Nonholonomic

\ \ \ \ \ \ \ \ Transformations from Flat to Curved Field Space.

\ \ \ \ \ \ \ \ http://arxiv.org/abs/hep-th/9606065v1

\item \lbrack 49] \ Fiziev P.,Kleinert H., New Action Principle for
Classical Particle \ 

\ \ \ \ \ \ \ \ \ Trajectories In Spaces with Torsion, (hep-th/9503074).

\item \lbrack 50] \ Fiziev P.,Kleinert H.,Euler Equations for Rigid-Body - A
Case for

\ \ \ \ \ \ \ \ \ Autoparallel Trajectories in Spaces with
Torsion,(hep-th/9503075).

\item \lbrack 51] \ Deser S.,Jackiw R.,'t Hooft G.,Three-Dimensional
Einstein Gravity: \ 

\ \ \ \ \ \ \ \ \ Dynamics of Flat Space.Annals of physics 152. 220-235
(1984).

\item \lbrack 52] \ Liebscher D.-E.,Yourgrau W.,Classical Spontaneous
Breakdown of \ 

\ \ \ \ \ \ \ \ Symmetry and Induction of Inertia. Annalen der Physik.Volume
491

\ \ \ \ \ \ \ \ \ Issue 1, Pages 20 - 24.

\item \lbrack 53] Haisch B.,Rueda A.,Puthoff H. E.,"Inertia as a zero-point
Field

\ \ \ \ \ \ \ Lorentz Force," Phys. Rev. A 49, 678 (1994).

\item \lbrack 54] Bini D.,Cherubini C., Chicone C.,Mashhoon B.,Gravitational

\ \ \ \ \ \ \ induction.Class. Quantum Grav. 25 (2008) 225014,14pp.

\item \lbrack 55] Sciama D.W., On the origin of inertia.Monthly Notices of
the

\ \ \ \ \ \ \ Royal Astronomical Society, Vol. 113, p.34.

\item \lbrack 56] William D. McGlinn, Introduction to relativity.JHU Press,
2003,

\ \ \ \ \ \ \ \ 205 pages.

\item \lbrack 57] Kleyn A., Lorentz Transformation and General Covariance
Principle

\item \lbrack 58] Ortin T.,Gravity and Strings.

\item \lbrack 59] Eckehard W. Mielke, Affine generalization of the Komar
complex of \ 

\ \ \ \ \ \ \ general relativity, Phys. Rev. D 63, 044018 (2001)

\item \lbrack 60] Yu. N. Obukhov and J. G. Pereira, Metric-affine approach
to teleparallel

\ \ \ \ \ \ \ gravity, Phys. Rev. D 67, 044016 (2003), eprint
arXiv:gr-qc/0212080

\ \ \ \ \ \ \ (2002)

\item \lbrack 61] Giovanni Giachetta, Gennadi Sardanashvily, Dirac Equation
in Gauge \ 

\ \ \ \ \ \ \ \ and Affine-Metric Gravitation Theories, eprint
arXiv:gr-qc/9511035 (1995)

\item \lbrack 62] Frank Gronwald and FriedrichW. Hehl, On the Gauge Aspects
of

\ \ \ \ \ \ \ Gravity,eprint arXiv:gr-qc/9602013 (1996)

\item \lbrack 63] Yuval Neeman, Friedrich W. Hehl, Test Matter in a
Spacetime with \ \ \ 

\ \ \ \ \ \ \ \ Nonmetricity,eprint arXiv:gr-qc/9604047 (1996)

\item \lbrack 64] F. W. Hehl, P. von der Heyde, G. D. Kerlick, and J. M.
Nester, General

\ \ \ \ \ \ \ \ relativity with spin and torsion: Foundations and prospects,

\item \lbrack 65] Landau L.D., Lifshits E.M.,The classical theory of fields.
Fourth Edition: \ \ 

\ \ \ \ \ \ \ Volume 2. Course of Theoretical Physics Series.

\item \lbrack 66] Jones P.,Munoz G.,Ragsdale M., Singleton D.,The general
relativistic \ 

\ \ \ \ \ \ \ \ infinite plane.http://arxiv.org/abs/0708.2906v1

\item \lbrack 67] Kluber H.,The determination of Einstein's light-deflection
in the \ \ \ \ \ \ \ \ \ \ 

\ \ \ \ \ \ \ gravitational field of the sun.Vistas in Astronomy,Volume 3,
1960,

\ \ \ \ \ \ \ Pages 47-77.

\item \lbrack 68] Vargashkin V.Ya., LIGHT BEAM PRECESSION EFFECT IN \ 

\ \ \ \ \ \ \ \ CIRCUMSOLAR SPACE, Gravitation \& Cosmology, Vol. 2

\ \ \ \ \ \ \ (1996), No. 2 (6), pp.161--166.

\item \lbrack 69] \ Nieuwenhuizen T. M., Einstein vs Maxwell: Is gravitation
a

\ \ \ \ \ \ \ \ \ curvature of space, a field in flat space, or both?

\ \ \ \ \ \ \ \ \ EPL (Europhysics Letters) Volume 78, Number 1.

\ \ \ \ \ \ \ \ \ http://arxiv.org/abs/0704.0228v1

\item \lbrack 70] \ Podosenov S. A., Space-Time and Classical Fields of
Bound 

\ \ \ \ \ \ \ \ Structures (Sputnikpu blishers, Moscow,2000).
\end{itemize}

\end{document}